\documentclass[prd,aps,superscriptaddress,showpacs,preprintnumbers,amsmath,amssymb,floatfix,12pt,manuscript]{revtex4}
\usepackage[dvips]{graphicx}
\usepackage{ulem}
\preprint{RBRC-742}
\begin{document}
\bibliographystyle{apsrev}

\title{On-shell $\Delta I = 3/2$ kaon weak matrix elements with 
non-zero total momentum}

\author{Takeshi~Yamazaki
\footnote{present address:
Yukawa Institute for Theoretical Physics,
Kyoto University, Kyoto, Kyoto 606-8502, Japan
}}
\affiliation{Physics Department, University of Connecticut, Storrs, 
Connecticut 06269-3046, USA}
\affiliation{RIKEN-BNL Research Center, Brookhaven National Laboratory,Upton, NY 11973}

\collaboration{RBC and UKQCD Collaborations}
\bibliographystyle{apsrev}

\date{
\today
}

\begin{abstract}
We present our results for the on-shell $\Delta I = 3/2$ kaon decay
matrix elements using domain wall fermions and the DBW2 gauge action
at one coarse lattice spacing corresponding to $a^{-1} = 1.31$ GeV
in the quenched approximation.
The on-shell matrix elements are evaluated in 
two different frames:
the center-of-mass frame and non-zero total-momentum frame. 
We employ the formula proposed by Lellouch and L\"uscher in
the center-of-mass frame, and its extension for non-zero total momentum
frame to extract the infinite volume, on-shell, 
center-of-mass frame decay amplitudes.
We determine the decay amplitude at the physical pion mass and momentum
from the chiral extrapolation and an interpolation of the relative 
momentum using the results calculated in the two frames.
We have obtained 
$\mathrm{Re}A_2 = 1.66(23)(^{+48}_{-03})(^{+53}_{-0})\times 10^{-8}$ GeV
and $\mathrm{Im}A_2 = -1.181(26)(^{+141}_{-014})(^{+44}_{-0})\times 10^{-12}$ GeV
at the physical point, using the data at the relatively large pion mass,
$m_\pi > 0.35$ GeV.
The first error is statistic, and the second and third are systematic.
The second error is estimated with several fits of the chiral 
extrapolation including (quenched) chiral perturbation
formula at next to leading order using only lighter pion masses. 
The third one is estimated 
with an analysis using the lattice dispersion relation.
The result of $\mathrm{Re}A_2$ is reasonably consistent with experiment.
\end{abstract}

\maketitle

%
%
%
\section{Introduction}
\label{sec:introduction}

There are longstanding problems in 
non-leptonic kaon decays, such as, 
$\Delta I = 1/2$ selection rule and the ratio of the direct 
and indirect CP violation parameters
$\varepsilon^\prime / \varepsilon$~\cite{Batley:2002gn,AlaviHarati:2002ye},
which need to be resolved for precision tests of the standard model.
The major difficulty is to evaluate the non-perturbative strong
interaction effect.
Although such non-perturbative effect in principle can be quantified by 
lattice QCD, there are many technical difficulties to calculate
the $K \to \pi\pi$ decay process directly on the lattice.
This is because there are problems to deal with the two-pion state 
in finite volume~\cite{Maiani:1990ca}.
To avoid the difficulties, Bernard {\it et al.} proposed one
possible approach, called as the indirect method,~\cite{Bernard:1985wf} where
$K\to\pi\pi$ process is reduced to $K\to\pi$ and 
$K\to 0$ processes through chiral perturbation theory (ChPT).
So far several groups have reported results obtained by this
method~\cite{Bernard:1985tm,Pekurovsky:1998jd,Lellouch:1998sg,Donini:1999nn,Lee:2005sr,Noaki:2001un,Blum:2001xb}.

Especially, it is worth mentioning that CP-PACS~\cite{Noaki:2001un} and RBC~\cite{Blum:2001xb} collaborations 
have independently calculated 
the full weak matrix elements of the $K\to\pi\pi$ decay with the indirect method by using 
domain wall fermion action~\cite{Kaplan:1992bt,Shamir:1993zy,Furman:1994ky}, 
which has good chiral symmetry on the lattice.
Their final results of $\varepsilon^\prime / \varepsilon$, however,
have the opposite sign to that of the experiment.
Indeed, in their calculations there are many systematic uncertainties since 
those calculations have been performed at a single
finite lattice spacing using the indirect method based on tree level ChPT
within the quenched approximation.
The indirect method might cause the larger systematic error
than other sources, {\it i.e.}, the quenching effect and
the scaling violation,
because the final state interaction of the two-pion
is expected to play an important role in this decay process.
Therefore we have to directly calculate the scattering effect of 
the two-pion state on the lattice to eliminate this particular systematic error.
For this purpose we attempt to carry out a calculation with direct method, 
where the two-pion state is properly treated on the lattice, 
in the $\Delta I=3/2$ $K\to\pi\pi$ decay process.

There are two main difficulties for the direct method.
The first problem is that it is hard to extract two-pion state
with relatively large momentum, {\it e.g.}, about $200$ MeV, on the lattice by 
a traditional single exponential analysis. 
The problem was pointed out by Maiani and Testa~\cite{Maiani:1990ca}.
To avoid this problem two groups~\cite{Bernard:1988zj,Aoki:1997ev} calculated 
the two-pion at rest, but
allowing a non-zero energy transfer in the weak operator.
We however cannot obtain physical amplitudes from the calculation, 
unless we use some effective theory, such as ChPT, to extrapolate unphysical amplitudes to physical one.

There are several ideas for solving the problem.
One of the ideas is to employ a proper projection of the $K\to\pi\pi$
four-point functions~\cite{Ishizuka:2002nm}.
In this approach we need complicated calculations and analyses, 
{\it e.g.}, diagonalization of a matrix of 
the two-pion correlation functions~\cite{Luscher:1990ck},
to treat the two-pion state with non-zero relative momentum on the lattice.

A simpler idea, where such complicated analyses are not required,
is to prohibit the two-pion state with zero momentum.
Recently Kim~\cite{Kim:2004mb,Kim:2004mk} reported an exploratory study 
with H-parity (anti-periodic) boundary conditions
in the spatial direction.
The method works to extract the two-pion state with non-zero relative momentum
from the ground state contribution of correlation functions,
because the two-pion state with zero momentum is prohibited by 
the boundary condition.

Alternatively, if we perform the calculation in non-zero total momentum (Lab) frame,
we will be able to forbid the zero momentum two-pion state, which 
certainly appears in center-of-mass (CM) frame.
In the simplest Lab frame the ground state of the two-pion 
is $|\pi(0)\pi(\vec{P})\rangle$ with $\vec{P}$ being non-zero total momentum,
which is related to the two-pion state with the non-zero relative momentum
in the CM frame.
Therefore, we can extract the two-pion state with non-zero momentum from
the ground state contributions~\cite{Rummukainen:1995vs,Boucaud:2004aa,Yamazaki:2004qb}, as well as
in the H-parity boundary case.
In this work we mainly employ the latter, namely the Lab frame method.

The other difficulty of the direct method is the finite volume correction due to
the presence of the two-pion interaction.
We have to estimate this finite volume effect to 
obtain desired matrix elements in the infinite volume,
because the effect is much larger than that of one-particle state.
Lellouch and L\"uscher (LL)~\cite{Lellouch:2000pv} suggested a solution of 
this difficulty,
which is a relation of the on-shell, CM frame decay amplitudes 
in finite and infinite volumes.
However, their derived relation is valid only in the CM frame with
periodic boundary condition in the spatial direction,
so that we need a modified formula when we utilize H-parity
boundary condition~\cite{Kim:2004mk} or Lab frame method.
Recently, two groups, Kim {\it et al.}~\cite{Kim:2005gf} and 
Christ {\it et al.}~\cite{Christ:2005gi}, generalized the formula that 
is an extension 
of the LL formula for the Lab frame calculation.
The generalized formula is based on the finite volume method to extract
the scattering phase shift from the Lab frame calculation~\cite{Rummukainen:1995vs} 
derived by Rummukainen and Gottlieb.
Here we attempt to apply this generalized formula to the calculation of
the $\Delta I = 3/2$ kaon weak matrix elements with 
domain wall fermions
and quenched DBW2 gauge action~\cite{Takaishi:1996xj,deForcrand:1999bi} at a 
single coarse lattice spacing.

It is worth noting that there are some (dis)advantages of 
the two methods, the Lab frame
and H-parity boundary condition calculations.
In the H-parity method, the statistical error of the matrix elements
is smaller than those in the Lab calculation when same simulation parameters,
{\it e.g.}, lattice size and quark mass, are utilized.
This is because the unit of the momentum in the H-parity method 
is a half of that 
in the Lab frame calculation due to the anti-periodic boundary condition in the spatial
direction.
However, the H-parity method is not effective in the calculation of
$\Delta I = 1/2$ $K\to\pi\pi$ decay, because under the H-parity boundary condition
only $\pi^\pm$ states satisfy the anti-periodic boundary condition~\cite{Kim:2004mb,Sachrajda:2004mi}.
The $\pi^0$ state does not satisfy the boundary condition, unless
we introduce another discrete boundary condition, such as G-parity boundary condition~\cite{Kim:2003xt}.
On the other hand, the Lab frame method does not 
break the isospin symmetry, so that we can apply 
the method to the calculation of $\Delta I = 1/2$ channel
as well as that of $\Delta I = 3/2$.

The organization of the article is as follows.
In Sec.~\ref{sec:methods} we give a brief explanation of the generalized formula as well as
the original LL formula.
We also explain the calculation method of the correlation functions, from which
we evaluate the decay amplitudes.
The parameters of our simulation are given in Sec.~\ref{sec:simu_para}.
We show the result of the $I=2$ two-pion scattering length and phase shift 
in Sec.~\ref{sec:results}.
Then we present the results for the off-shell and on-shell decay amplitudes,
weak matrix elements, Re$A_2$, and Im$A_2$.
Finally, we briefly summarize this work in Sec.~\ref{sec:conclusions}.
Preliminary result of this work was presented in 
Refs.~\cite{Yamazaki:2005eg,Yamazaki:2006ce}.

%
%
%
\section{Methods}
\label{sec:methods}

%
%
%
\subsection{LL formula}
\label{subsec:LL_formula}

Let us briefly review the method suggested by 
Lellouch and L\"uscher~\cite{Lellouch:2000pv} (LL).
They derived a formula, which connects a CM
decay amplitude $|A|$ defined in infinite volume
to the one given on finite volume $|M|$ as,
\begin{equation}
| A |^2 = 8\pi \left(\frac{E_{\pi\pi}}{p}\right)^3
\left\{ p \frac{ \partial \delta(p) }{ \partial p }
      + q \frac{ \partial \phi(q) }{ \partial q }
\right\} | M |^2,
\label{eq:CM_Formula}
\end{equation}
where $E_{\pi\pi}$ is the two-pion energy in the CM frame
and the scattering phase shift 
$\delta$ is responsible for the $\pi\pi$ final state interactions.
The relative momentum of the two pions $p$ is 
determined from measured $E_{\pi\pi}$,
\begin{equation}
p^2 = E_{\pi\pi}^2/4 - m_\pi^2,
\label{eq:relative_mom}
\end{equation}
and its normalized momentum $q$ is defined by
\begin{equation}
q^2 = ( p L / 2\pi )^2.
\label{eq:relative_mom_q}
\end{equation}
with the spatial extent $L$. 
The function $\phi(q)$, derived by L\"uscher~\cite{Luscher:1990ux},
is defined by
\begin{equation}
\tan \phi(q) = 
- \frac{q \pi^{3/2}}{Z_{00}(1;q^2)},
\end{equation}
where 
\begin{equation}
Z_{00}(1;q^2) = 
\frac{1}{\sqrt{4\pi}}
\sum_{\vec{n}\in \mathbb{Z}^3}\frac{1}{\vec{n}^2-q^2}.
\label{eq:phi}
\end{equation}
Using the function $\phi(q)$, we can also determine the scattering phase shift
at the momentum $p$ through the relation
\begin{equation}
\delta(p) = n \pi - \phi(q),
\label{eq:finite_vol_method_L}
\end{equation}
where $n$ is an integer, and $\phi(0) = 0$.
Note that the formula eq.~(\ref{eq:CM_Formula}) is valid only for on-shell decay amplitude,
{\it i.e.}, $E_{\pi\pi} = m_K$.

%
%
%
\subsection{Extended LL formula for non-zero total momentum frame}
\label{subsec:LL_formula_non-zero_total_mom}

Next, we describe the extended LL formula for the 
the non-zero total momentum frame 
(denoted in the following by the Lab frame).
Recently two groups~\cite{Kim:2005gf,Christ:2005gi} have derived a formula,
which connects a CM decay amplitude $|A|$ in the infinite volume 
to a Lab frame decay amplitude $|M^P|$ given on finite volume
\begin{equation}
| A |^2 = 8\pi \gamma^2 \left(\frac{E_{\pi\pi}}{p}\right)^3
\left\{ p \frac{ \partial \delta (p)}{ \partial p }
      + q \frac{ \partial \phi_{\vec{P}} (q)}{ \partial q }
\right\} | M^P |^2,
\label{eq:Lab_Formula}
\end{equation}
where $\gamma$ is a boost factor given by
\begin{eqnarray}
\gamma &=& E^P_{\pi\pi}/E_{\pi\pi},
\end{eqnarray}
with $E^P_{\pi\pi}$ being the Lab frame value, which is determined by 
two-pion energy with the total momentum ${\vec P}$.
The total CM frame energy of two-pion states $E_{\pi\pi}$ is evaluated through the energy-momentum
conservation,
\begin{equation}
E_{\pi\pi}^2 = (E^P_{\pi\pi})^2 - P^2.
\label{eq:boost}
\end{equation}
In~eq.(\ref{eq:Lab_Formula}), 
$p$ and $q$ are defined as the same in eqs.(\ref{eq:relative_mom})
and (\ref{eq:relative_mom_q}),
except the determination of $E_{\pi\pi}$, which is given by eq.(\ref{eq:boost}).
In the following the momentum denoted by the capital letter like $P$ represents
the total momentum, while the small one $p$ represents the relative momentum,
unless explicitly indicated otherwise.
In addition quantities with a superscript $P$, {\it e.g.} $E^P_{\pi\pi}$, indicate the Lab frame values.
The function $\phi_{\vec{P}}(q)$, derived by Rummukainen and Gottlieb~\cite{Rummukainen:1995vs},
is given by
\begin{equation}
\tan \phi_{\vec{P}}(q) = 
- \frac{\gamma q\pi^{3/2}}{Z^{\vec{P}}_{00}(1;q^2;\gamma)},
\end{equation}
where 
\begin{equation}
Z_{00}^{\vec{P}}(1;q^2;\gamma) = 
\frac{1}{\sqrt{4\pi}}
\sum_{\vec{n}\in \mathbb{Z}^3}\frac{1}{n_1^2+n_2^2+\gamma^{-2}(n_3+1/2)^2-q^2},
\label{eq:phi_P}
\end{equation}
in the case for $\vec{P} = (0, 0, 2\pi/L)$.
The formula given in eq.~(\ref{eq:Lab_Formula}) is valid only for the case
$E^P_{\pi\pi} = E_K^P$ where $E_K^P$ is kaon energy with the momentum $\vec{P}$, in other words,
$E_{\pi\pi} = m_K$ as in the LL formula given in eq.(\ref{eq:CM_Formula}).
When we set $P^2 = 0$ and $\gamma = 1$, the formula eq.(\ref{eq:Lab_Formula}) reproduces the original LL formula 
eq.~(\ref{eq:CM_Formula}).
Of course,
we can determine the scattering phase shift at the momentum $p$ by using $\phi_{\vec{P}}(q)$ as
\begin{equation}
\delta(p) = n \pi - \phi_{\vec{P}}(q),
\label{eq:finite_vol_method_RG}
\end{equation}
where $n$ is an integer, and $\phi_{\vec{P}}(0) = 0$
as the same in eq.(\ref{eq:finite_vol_method_L}).

%
%
%
\subsection{Calculation of $K\to\pi\pi$ four-point function}
\label{subsec:calc_four-point_func}

We calculate four-point functions for the $\Delta I = 3/2$ $K\to\pi\pi$ decay
in zero (CM) and non-zero (Lab) total momentum frames.
In actual simulations, we employ $\vec{P} = (0, 0, 2\pi/L)$ for the non-zero total momentum.
The four-point functions in the CM and Lab frames are defined by
\begin{eqnarray}
G_i(t,t_\pi,t_K) &=& 
\langle 0 | 
\pi^+\pi^0(\vec{0},t_\pi) O^{3/2}_i(t) [K^+(\vec{0},t_K)]^\dagger 
| 0 \rangle,
\label{eq:kpipi4point_CM}\\
G_i^P(t,t_\pi,t_K) &=& 
\langle 0 | 
\pi^+\pi^0(\vec{P},t_\pi) O^{3/2}_i(t) [K^+(\vec{P},t_K)]^\dagger 
| 0 \rangle,
\label{eq:kpipi4point_Lab}
\end{eqnarray}
where the two-pion operator in $G_i^P$ is averaged with the following 
two types of products of pion operators
in order to extract the $I=2$ part of the two-pion state,
\begin{equation}
\pi^+\pi^0(\vec{P},t_\pi) = ( \pi^+(\vec{P},t_\pi)\pi^0(\vec{0},t_\pi)
                            + \pi^+(\vec{0},t_\pi)\pi^0(\vec{P},t_\pi) ) / 2.
\end{equation}
$O^{3/2}_i$ are lattice operators entering
$\Delta I = 3/2$ weak decays:
\begin{eqnarray}
O^{3/2}_{27} &=& 
\sum_{\vec x}\left\{
(\overline{s}^a({\vec x})d^a({\vec x}))_L
\left[(\overline{u}^b({\vec x})u^b({\vec x}))_{L}-(\overline{d}^b({\vec x})d^b({\vec x}))_{L}\right]
       + (\overline{s}^a({\vec x})u^a({\vec x}))_L(\overline{u}^b({\vec x})d^b({\vec x}))_{L}
\right\},\nonumber\\
\label{eq:operator_27}
\\
O^{3/2}_{88}  &=& 
\sum_{\vec x}\left\{
(\overline{s}^a({\vec x})d^a({\vec x}))_L
\left[(\overline{u}^b({\vec x})u^b({\vec x}))_{R}-(\overline{d}^b({\vec x})d^b({\vec x}))_{R}\right]
       + (\overline{s}^a({\vec x})u^a({\vec x}))_L(\overline{u}^b({\vec x})d^b({\vec x}))_{R}
\right\},\nonumber\\
\label{eq:operator_88}
\\
O^{3/2}_{m88}  &=& 
\sum_{\vec x}\left\{
(\overline{s}^a({\vec x})d^b({\vec x}))_L
\left[(\overline{u}^b({\vec x})u^a({\vec x}))_{R}-(\overline{d}^b({\vec x})d^a({\vec x}))_{R}\right]
       + (\overline{s}^a({\vec x})u^b({\vec x}))_L(\overline{u}^b({\vec x})d^a({\vec x}))_{R}
\right\}\nonumber\\
\label{eq:operator_m88}
\end{eqnarray}
where $(\overline{q}q)_L = \overline{q}\gamma_\mu(1-\gamma_5)q$,
$(\overline{q}q)_R = \overline{q}\gamma_\mu(1+\gamma_5)q$,
and $a,b$ are color indices.
The kaon weak decay operators are classified into the (8,1), (27,1), and (8,8) 
representations of $SU(3)_L\otimes SU(3)_R$,
but the $\Delta I = 3/2$ part has only the (27,1) and (8,8) representations.
$O^{3/2}_{27}$ and $O^{3/2}_{88}$
are the operators in the (27,1) and (8,8) representations with $I=3/2$, respectively.
$O^{3/2}_{m88}$ equals $O^{3/2}_{88}$ with its color summation
changed to cross the two currents.

We employ the momentum projection source for the quark operator with the Coulomb gauge fixing
to obtain better overlap to a state with each momentum.
The pion operators with the source are constructed as,
\begin{eqnarray}
\pi^+({\vec k}, t) &=& 
[ \sum_{\vec x}\overline{d}({\vec x},t) e^{-i{\vec k}_1 \cdot {\vec x}} ]
\gamma_5
[\sum_{\vec y}u({\vec y},t) e^{-i{\vec k}_2 \cdot {\vec y}} ]
\label{eq:mom_wall}\\
\pi^0({\vec k}, t) &=& 
\left( 
[ \sum_{\vec x}\overline{u}({\vec x},t) e^{-i{\vec k}_1 \cdot {\vec x}}]
\gamma_5
[\sum_{\vec y}u({\vec y},t) e^{-i{\vec k}_2 \cdot {\vec y}} ]
-
[ \sum_{\vec x}\overline{d}({\vec x},t) e^{-i{\vec k}_1 \cdot {\vec x}}]
\gamma_5
[\sum_{\vec y}d({\vec y},t) e^{-i{\vec k}_2 \cdot {\vec y}} ]
\right)/\sqrt{2},\nonumber\\
\label{eq:mom_wall_0}
\end{eqnarray}
with ${\vec k} = {\vec k}_1 + {\vec k}_2 = \vec{P}$ or $\vec{0}$,
where the momentum ${\vec k}$ and ${\vec k}_i$ represent 
the momentum of the pion and each quark, respectively.
In the zero momentum case, these operators are nothing but the wall source operator.
The $K^+$ operator with each momentum is calculated in the same way with
changing $\overline{d}$ to $\overline{s}$ in eq.(\ref{eq:mom_wall}).

The four-point function in the Lab frame is much noisier than that in the CM frame.
In order to improve the statistics,
we calculate the four-point function $G_i^P(t,t_\pi,t_K)$ with two possible momentum insertions,
${\vec k}_1 = {\vec 0}, {\vec k}_2 = {\vec P}$ and ${\vec k}_1 = {\vec P}, {\vec k}_2 = {\vec 0}$ 
in eqs.(\ref{eq:mom_wall}) and (\ref{eq:mom_wall_0}) for the pion operators,
and then we average them on each  configuration.
On the other hand, we fix the momentum of the kaon operator 
in the four-point function as ${\vec k}_1 = {\vec P}, {\vec k}_2 = {\vec 0}$.

We also calculate four-point function for the $I=2$ two-pion and the 
two-point function for the kaon and pion 
with zero momentum,
\begin{eqnarray}
G_{\pi\pi}(t,t_\pi) &=& \langle 0 | \pi^+\pi^+(\vec{0},t) [\pi^+\pi^+(\vec{0},t_\pi)]^\dagger | 0 \rangle,\\
G_K(t,t_K) &=& \langle 0 | K^+(\vec{0},t) [K^+(\vec{0},t_K)]^\dagger | 0 \rangle,\\
G_\pi(t,t_\pi) &=& \langle 0 | \pi^+(\vec{0},t) [\pi^+(\vec{0},t_\pi)]^\dagger | 0 \rangle,
\end{eqnarray}
and with non-zero total momentum,
\begin{eqnarray}
G_{\pi\pi}^P(t,t_\pi) &=& \langle 0 | \pi^+\pi^+(\vec{P},t) [\pi^+\pi^+(\vec{P},t_\pi)]^\dagger | 0 \rangle,\\
G_K^P(t,t_K) &=& \langle 0 | K^+(\vec{P},t) [K^+(\vec{P},t_K)]^\dagger | 0 \rangle,\\
G_\pi^P(t,t_\pi) &=& \langle 0 | \pi^+(\vec{P},t) [\pi^+(\vec{P},t_\pi)]^\dagger | 0 \rangle.
\end{eqnarray}
We employ the operator
with ${\vec k}_1 = {\vec 0}, {\vec k}_2 = {\vec P}$ in eq.(\ref{eq:mom_wall}) for 
$G_\pi^P(t,t_\pi)$ and $G_{\pi\pi}^P(t,t_\pi)$,
while with ${\vec k}_1 = {\vec P}, {\vec k}_2 = {\vec 0}$ in the kaon operator of $G_K^P(t,t_K)$.
At the sink $t$ of 
$G_{\pi\pi}^{(P)}, G_K^{(P)}$, and $G_\pi^{(P)}$, we use the same operators as 
the source operator, 
and also the momentum projected operator for the meson field.
The $\pi^+(\vec{k},t)$ field with the latter operator is given by
\begin{equation}
\sum_{\vec x}\overline{d}({\vec x},t) 
\gamma_5
u({\vec x},t) e^{-i{\vec k} \cdot {\vec x}}
\label{eq:mom_point}
\end{equation}
with $\vec{k} = \vec{P}$ or $\vec{0}$.
In the zero momentum case, the operator corresponds to the point sink operator of the pion.
We shall call the symmetric correlator under exchange of the source and 
sink operators as ``wall'' sink correlator,
while the one calculated by the different source and sink operators as ``point'' sink correlator.
We determine the energy for each ground state from the point sink correlator,
and the amplitude from the wall sink operator.

We use linear combinations of quark propagators with
the periodic and anti-periodic boundary conditions in the temporal direction
to have a quark propagator with $2T$ periodicity in the time direction.

%
%
%
\section{Simulation parameters}
\label{sec:simu_para}

Our simulation is carried out in quenched lattice QCD employing a renormalization group improved
gauge action for gluons,
\begin{equation}
  S_G[U] = \frac{\beta}{3}
          \left[(1-8\,c_1) \sum_{x;\mu<\nu} P[U]_{x,\mu\nu}
        + c_1 \sum_{x;\mu\neq\nu} R[U]_{x,\mu\nu}\right]
\end{equation}
where $P[U]_{x,\mu\nu}$ and $R[U]_{x,\mu\nu}$ represent the real part of the trace of the path
ordered product of links around the $1\times 1$ plaquette and $1\times 2$ rectangle, 
respectively, in the $\mu,\nu$ plane at the point $x$ and $\beta \equiv 6/g^2$ with
$g$ being the bare coupling constant.  For the DBW2 gauge action~\cite{Takaishi:1996xj,deForcrand:1999bi}, the
coefficient $c_1$ is chosen to be $-1.4069$, using a renormalization group flow for 
lattices with $a^{-1} \simeq 2$ GeV~\cite{Takaishi:1996xj,deForcrand:1999bi}.
Gauge configuration was previously generated~\cite{Aoki:2002vt} at $\beta = 0.87$ with
the heat bath algorithm and the over-relaxation algorithm mixed in the ratio 1:4.
The combination is called a sweep and physical quantities are measured every 200 sweeps.

We employ the domain wall fermion action~\cite{Kaplan:1992bt,Shamir:1993zy,Furman:1994ky}
with the domain wall height $M=1.8$ and the fifth dimension length $L_s = 12$.
The residual mass is reasonably small in these parameters, 
$m_{\mathrm{res}} = 0.00125(3)$~\cite{Aoki:2006ib}.
Our conventions for the domain wall fermion operator are given in Ref.~\cite{Blum:2000kn}.
The inverse lattice spacing is 1.31(4) GeV~\cite{Aoki:2002vt} determined by the $\rho$ meson mass.
The lattice size is $L^3 \times T = 16^3 \times 32$, 
where the physical spatial extent corresponds to about 2.4 fm.

We fix the two-pion operator at $t_\pi = 0$,
while we employ three source points $t_K = 16, 20$ and $25$ 
for the kaon operator to investigate $t_K$ dependence of 
the statistical error of the Lab frame decay amplitude, 
and to check the consistency of these results.
We employ four $u,d$ quark masses, $m_u = 0.015, 0.03, 0.04$ and 0.05
corresponding to $m_\pi = 0.354(2), 0.477(2), 0.545(2)$ and $0.606(2)$ GeV,
for the chiral extrapolation of the decay amplitudes.

To utilize the LL formula eq.(\ref{eq:CM_Formula}) and the extended formula eq.(\ref{eq:Lab_Formula}), 
it is important to obtain a decay amplitude at on-shell, where the energies of the initial and final
states are equal.
In order to obtain the decay amplitude,
we vary the kaon energy with several strange quark masses at a fixed light quark mass.
Then we carry out an interpolation of the amplitudes to the on-shell point.
For the interpolations,
six strange quark masses are employed, $m_s = 0.12, 0.18, 0.24, 0.28, 0.35$ and 0.44,
except the lightest $m_u$ for the $t_K = 16$ and 20 cases
where we use the three lighter $m_s$.
We will see that the three strange quark masses are enough for the interpolation of the case in a later section.

In our simulation we employ relatively heavy strange quark masses to obtain
the kaon energy closely satisfying to the on-shell condition, $E_{\pi\pi} \sim m_K$. 
This might cause a systematic error, because lower modes of
the hermitian dirac operator would not be well separated to left- and right-hand in a heavy quark mass
as discussed in Ref.~\cite{Lin:2006vc}.
The systematic error can be removed 
by simulations with lighter pion mass, larger volume,
and finer lattice spacing~\cite{Lin:2006vc}.
However, we will not estimate such effect in this work, which may be accepted as an exploratory work.  
Thus, it should be reminded that our result might suffer from above mentioned systematic error at least at the heavier strange quark masses.

The LL formula and its extension requires a relatively
large spatial volume.
This requirement stems from both the 
on-shell condition and size of the scattering range $R$.
$R$ is defined by 
\begin{equation}
V(\vec{r}) = 0\ \mathrm{at}\ |\vec{r}| \ge R,
\end{equation}
where $V(\vec{r})$ is the effective potential of the two-pion scattering
at the relative coordinate $\vec{r}$ of the two pions.
The former is much essential here. The on-shell condition requires 
a large spatial extent, {\it e.g.,}
$L = 6$(3.2) fm for the CM(Lab) frame, which is evaluated 
at the physical pion and kaon masses. 
The required spatial size, indeed, changes
in accordance with simulated masses of the pion and kaon states,
as $L^2 = 4\pi^2/(m_K^2/4-m_\pi^2)$ for the CM frame and
$L^2 = 4\pi^2/(m_K^2/4-m_\pi^2)\cdot m_\pi^2/m_K^2$ for the Lab frame.
Thus, it is possible to fulfill the on-shell condition on
a smaller volume than the one required at the physical point.
In this work we tune $m_K$ at the simulated $m_\pi$ and the fixed $L$ to
satisfy the on-shell condition in each frame as we explained above.

Another constraint on the spatial volume is determined from
the scattering range $R$, because $R < L / 2$ is an important assumption~\cite{Luscher:1990ux,Rummukainen:1995vs} to derive the two relations 
eqs.(\ref{eq:finite_vol_method_L}) and (\ref{eq:finite_vol_method_RG}),
where we can evaluate the scattering phase 
through the two-pion energy on finite volume. 
The LL formula and its extension are based on those
relations. If the assumption is not satisfied, 
the scattering effect is contaminated by unwanted finite spatial size effect, in
other words, the effective potential is distorted by the boundary condition.
The determinations of $R$ were carried out numerically in both the
CM~\cite{Aoki:2005uf} and Lab~\cite{Sasaki:2008sv} frame calculations in the $I=2$ two-pion channel.
These references reported that the required spatial extent was 
estimated as $L = 2.4$--3.2 fm in the range of $m_\pi = 0.42$--0.86 GeV.
The spatial volume $\sim(2.4\;{\rm fm})^3$ in this calculation is not fully
justified in this sense.
Thus, we simply assume that effects stemming from the distortion of the effective potential are small or negligible in our simulation.

We use 252 configurations in the case of $t_K=16$ and 20, 
except at the lightest pion mass
where 371 configurations are employed in order to improve statistics,
while we employ 100 configurations in the case of $t_K=25$ at all 
the quark masses.
The numbers of the gauge configuration used in our simulations 
are summarized in table~\ref{tab:sim_para}.

In the following analysis we evaluate the statistical error of
all the measured quantities by the single elimination jackknife method.
For chiral extrapolations in the $t_K=16$ and 20 cases, 
we employ the modified jackknife method~\cite{AliKhan:2001tx}
to take into account the different numbers of the configurations
between in the lightest and other heavier $u,d$ quark masses.

%
%
%
\section{Results}
\label{sec:results}

%
%
%
\subsection{Physical quantities for $I=2$ $\pi\pi$ scattering}
\label{subsec:pipi_scattering}

The calculation of the two-pion scattering is important not only to employ the LL type methods,
but also to understand hadronic scattering from lattice QCD.
So far many groups calculated the scattering length of 
the S-wave $I=2$ $\pi\pi$ channel~\cite{Sharpe:1992pp,Gupta:1993rn,Kuramashi:1993ka,Fukugita:1994ve,Alford:2000mm,Juge:2003mr,Gattringer:2004wr,Aoki:2002ny,Aoki:2005uf}
with the finite volume method~\cite{Luscher:1986pf,Luscher:1990ux}.
Some of the works obtained the result in the continuum limit~\cite{Aoki:2002in,Liu:2001ss,Du:2004ib,Li:2007ey,Yamazaki:2004qb}.
Recently, the result using domain wall valence quark on the 2+1 flavor 
improved staggered sea quark was reported by 
Beane {\it et al.}~\cite{Beane:2005rj,Beane:2007xs}.
There are a few papers to calculate the $I=2$ $\pi\pi$ scattering phase shift,
with the finite volume method and its extension~\cite{Rummukainen:1995vs},
at only a single lattice spacing~\cite{Aoki:2002ny,Kim:2004mb,Kim:2004mk,Sasaki:2008sv} and in the continuum limit with quenched 
approximation~\cite{Li:2007ey} and two-flavor dynamical
quark effect~\cite{Yamazaki:2004qb}.
It is worth remarking that 
there is the recent work for the $I=1$ $\pi\pi$ scattering phase shift with
$\rho$ meson resonance~\cite{Aoki:2007rd}.
In this section we present our results of 
the $I=2$ $\pi\pi$ scattering length and scattering phase shift.

%
%
\subsubsection{Scattering length}
\label{subsucsec:a_0}

We evaluate the scattering phase shift $\delta(p)$ through 
the finite volume method eq.(\ref{eq:finite_vol_method_L})
of L\"uscher~\cite{Luscher:1990ux} in the CM calculation.
In the CM calculation we extract the two-pion energy $E_{\pi\pi}$ to fit
the point sink two-pion correlator using the fit form~\cite{Kim:2004mk},
\begin{equation}
A \cdot 
\left(
e^{-E_{\pi\pi} t} + e^{-E_{\pi\pi}(2T-t)}
+
C
\right).
\label{eq:CM_two_pi_ene_fit}
\end{equation}
We should notice that we employ linear combinations of quark propagators with
the periodic and anti-periodic boundary conditions in the temporal direction,
so that the periodicity of the correlators is $2T$.
The fitting analysis is carried out with 
three parameters $A$, $C$, and $E_{\pi\pi}$ in the fit range of $t = $ 6--31.
The constant $C$ stems from two pions propagating opposite way
in the temporal direction due to the periodic boundary condition.

Using the finite volume method eq.(\ref{eq:finite_vol_method_L}),
we obtain the scattering phase shift from the relative momentum which is determined
from $E_{\pi\pi}$ through eq.(\ref{eq:relative_mom}).
The results for $E_{\pi\pi}$, $p$, and $\delta(p)$
in the CM calculation are summarized in table~\ref{tab:twopi_CM}.

\begin{figure}[!h]
\centering
\scalebox{0.5}[0.5]{
\includegraphics*{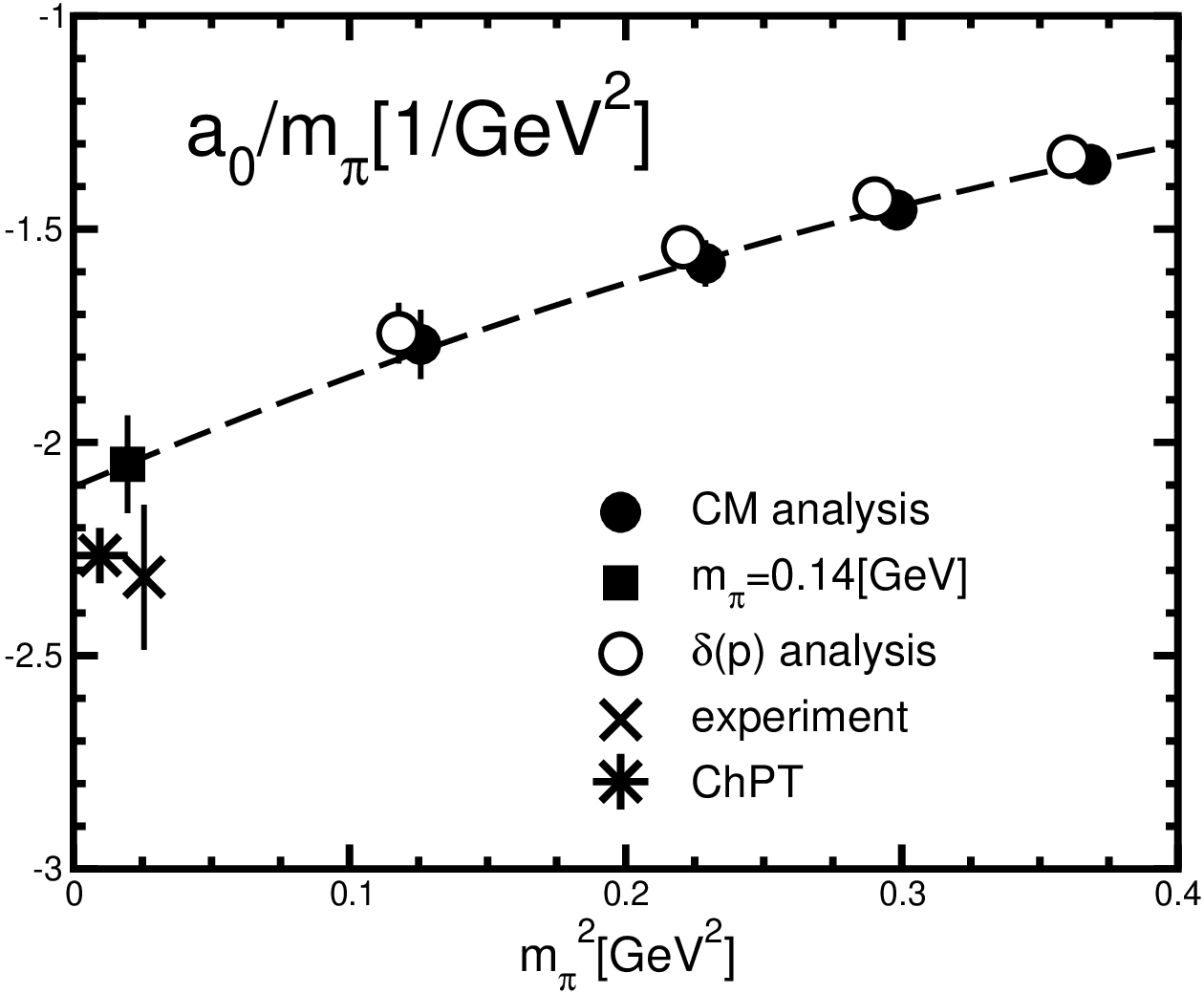}
}
\caption{
Scattering length over pion mass and its chiral extrapolation.
Star and cross symbols are prediction of ChPT and experimental result, respectively.
Open circles are obtained from analysis of scattering phase shift,
and are slightly shifted to minus direction in x-axis.
}
\label{fig:a0}
\end{figure}

We estimate the scattering length $a_0$ with the 
measured $\delta(p)$ by using an assumption,
\begin{equation}
a_0 \approx \frac{ \delta( p ) }{ p },
\end{equation}
where the scattering length is defined by
\begin{equation}
a_0 = \lim_{p\to 0} \frac{ \delta( p ) }{ p }.
\end{equation}
This assumption is valid in our CM calculation, because $p$ is small
enough.
The measured value of $a_0 / m_\pi$ is presented in table~\ref{tab:twopi_CM}.
Figure~\ref{fig:a0} shows that the measured value of $a_0 / m_\pi$, 
denoted by ``CM analysis'', has an appreciable curvature
for the pion mass squared.
The scattering length $a_0 / m_\pi$ at the physical pion mass,
$m_\pi = 140$ MeV, 
is determined by a chiral extrapolation with the fit form,
\begin{equation}
A + B m_\pi^2 + C m_\pi^4,
\end{equation}
whose result is summarized in table~\ref{tab:fit_a0}.
In Fig.~\ref{fig:a0}, the result at the physical pion mass 
is compared with the experiment~\cite{Pislak:2003sv}
and a prediction of ChPT~\cite{Colangelo:2001df}.

We also fit the data using the prediction of the next to leading order (NLO) 
ChPT~\cite{Gasser:1983yg},
\begin{equation}
-\frac{1}{8\pi f^2}\left[ 1 - \frac{m_\pi^2}{8 \pi^2 f^2}\left(  l(\mu) 
- c_l \log\left(\frac{m_\pi^2}{\mu^2}\right)
\right)\right],
\label{eq:a0_chpt}
\end{equation}
where $\mu$ is scale, $c_l = 7/2$, and $l(\mu)$ is a low energy constant.
$f$ is the pion decay constant at the chiral limit.
The scale is fixed at $\mu = 1$ GeV for simplicity.
We cannot obtain a reasonable $\chi^2/$d.o.f. from a fit with one free parameter $l(\mu)$,
where we use $f=0.133$ GeV~\cite{Aoki:2006ib}.
As summarized in table~\ref{tab:fit_a0_chpt},
a three-parameter fit with $f$, $l(\mu)$, and $c_l$, gives
a reasonable value of $\chi^2/$d.o.f.
While in the fit $f$ is consistent with the one at the chiral limit~\cite{Aoki:2006ib},
the coefficient $c_l$ differs from the prediction of ChPT. 
Although the quenched ChPT formula~\cite{Bernard:1995ez,Colangelo:1997ch} is also available,
the quality of the fit is similar to what we obtained with eq.(\ref{eq:a0_chpt}).
At the physical pion mass the ChPT fit result agrees with the simple polynomial fit in the above,
so that we choose the polynomial one in the following analysis.

%
%
\subsubsection{Scattering phase shift}
\label{subsucsec:delta_0}

The two-pion energy in the Lab frame is noisier than the one obtained in the CM calculation,
so that we determine the two-pion energy in the Lab frame $E_{\pi\pi}^P$ in the following way 
to reduce the statistical error.
First, we fit the point sink two-pion correlator with the non-zero total momentum
using the fit form~\cite{Boucaud:2004aa},
\begin{equation}
A \cdot 
\left(
e^{-W t} + e^{-W(2T-t)}
+
C\left[
e^{-(E_\pi^P-m_\pi)t} + e^{-(E_\pi^P-m_\pi)(2T-t)}
\right]
\right),
\end{equation}
where $E_\pi^P$ is the measured value of the single pion energy with 
the momentum.
The terms multiplied by $C$ come from two pions propagating in different temporal direction
due to periodic boundary condition as in eq.(\ref{eq:CM_two_pi_ene_fit}).
The fit is carried out with three parameters $A$, $C$, and $W$
in the range of $t=$ 6--31.
Using the parameter $W$, we determine the energy shift 
$\Delta E^P_{\pi\pi} = W - (E_\pi^P+m_\pi)$
from the non-interacting two-pion energy.
Finally we reconstruct $E_{\pi\pi}^P$ using $m_\pi$, $\Delta E^P_{\pi\pi}$, and the total
momentum squared $P^2$
\begin{equation}
E_{\pi\pi}^P = \Delta E^P_{\pi\pi} + m_\pi + \sqrt{m_\pi^2 + P^2}.
\end{equation}

We determine the CM frame, two-pion energy $E_{\pi\pi}$
from $E_{\pi\pi}^P$ with eq.~(\ref{eq:boost}).
We also determine the relative momentum $p$ from $E_{\pi\pi}$ as in the CM case through eq.(\ref{eq:relative_mom}),
and then obtain the scattering phase shift by the extension of the finite volume method
to the Lab frame, eq.(\ref{eq:finite_vol_method_RG})~\cite{Rummukainen:1995vs}.
The results for $E^P_{\pi\pi}$, $\Delta E^P_{\pi\pi}$, $E_{\pi\pi}$, 
$\gamma = E^P_{\pi\pi} / E_{\pi\pi}$, $p$, and $\delta(p)$
in the Lab calculation are summarized in table~\ref{tab:twopi_Lab}.

\begin{figure}[!h]
\centering
\scalebox{0.5}[0.5]{
\includegraphics*{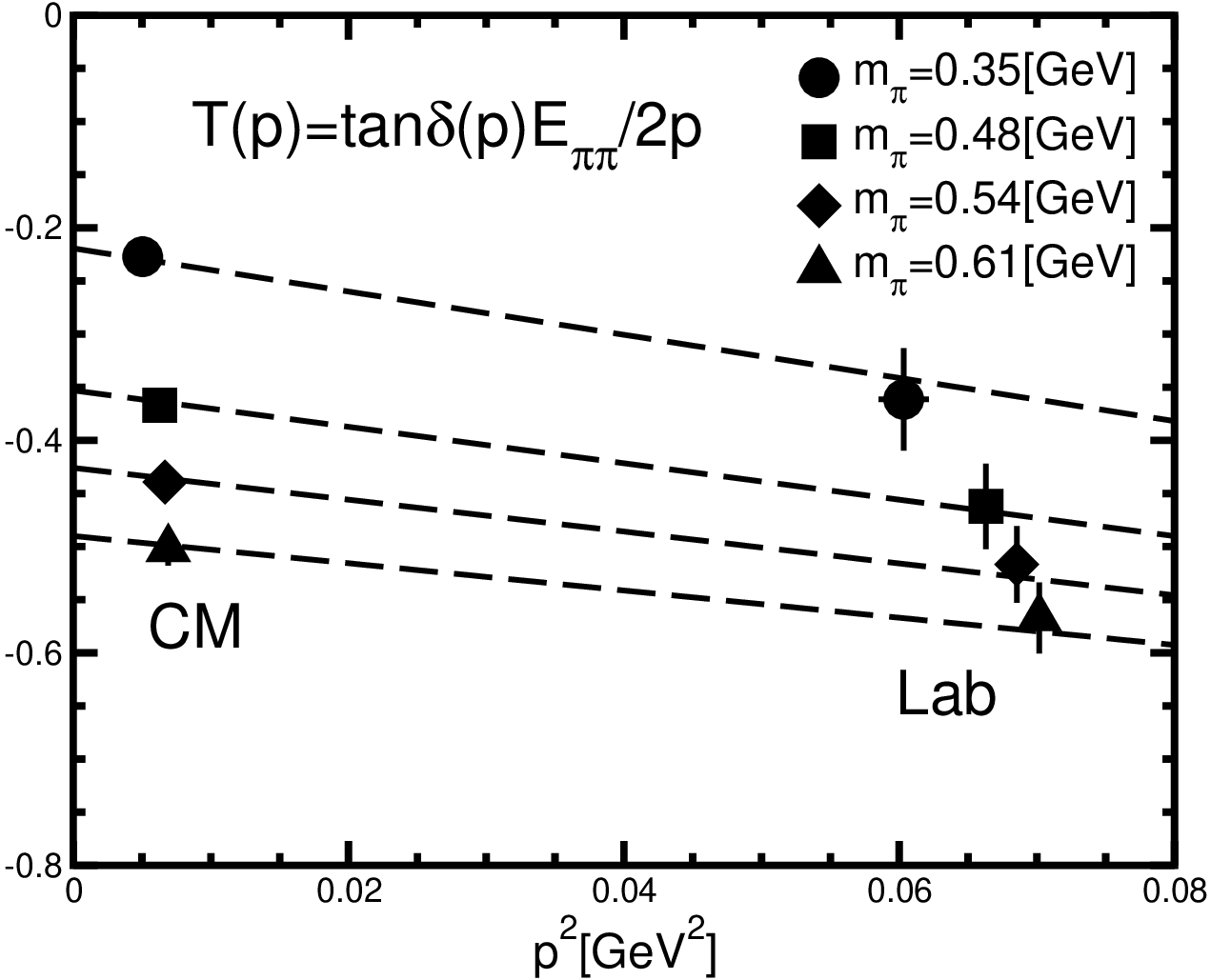}
}
\caption{
Scattering amplitude $T(p)$ defined in eq.(\ref{eq:T}).
Dashed lines are fit results with polynomial function eq.(\ref{eq:fit_sctamp}).
Results at smaller(larger) $p^2$ are obtained from CM(Lab) calculation.
}
\label{fig:sctamp}
\end{figure}

To investigate the momentum dependence of the scattering phase shift,
we use the results obtained from both the frames.
We define ``scattering amplitude'' $T(p)$ as in Refs.~\cite{Aoki:2002ny,Yamazaki:2004qb,Li:2007ey,Sasaki:2006jn}
\begin{equation}
T(p) = \frac{ \tan\delta(p) }{ p } \cdot \frac{ E_{\pi\pi} }{ 2 },
\label{eq:T}
\end{equation}
which is normalized by $a_0 m_\pi$ at the zero momentum.
The results of $T(p)$ in each frame are tabulated in tables~\ref{tab:twopi_CM} and 
\ref{tab:twopi_Lab}.
Figure~\ref{fig:sctamp} shows the measured value of $T(p)$ as a function
of the relative momentum squared.
The amplitude $T(p)$ is fitted with a naive polynomial function for $m_\pi^2$ and $p^2$
\begin{equation}
A_{10} m_\pi^2 + A_{20} m_\pi^4 + A_{30} m_\pi^6 + A_{01} p^2 + A_{11} m_\pi^2 p^2,
\label{eq:fit_sctamp}
\end{equation}
where the indices of the parameter $A_{ij}$ denote the powers of $m_\pi^2$ and $p^2$, respectively.
This fit form does not include $p^4$ term,
because our calculation is carried out
with only two different total momentum frames.
In order to include a $p^4$ term in the fit form,
we need several data with different relative momentum.
Since the scattering amplitude at $p=0$ is nothing but the scattering length $a_0 m_\pi$,
we need the $A_{30}$ term to coincide with the pion mass dependence 
of the scattering length in the previous section.
The result of the fit parameters is presented in table~\ref{tab:fit_sctamp},
and the fit lines for each $m_\pi^2$ are plotted in Fig.~\ref{fig:sctamp}.

The $a_0/m_\pi$ estimated with the fit result,
shown in Fig.~\ref{fig:a0} 
as denoted by ``$\delta(p)$ analysis'',
is reasonably consistent with those of the CM analysis at each pion mass.
We obtain $a_0 / m_\pi = -1.99(12)$ GeV$^{-2}$ at the physical pion mass,
which agrees with the fit result given in the previous section (see table~\ref{tab:fit_a0}).

\begin{figure}[!h]
\centering
\scalebox{0.5}[0.5]{
\includegraphics*{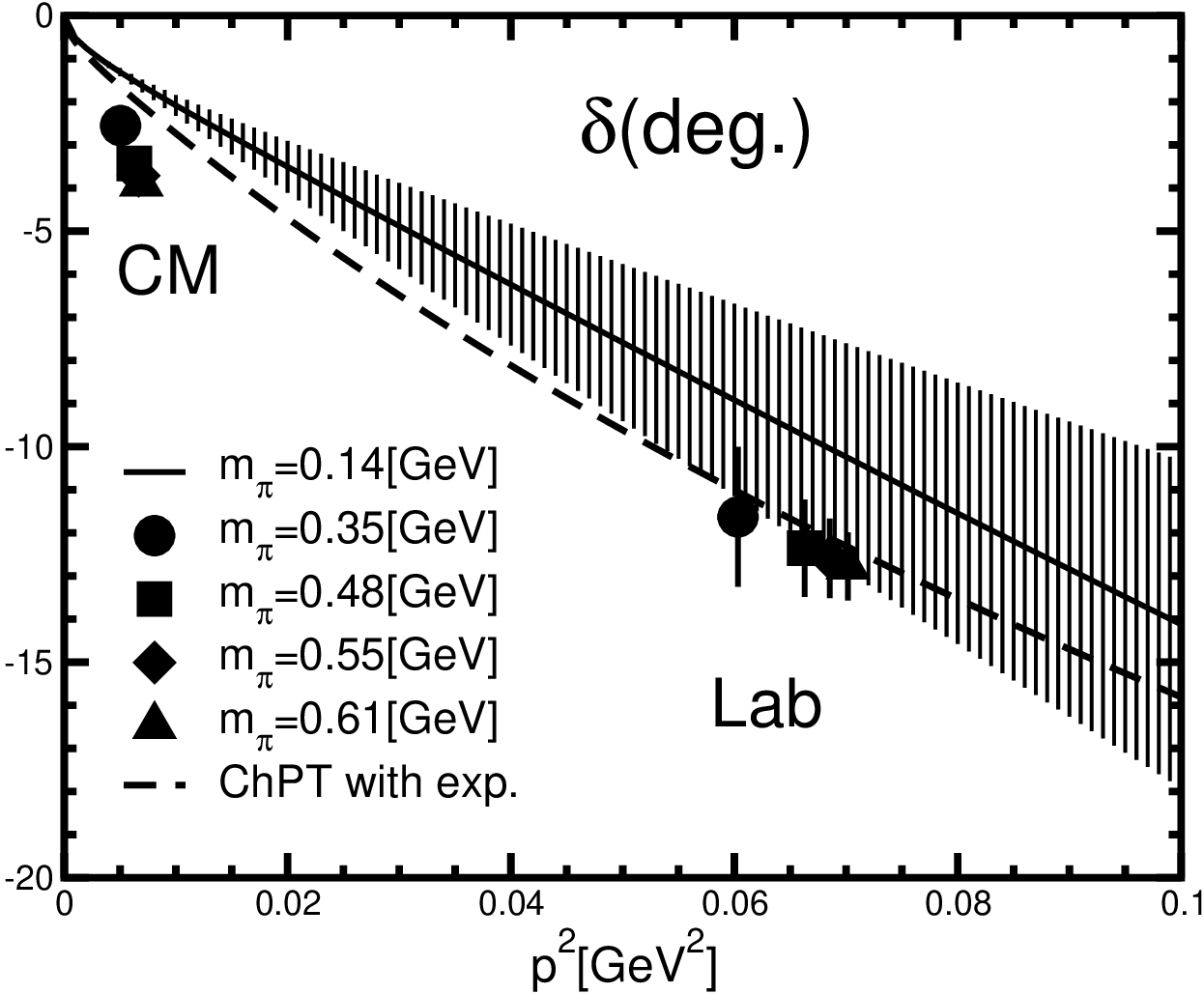}
}
\caption{
Measured scattering phase shift.
Solid line with error band is result at physical pion mass.
Dashed line is prediction of ChPT with experiment.
Results at smaller(larger) $p^2$ are obtained from CM(Lab) calculation.
}
\label{fig:phsh}
\end{figure}

The measured scattering phase shift is plotted in Fig.~\ref{fig:phsh}
as well as the result at the physical pion mass estimated from 
the $T(p)$ fitting.
The result is compared with the prediction of 
ChPT~\cite{Colangelo:2001df} estimated with experiment in the figure.

%
%
\subsubsection{Finite volume effect from final state interaction}
\label{subsucsec:FVFSI}

We estimate the derivative of the scattering phase shift from
the fit results eq.(\ref{eq:fit_sctamp}),
and evaluate the derivatives for the functions $\phi(q)$ and $\phi_{\vec{P}}(q)$ numerically
to utilize the LL formula eq.(\ref{eq:CM_Formula}) and its extension eq.(\ref{eq:Lab_Formula}), respectively.

For convenience, we define the conversion factors $F$ and $F^P$,
\begin{eqnarray}
F &=&
\sqrt{
8\pi \left(\frac{E_{\pi\pi}}{p}\right)^3
\left\{ p \frac{ \partial \delta }{ \partial p }
      + q \frac{ \partial \phi }{ \partial q }
\right\}
},
\label{eq:F}\\
F^P &=&
\sqrt{
8\pi \gamma^2 \left(\frac{E_{\pi\pi}}{p}\right)^3
\left\{ p \frac{ \partial \delta }{ \partial p }
      + q \frac{ \partial \phi_{\vec{P}} }{ \partial q }
\right\}
},
\label{eq:FP}
\end{eqnarray}
which are the factors connecting the finite volume, decay amplitude in CM and Lab frames to the CM
one in the infinite volume, respectively.
The results of $F$ and $F^P$ are shown in tables~\ref{tab:twopi_CM} and 
\ref{tab:twopi_Lab}.
To investigate the size of the interaction effect, we evaluate a ratio of 
the conversion factor to the one in the non-interacting case.
We estimate the factors without the interaction $\overline{F}$ and 
$\overline{F}^P$ in each frame as,
\begin{eqnarray}
\overline{F} &=&
\sqrt{
8\pi \left(\frac{\overline{E}_{\pi\pi}}{\overline{p}}\right)^3
\left\{ \overline{q} \frac{ \partial \phi }{ \partial \overline{q} }
\right\}
},
\label{eq:F0}\\
\overline{F}^P &=&
\sqrt{
8\pi \gamma^2 \left(\frac{\overline{E}_{\pi\pi}}{\overline{p}}\right)^3
\left\{ \overline{q} \frac{ \partial \phi_{\vec{P}} }{ \partial \overline{q} }
\right\}
},
\label{eq:FP0}
\end{eqnarray}
where the quantities with the overline are the ones without two-pion
interaction.
In eq.~(\ref{eq:FP0}), $\overline{E}_{\pi\pi}$ and $\overline{p}$
are determined by
\begin{equation}
\overline{E}_{\pi\pi}^2 = 4(m_\pi^2 + \overline{p}^2) = (\overline{W}^P_+)^2 - P^2,
\end{equation}
with
\begin{equation}
\overline{W}^P_\pm = \sqrt{m_\pi^2 + P^2} \pm m_\pi.
\end{equation}
In the CM frame case, $\overline{p}=0$, and then $\overline{F} = \sqrt{4(\overline{E}_{\pi\pi} L)^3}$ as in
Ref.~\cite{Lellouch:2000pv},
while in the Lab frame case,
\begin{equation}
\overline{F}^P = \sqrt{ 2 \overline{W}^P_+ \left( (\overline{W}^P_+)^2 - (\overline{W}^P_-)^2 \right) L^3 }.
\end{equation}

The ratios $F/\overline{F}$ and $F^P/\overline{F}^P$ are summarized in tables~\ref{tab:twopi_CM} and 
\ref{tab:twopi_Lab}.
The results show that the interaction effect is relatively large, {\it e.g.,} 13 -- 17\%.
The effect depends on the pion mass in the CM case, while there is no large dependence in the Lab case.
In the CM case we find that the ratio decreases as the pion mass increases.
This trend was also seen in the previous calculation~\cite{Aoki:1997ev} where the interaction effect
was estimated by one-loop ChPT.
It might be described by the fact that the scattering effect of the two-pion
decreases in the smaller pion mass, because $a_0 \propto m_\pi$.

%
%
%
\subsection{Decay amplitudes}

%
%
%
\subsubsection{Off-shell decay amplitudes in finite volume}
\label{subsec:off-shell_decay_amplitudes_finite_v}

To determine off-shell decay amplitudes in finite volume,
we define ratios of correlation functions $R_i(t)$ in the CM frame and $R^P_i(t)$ in the Lab frame
as
\begin{eqnarray}
R_i(t) &=& 
\sqrt{\frac{2}{3}}
\frac{ G_i(t,t_\pi,t_K) Z_{\pi\pi} Z_K }{ G_{\pi\pi}(t,t_\pi) G_K(t,t_K) },
\label{eq:ratio_CM}\\
R_i^P(t) &=& 
\sqrt{\frac{2}{3}}
\frac{ G_i^P(t,t_\pi,t_K) Z^P_{\pi\pi} Z^P_K }{ G^P_{\pi\pi}(t,t_\pi) G^P_K(t,t_K) },
\label{eq:ratio_Lab}
\end{eqnarray}
where $G_{\pi\pi} (G^P_{\pi\pi})$ and $G_K (G^P_K)$ are the $I=2$ two-pion four-point
function and the kaon two-point function in the CM (Lab) frame, respectively. 
These correlators are calculated with the wall sink operator as described 
in Sec.~\ref{subsec:calc_four-point_func}.
$G_i$ and $G_i^P$ are the $\Delta I = 3 / 2$ $K\to\pi\pi$ four-point functions 
defined in eqs.(\ref{eq:kpipi4point_CM}) and (\ref{eq:kpipi4point_Lab}).
The index $i$ denotes each operator $i=27,88$ and $m88$
defined in eqs.(\ref{eq:operator_27}), (\ref{eq:operator_88}), and (\ref{eq:operator_m88}).
$Z_{\pi\pi}^{(P)}$ and $Z_K^{(P)}$ are the overlaps for the relevant operators with each state.

\begin{figure}[!h]
\centering
\begin{tabular}{c}
\scalebox{0.5}[0.5]{
\includegraphics*{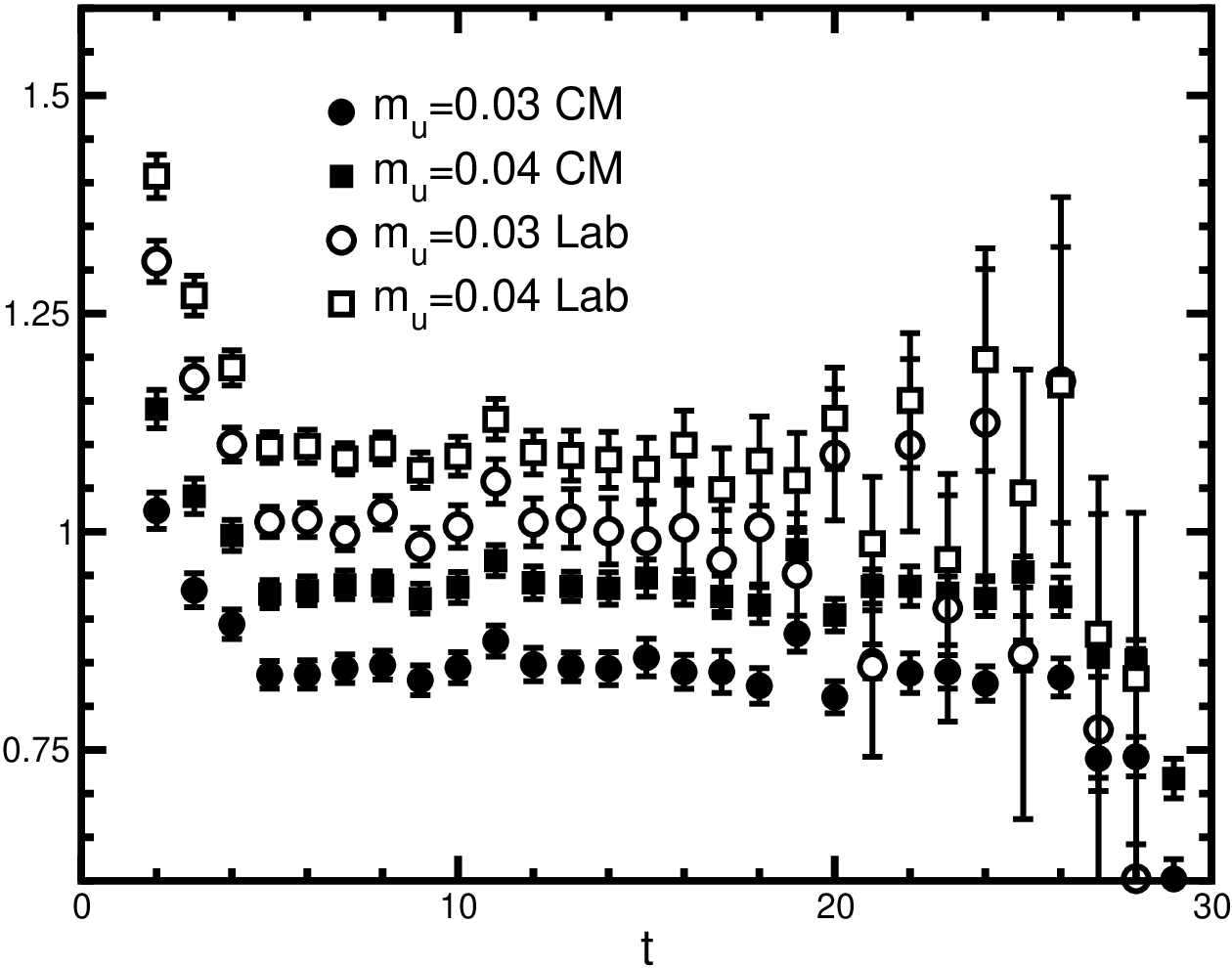}
}
\\
\scalebox{0.5}[0.5]{
\includegraphics*{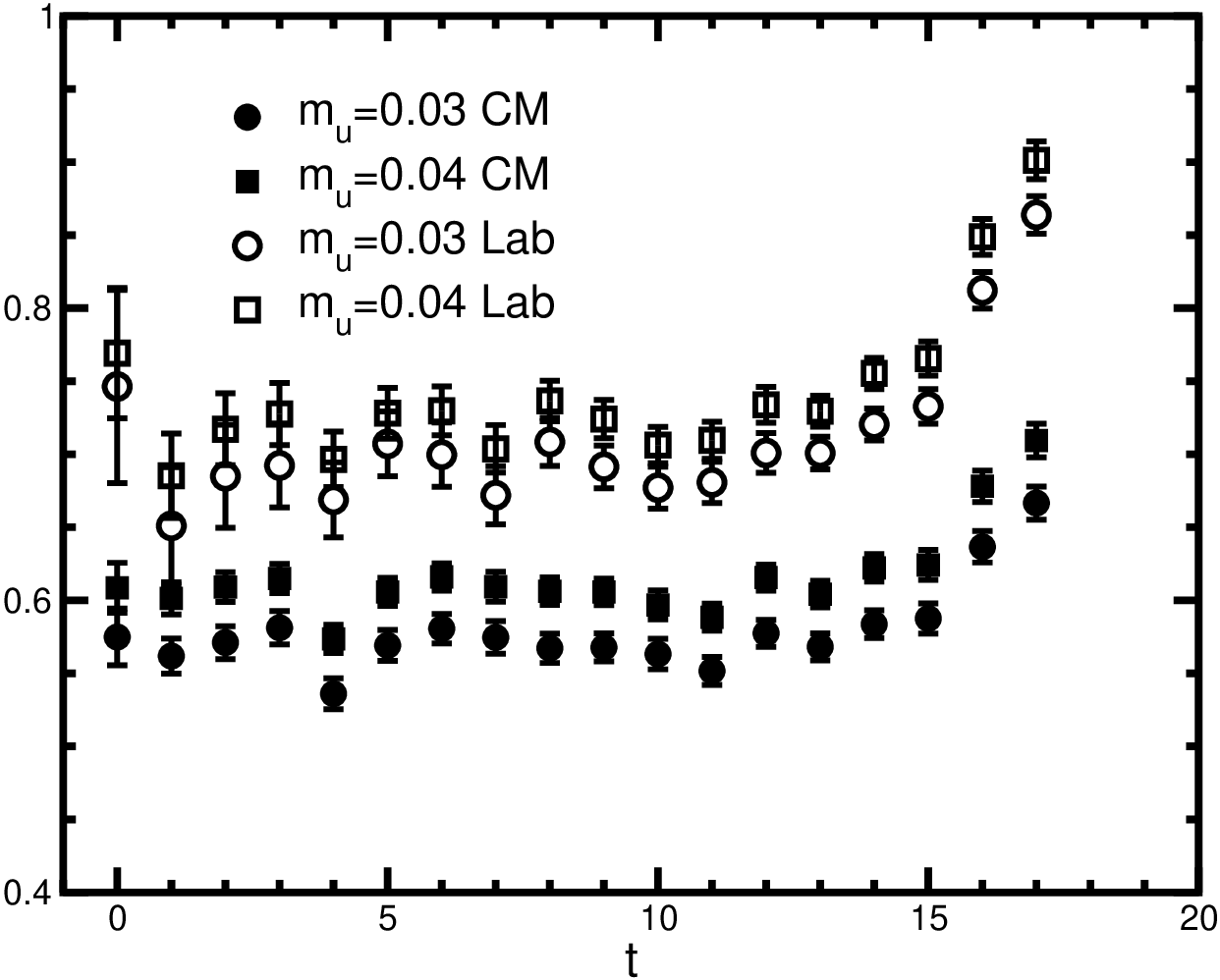}
}
\end{tabular}
\caption{
Top panel is effective mass and energy for $\pi\pi$ state with zero(CM)
and non-zero(Lab) momentum obtained from wall sink correlator.
Bottom panel is same as the top panel except for kaon state at $m_s = 0.12$ in $t_K = 20$ case.
}
\label{fig:eff_mass}
\end{figure}

When the correlation functions in eqs.(\ref{eq:ratio_CM}) and (\ref{eq:ratio_Lab}) 
are dominated by each ground state,
the ratios in principle will be a constant for those values of $t$,
which corresponds to off-shell decay amplitude.
We determine the desired amplitudes from the ratio in the flat region
where the effective energy of each state has a plateau.
The typical plateaus for the two-pion and kaon states are presented in Fig.~\ref{fig:eff_mass}.

The overlaps are determined by the wall sink correlators with the following fit forms,
\begin{eqnarray}
G_{\pi\pi}(t,t_\pi) &=& Z_{\pi\pi}^2 \cdot \Delta_{\pi\pi} ( E_{\pi\pi}, |t-t_\pi| ),\\
G_{\pi\pi}^P(t,t_\pi) &=& \left(Z_{\pi\pi}^P\right)^2 \cdot 
\Delta^P_{\pi\pi} ( E^P_{\pi\pi}, |t-t_\pi| ),\\
G_K(t,t_K) &=& Z_K^2 \cdot \Delta_K ( m_K, |t-t_K| ),\\
G_K^P(t,t_K) &=& \left(Z_K^P\right)^2 \cdot \Delta_K ( E_K^P, |t-t_K| ),
\end{eqnarray}
where $E_{\pi\pi}$, $E_{\pi\pi}^P$, $m_K$, and $E_K^P$ are 
the measured energies from the point sink correlators,
whose values are presented in tables.~\ref{tab:twopi_CM}, \ref{tab:twopi_Lab},
\ref{tab:mk}, and \ref{tab:ek}, respectively.
The $m_K$ and $E_K^P$ are in $t_K = 20$ case.
The kernels are defined by
\begin{eqnarray}
\Delta_{\pi\pi}( E, t ) &=& e^{-Et} + e^{-E(2T-t)} + C,\\
\Delta_{\pi\pi}^P( E, t ) &=& e^{-Et} + e^{-E(2T-t)} + C\left[
e^{-(E_\pi^P-m_\pi)t} + e^{-(E_\pi^P-m_\pi)(2T-t)}\right],\label{eq:del_P_pipi}\\
\Delta_K( E, t ) &=& e^{-Et} + e^{-E(2T-t)},
\end{eqnarray}
where $m_\pi$ and $E_\pi^P$ in eq.~(\ref{eq:del_P_pipi}) are the measured single pion mass and energy,
and $C$ is one of the fit parameters.
In the extraction of the overlaps,
we use the fitting range of $6 \le t \le 25$ for the two-pion correlators, and 
$0 \le t \le t_K-7$ for the kaon correlators in both the frames.

\begin{figure}[!h]
\centering
\scalebox{0.5}[0.4]{
\includegraphics*{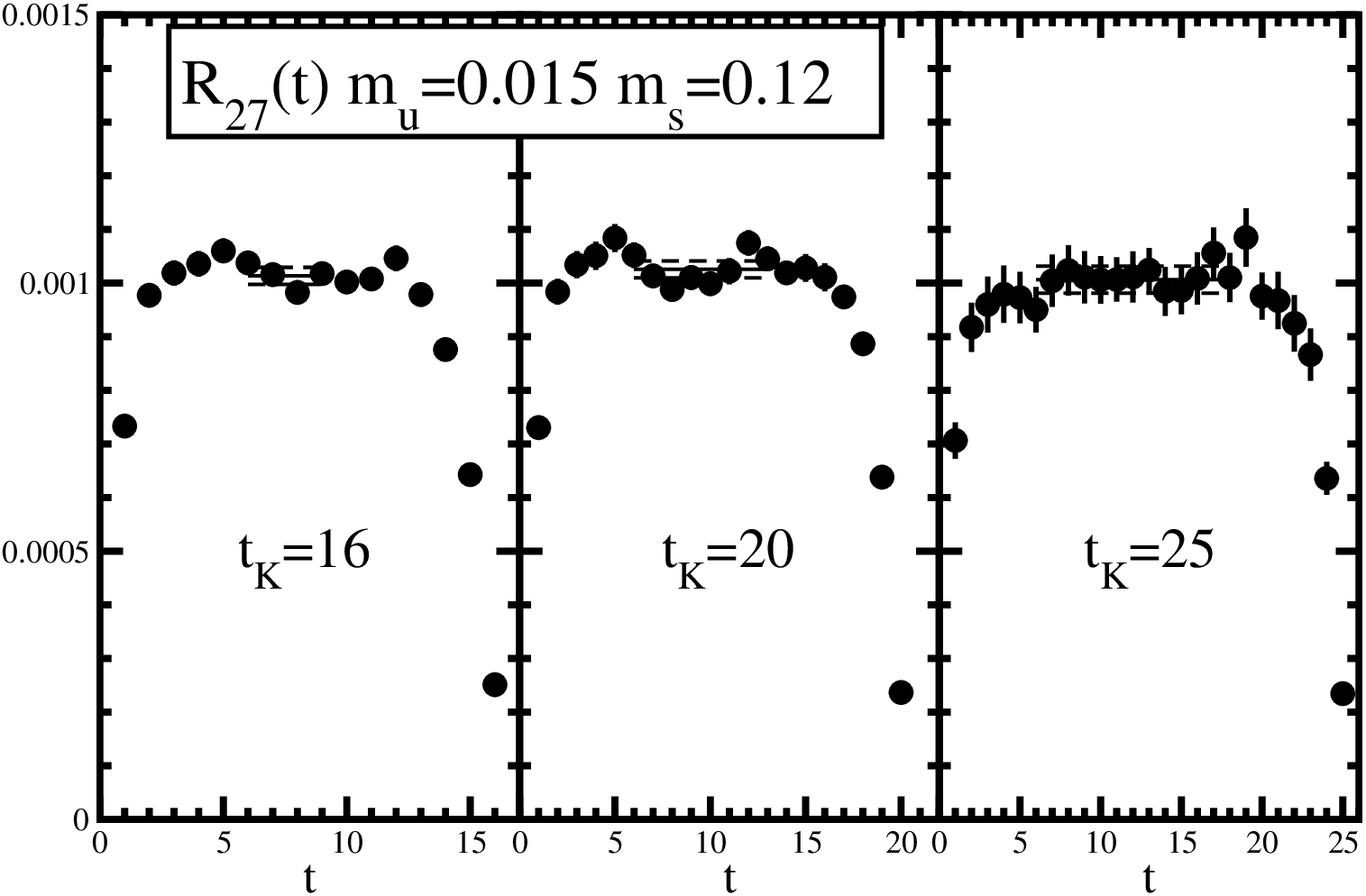}
}\\
\scalebox{0.5}[0.4]{
\includegraphics*{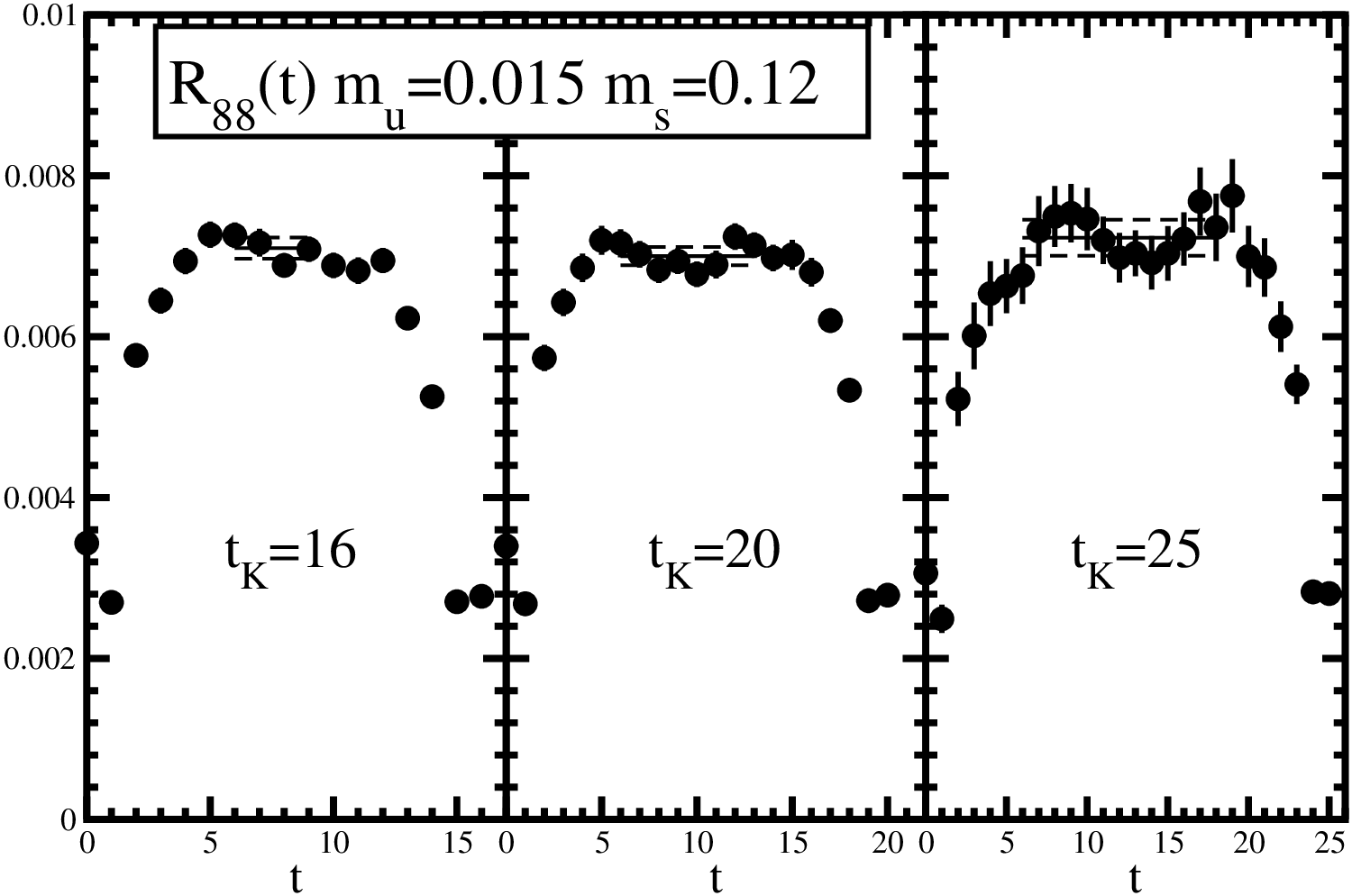}
}\\
\scalebox{0.5}[0.4]{
\includegraphics*{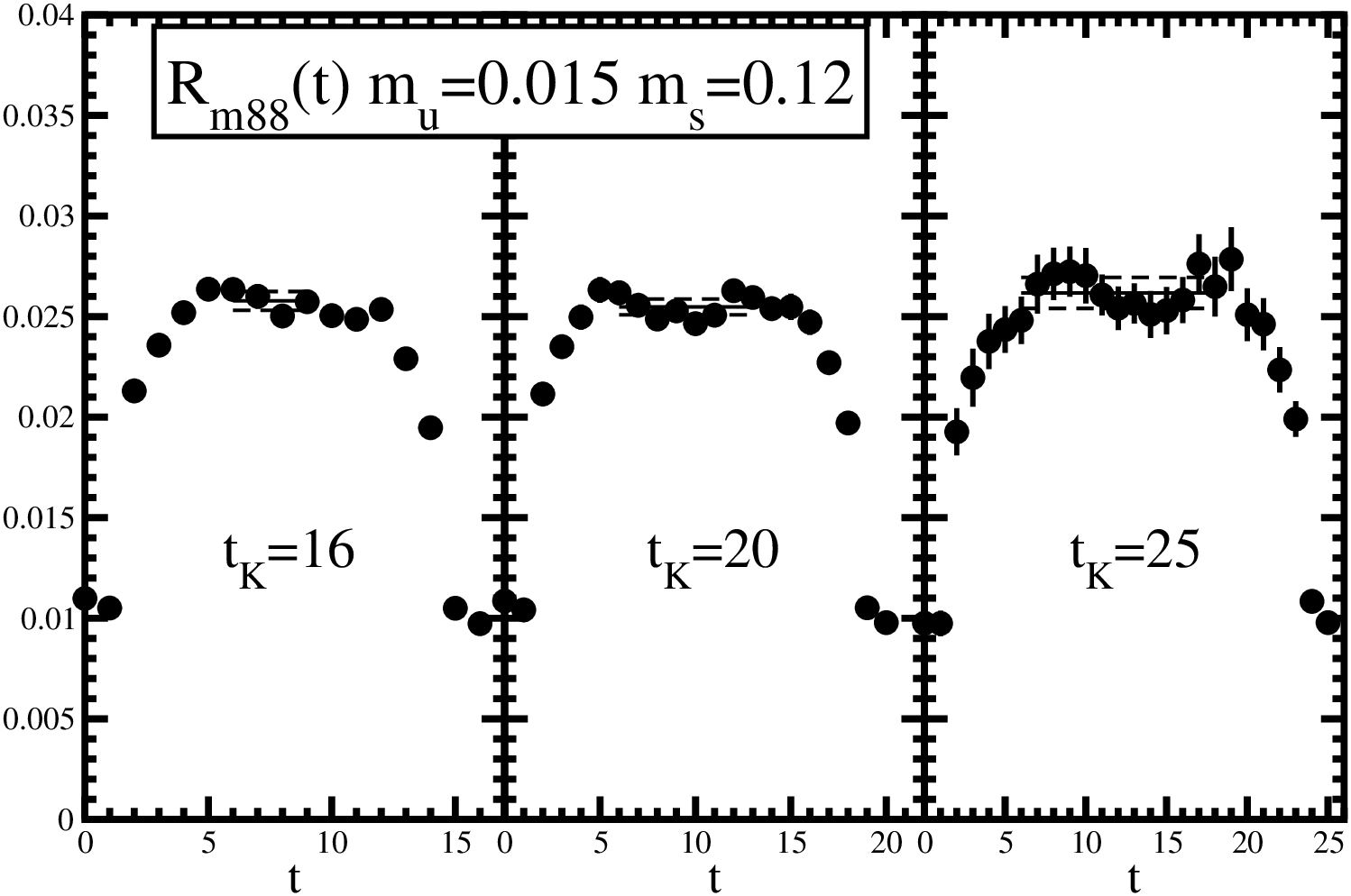}
}
\caption{
Ratio $R_i(t)$ for $27, 88, m88$ operators
defined in eq.(\ref{eq:ratio_CM}) obtained from CM calculation with
$m_u = 0.015$ and $m_s = 0.12$.
Solid and dashed lines are averaged value in flat region and its error,
respectively.
}
\label{fig:dcyamp_CM_u1}
\end{figure}

Figure~\ref{fig:dcyamp_CM_u1} shows $R_i(t)$ for all the operators 
with the different $t_K$ obtained in the CM calculation at
the lightest $m_u$ and $m_s$. 
All the ratios have clear signals, and are almost flat in the region of $t_{\pi} \ll t \ll t_K$.
Thus, the off-shell amplitudes are determined from averaged values in the 
flat region of $6 \le t \le t_K-7$.
The values with the error are presented in each panel by
the solid and dashed lines.
The averaged values in different $t_K$ are consistent with each other.

\begin{figure}[!h]
\centering
\scalebox{0.5}[0.4]{
\includegraphics*{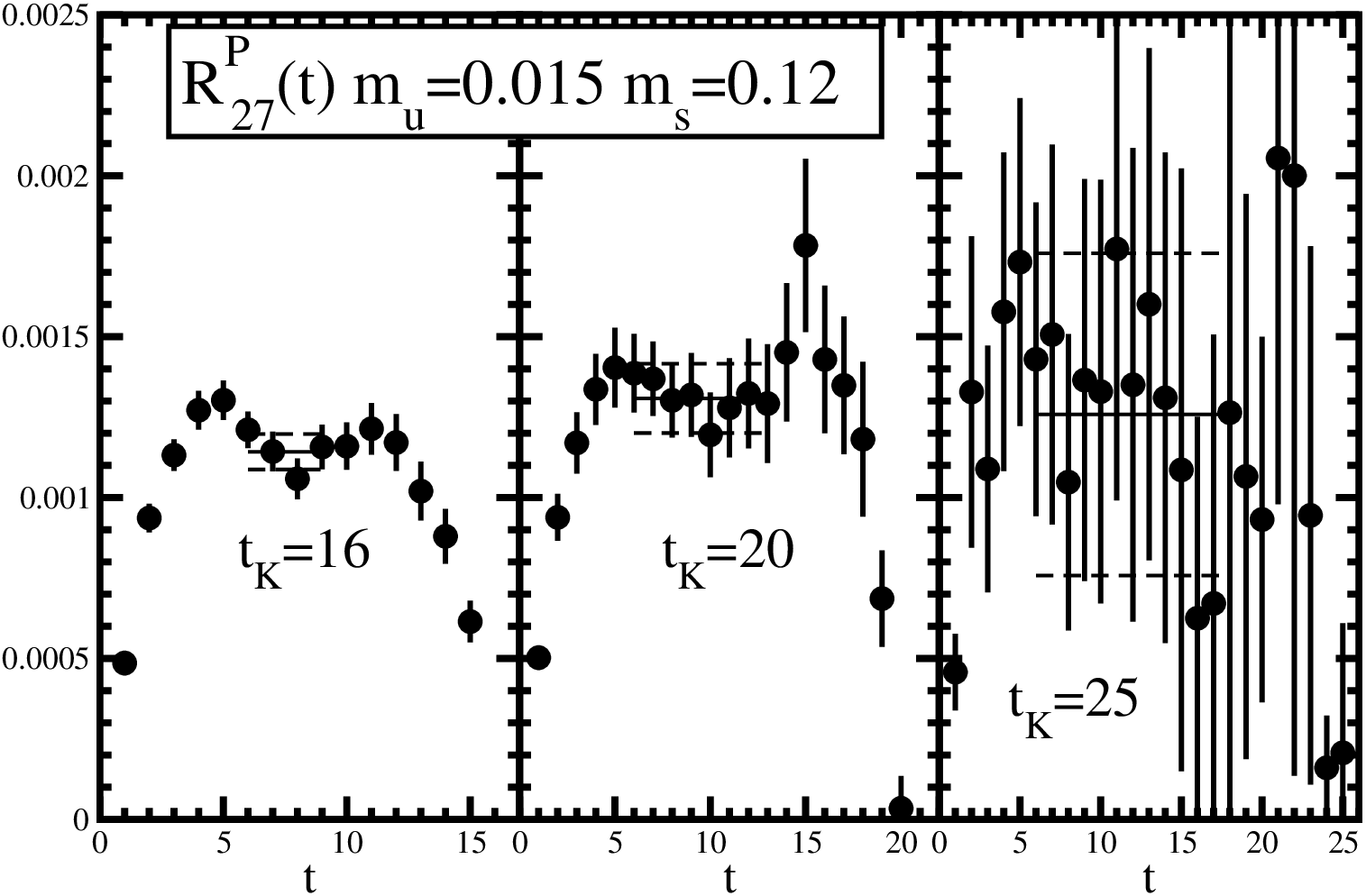}
}\\
\scalebox{0.5}[0.4]{
\includegraphics*{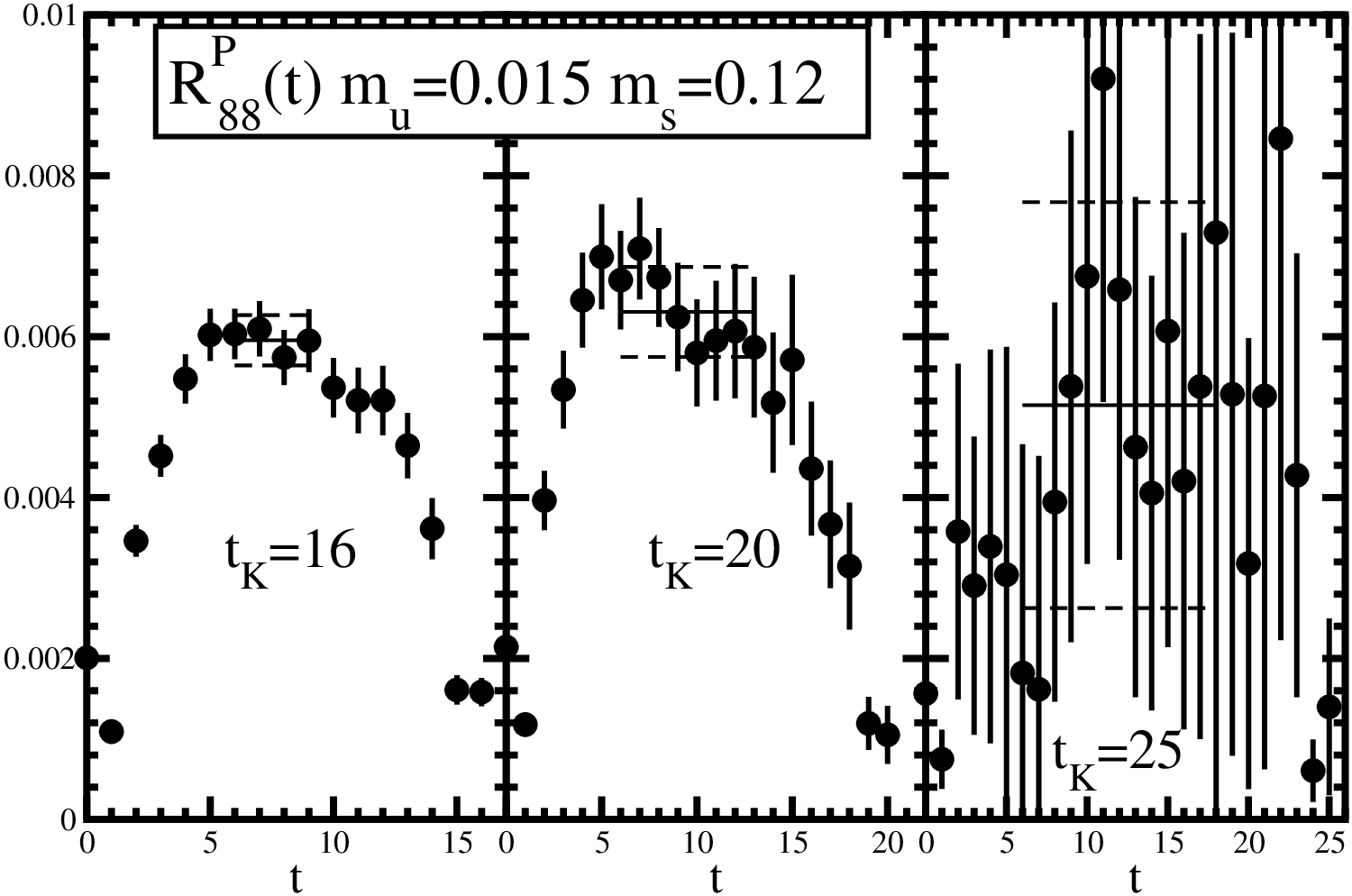}
}\\
\scalebox{0.5}[0.4]{
\includegraphics*{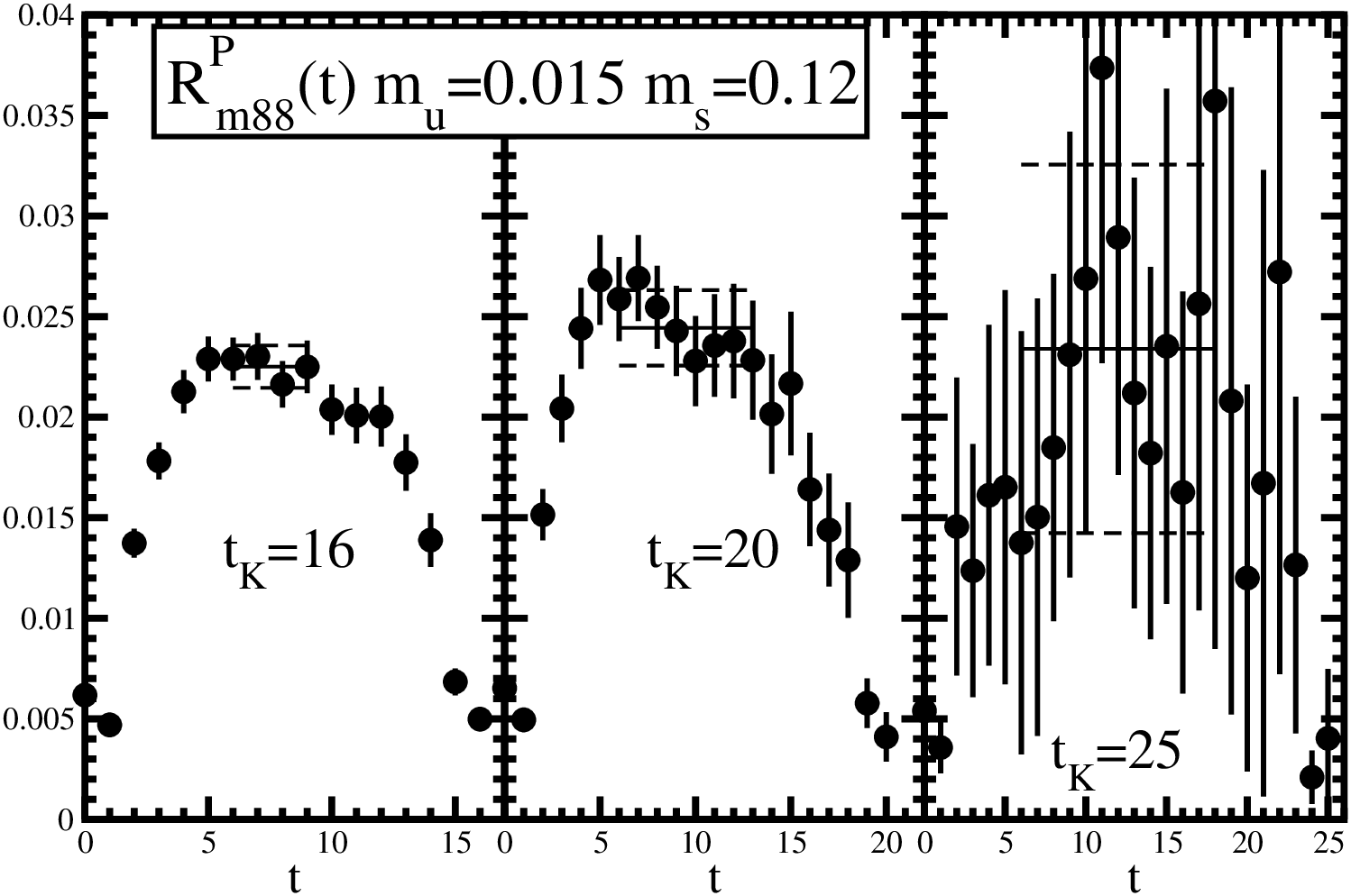}
}
\caption{
Ratio $R_i(t)$ for $27, 88, m88$ operators
defined in eq.(\ref{eq:ratio_Lab}) obtained from Lab calculation with
$m_u = 0.015$ and $m_s = 0.12$.
Solid and dashed lines are averaged value in flat region and its error,
respectively.
 }
\label{fig:dcyamp_Lab_u1}
\end{figure}

The ratios $R_i^P(t)$ are shown in Fig.~\ref{fig:dcyamp_Lab_u1}
at the same parameters as in the CM case.
For all operators, the statistical error of the ratio is much larger 
than those in the CM case,
while the error decreases as $t_K$ decreases.
In the case of $t_K = 25$ it is hard to determine a flat region due to the huge error,
while we can see a reasonable flat region for  $t_K = 16$ and 20 cases.
We choose the same time slice range for the averaged values as in the CM case.
The results of the off-shell amplitudes for $t_K=16$ and 20 reasonably
agree with each other.

According to the above discussion,
the result in the $t_K = 16$ case has the smallest statistical error in the Lab frame.
The separation between the two-pion and the kaon sources, however, seems to be small in this case.
Thus, we should worry about excited state contamination.
In order to avoid this systematic error as much as possible,
we choose the results with $t_K=20$ in the following analyses.
We tabulate the result of the off-shell amplitudes
for each $m_u$ and $m_s$ in tables~\ref{tab:dcyamp_27_CM}, \ref{tab:dcyamp_88_CM}, and \ref{tab:dcyamp_m88_CM} 
for the CM calculation, 
and in tables~\ref{tab:dcyamp_27_Lab}, \ref{tab:dcyamp_88_Lab}, and \ref{tab:dcyamp_m88_Lab} 
for the Lab calculation in the case of $t_K=20$.

%
%
%
\subsubsection{On-shell decay amplitudes in finite volume}
\label{subsec:on-shell_decay_amplitudes_finite_v}

We determine the finite volume, on-shell, amplitudes in both the frames, $|M_i|$ and $|M_i^P|$,
for each operator $i=27, 88$ and $m88$, by 
interpolating the off-shell amplitudes
with different strange quark masses at fixed $m_u$.
The on-shell kinematic point corresponds to $m_K = E_{\pi\pi}$ in the CM frame,
and $E_K^P = E_{\pi\pi}^P$ in the Lab frame.

\begin{figure}[!h]
\centering
\begin{tabular}{cc}
\scalebox{0.35}[0.35]{
\includegraphics*{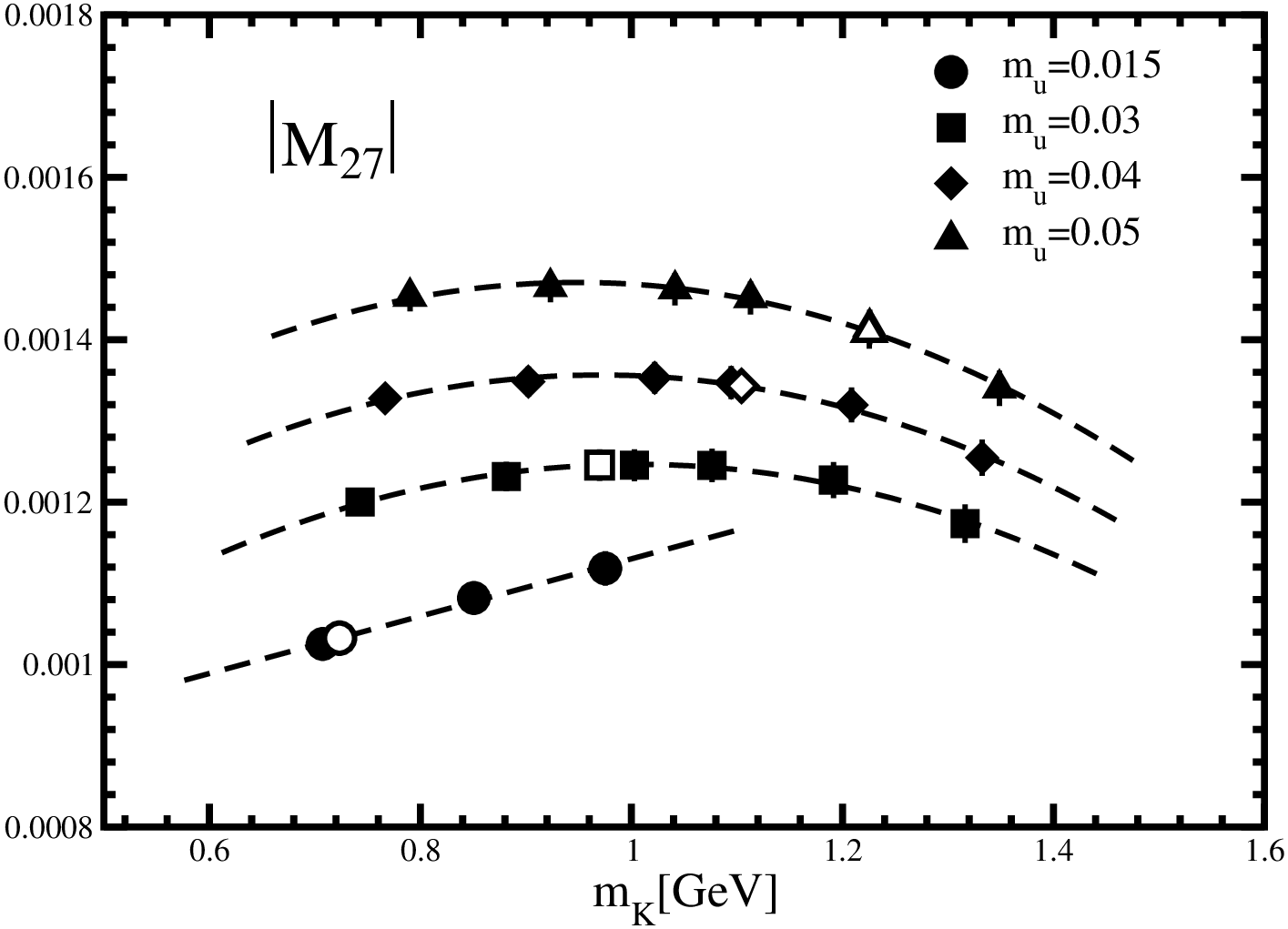}
}&
\scalebox{0.35}[0.35]{
\includegraphics*{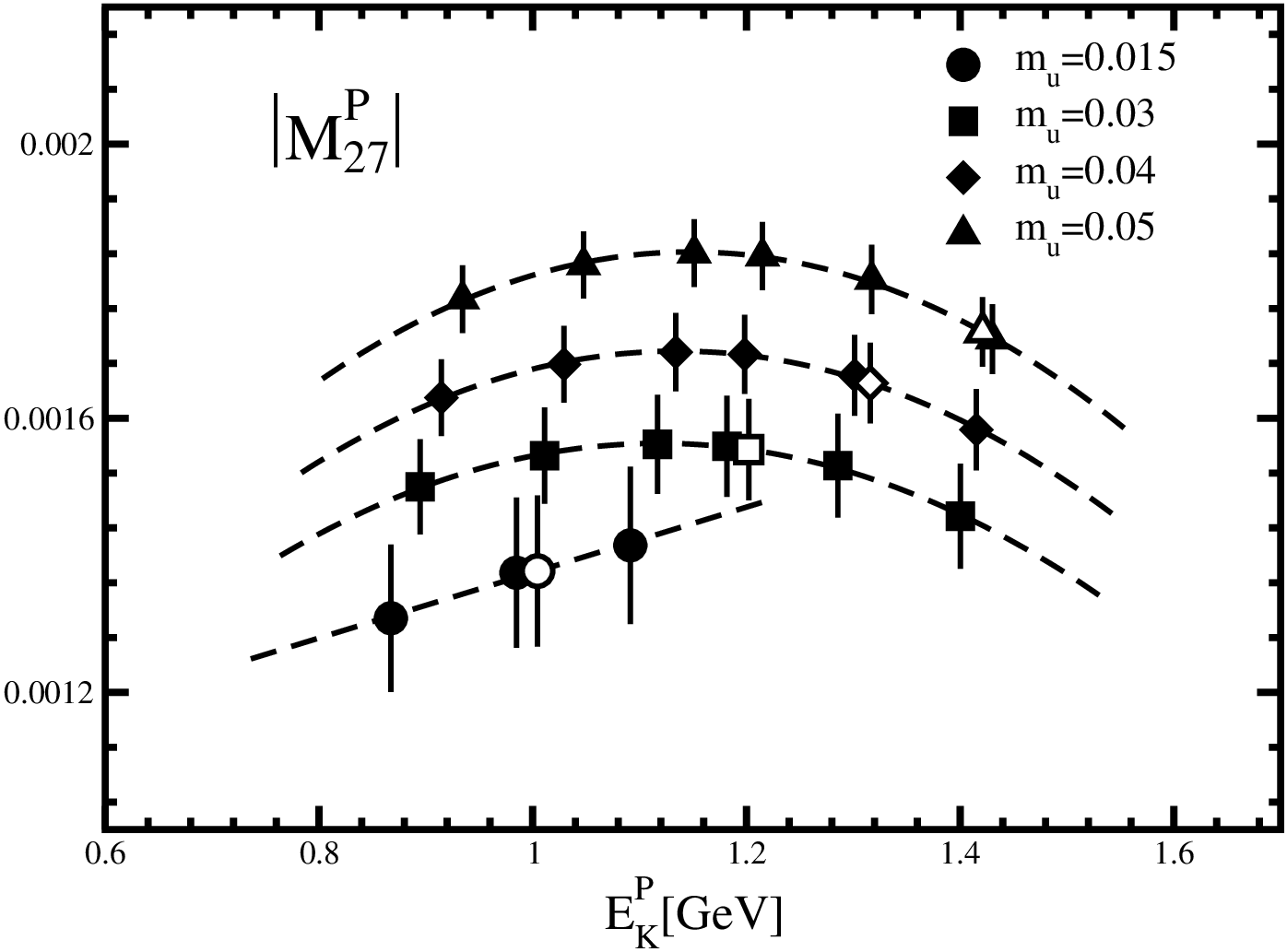}
}\\
\scalebox{0.35}[0.35]{
\includegraphics*{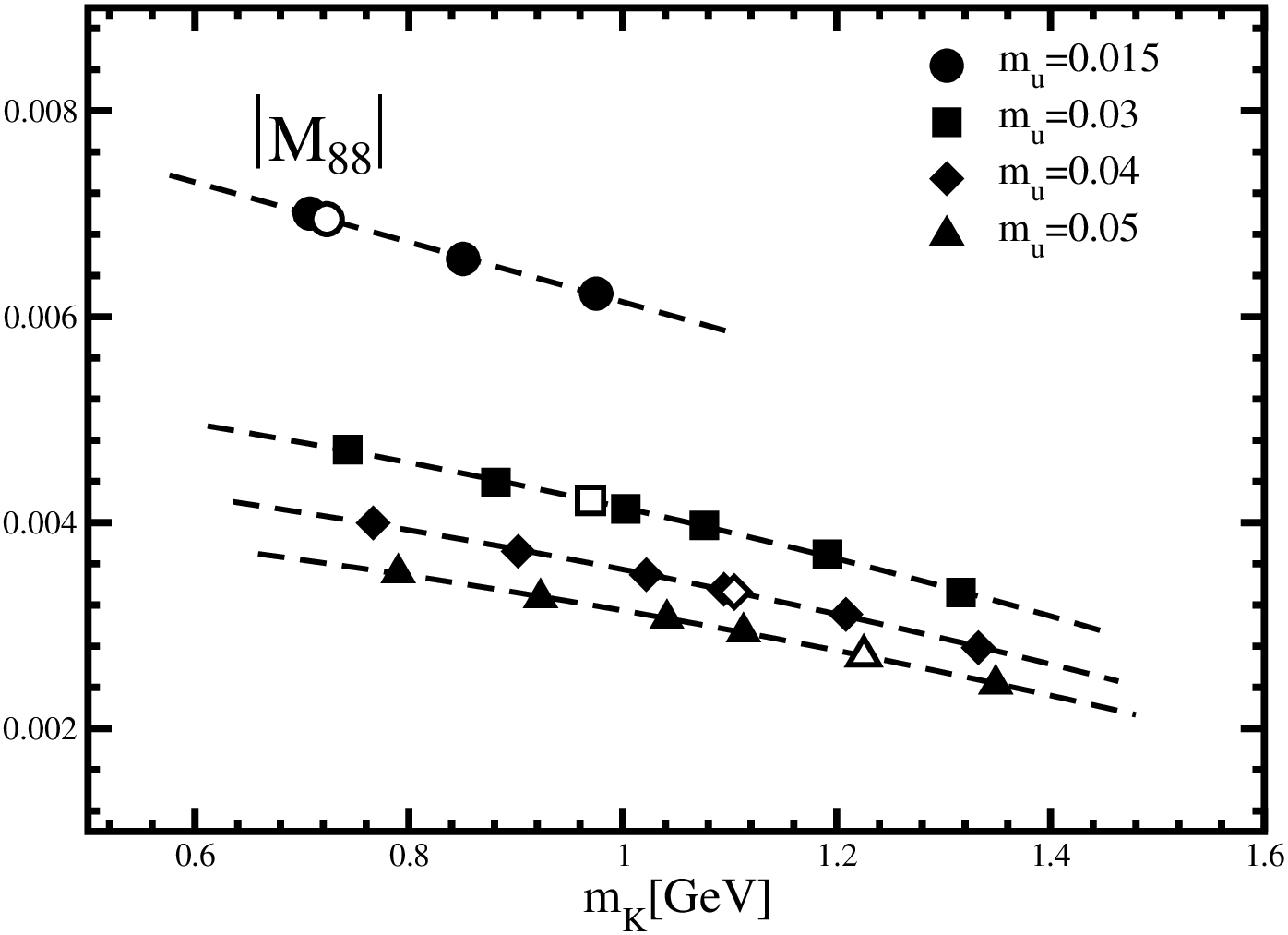}
}&
\scalebox{0.35}[0.35]{
\includegraphics*{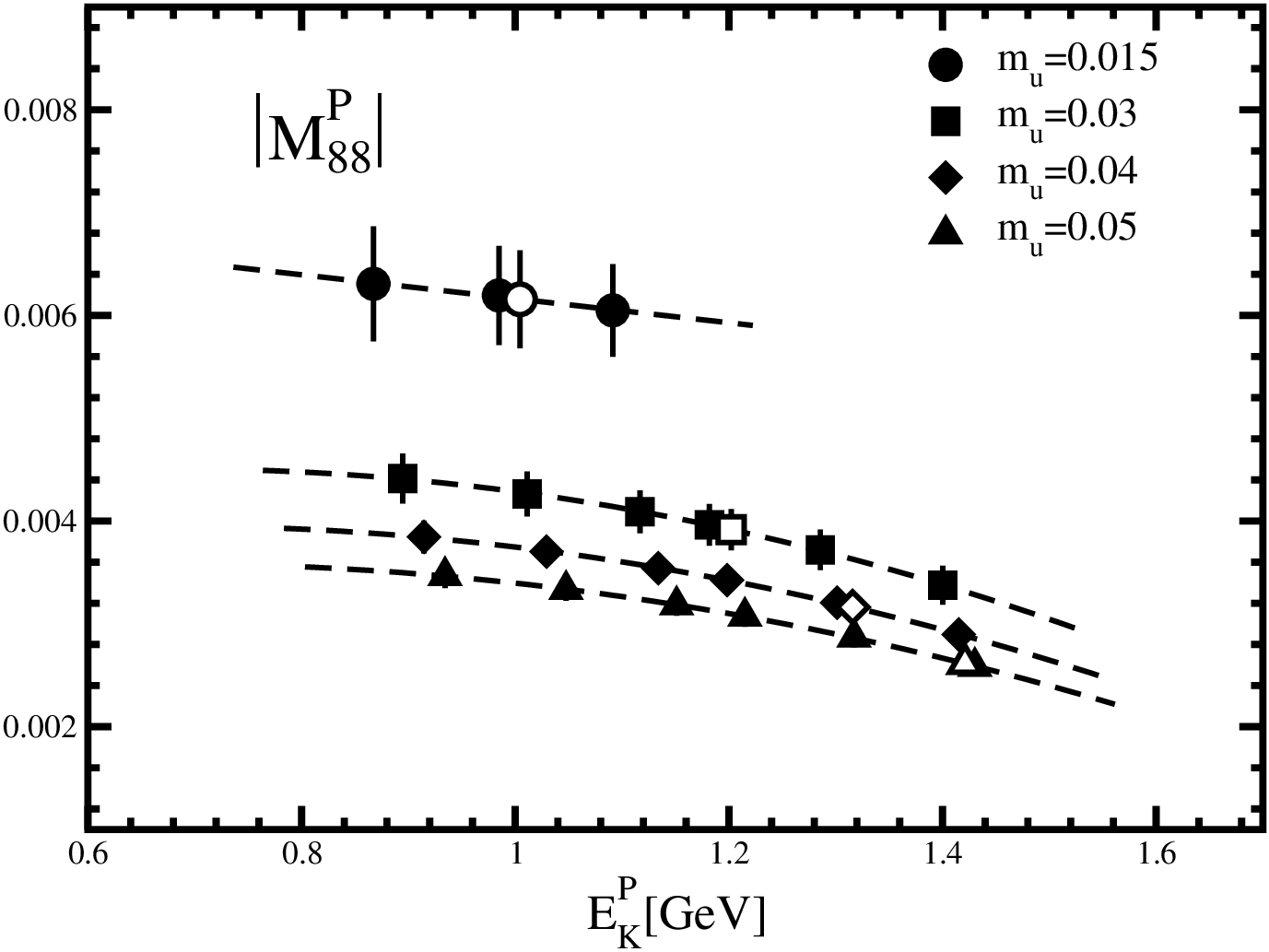}
}\\
\scalebox{0.35}[0.35]{
\includegraphics*{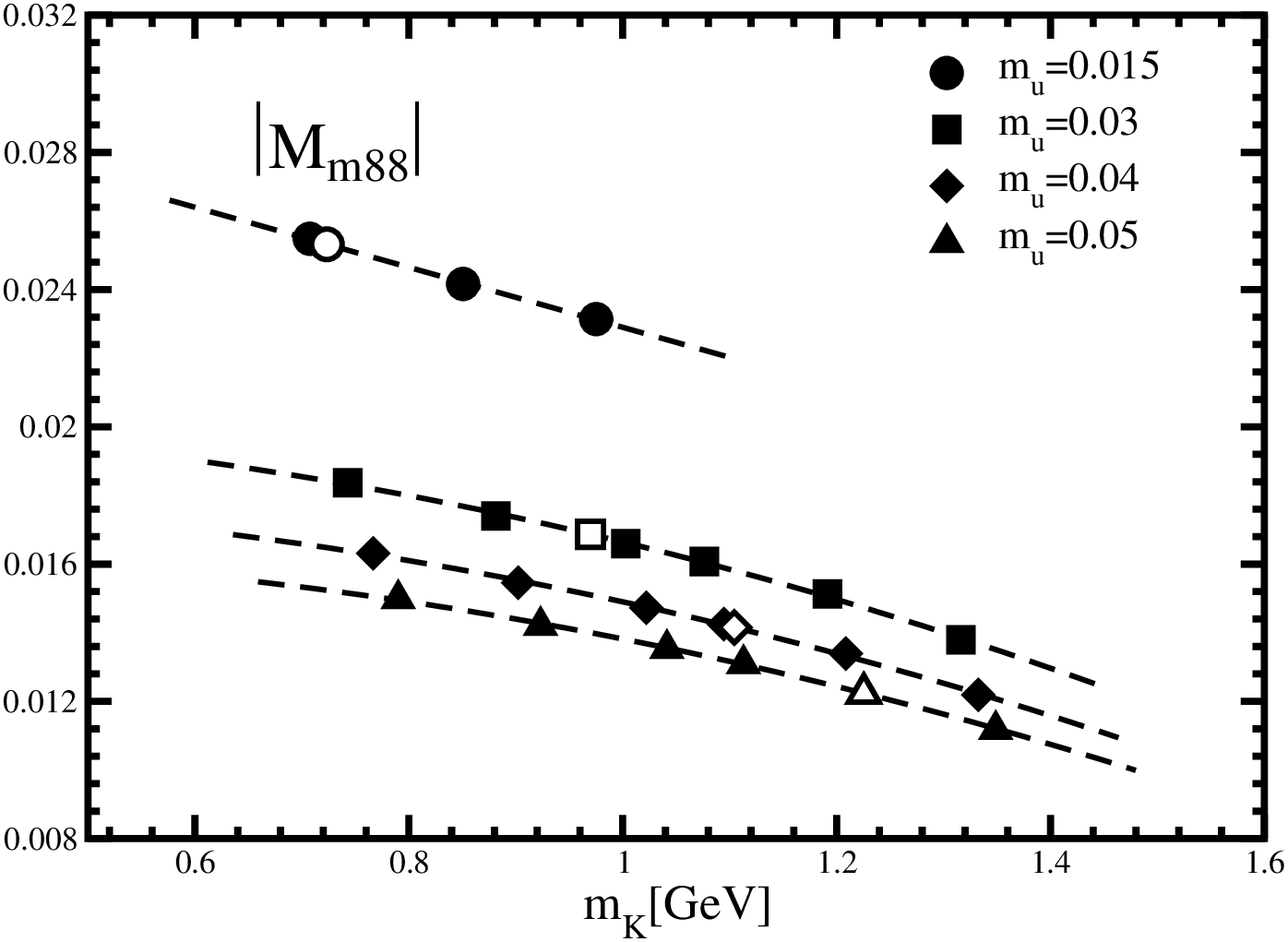}
}&
\scalebox{0.35}[0.35]{
\includegraphics*{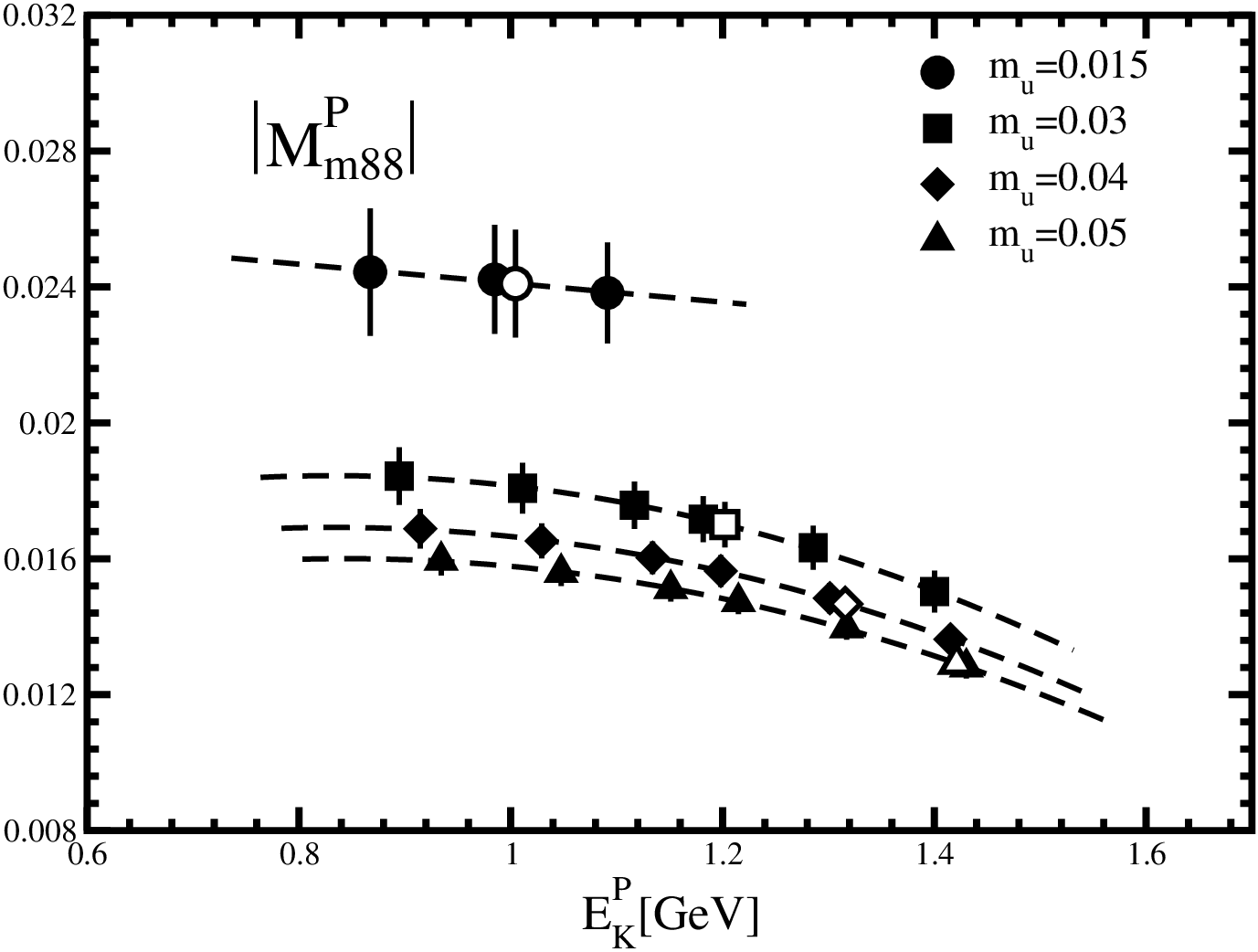}
}
\end{tabular}
\caption{
Interpolations of off-shell decay amplitude 
$|M_i|$ and $|M_i^P|$ for $27, 88, m88$ operators to on-shell.
Left(right) panel is CM(Lab) calculation.
Closed symbols are measured off-shell decay amplitudes,
dashed lines are interpolations, and
open symbols are on-shell decay amplitudes.
}
\label{fig:extrap_dcyamp}
\end{figure}

We plot the off-shell decay amplitude as a function of the kaon energy in
Fig.~\ref{fig:extrap_dcyamp}.
The figure shows that the amplitudes at $m_u = 0.015$ are
described by a linear function of the kaon energy 
using three strange quark masses.
On the other hand, the off-shell decay amplitudes for heavier $m_u$
show large curvature with the kaon energy, so we employ
a quadratic function of $m_K$ or $E_K^P$.
The results of the on-shell amplitudes are presented by open symbols in 
the figures, and also tabulated 
in tables~\ref{tab:dcyamp_27_CM}, \ref{tab:dcyamp_88_CM}, \ref{tab:dcyamp_m88_CM},
\ref{tab:dcyamp_27_Lab}, \ref{tab:dcyamp_88_Lab}, and \ref{tab:dcyamp_m88_Lab} 
denoted as ``On-shell'' for each operator and frame.

%
%
%
\subsubsection{On-shell decay amplitudes in infinite volume}
\label{subsec:on-shell_decay_amplitudes_infinite_v}

The on-shell decay amplitudes in infinite volume, $|A_i|$ for $i=27, 88$, and $m88$, are obtained for
the LL formula eq.~(\ref{eq:CM_Formula}) and its extension eq.~(\ref{eq:Lab_Formula}) by
combining the measured on-shell amplitudes $|M_i|$ and $|M_i^P|$ in finite volume and, 
the conversion factors $F$ and $F^P$ in tables~\ref{tab:twopi_CM} and \ref{tab:twopi_Lab}.
The results of the infinite volume, on-shell, decay amplitudes
are summarized in table~\ref{tab:dcyamp_infinit_v}.

\begin{figure}[!h]
\centering
\scalebox{0.5}[0.5]{
\includegraphics*{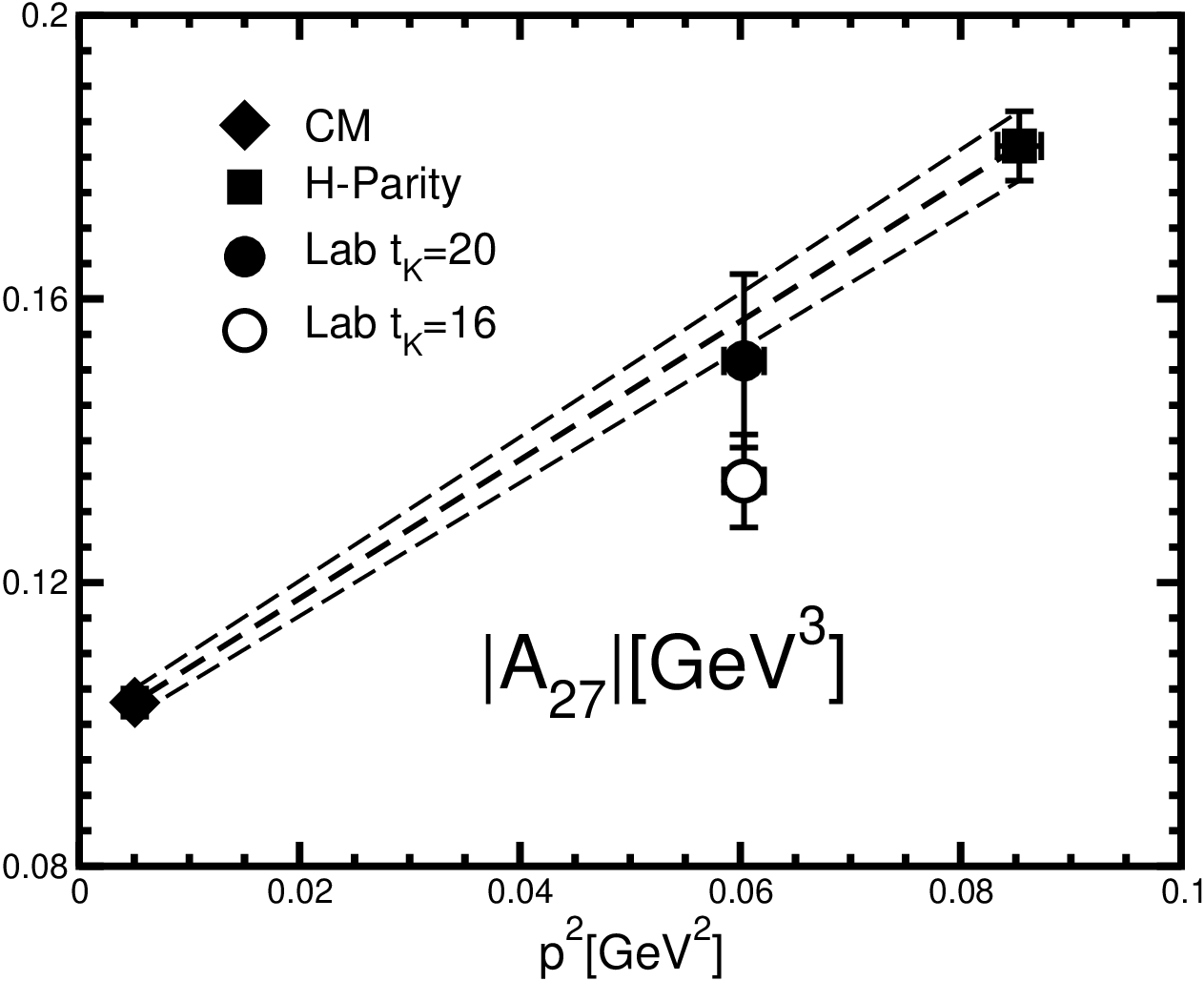}
}
\caption{
Comparison of decay amplitudes obtained from different calculations 
in infinite volume of $27$ operator at $m_u = 0.015$.
}
\label{fig:lwme_mom_27}
\end{figure}

We compare the infinite volume, on-shell decay amplitudes obtained from the different frame calculations. 
Figure~\ref{fig:lwme_mom_27} shows the decay amplitudes of the 27 operator with the CM and Lab
calculations at the lightest $m_u$ as a function of $p^2$.
We also plot previous results calculated with H-parity boundary
condition~\cite{Kim:2004mb,Kim:2004mk} in the figure.
The amplitude calculated in the Lab frame is consistent with the line
interpolated between 
those of the CM and H-parity boundary condition calculations.
This momentum dependence is consistent with a simple expectation that
the Lab result is smoothly connected to the results obtained from
the CM frame.
In contrast to this result, the one obtained with $t_K = 16$, 
also plotted in the figure, is below the linear fit. 
It may suggest that excited state contaminations are large in the 
$t_K = 16$ case.

%
%
%
\subsection{Weak matrix elements}
\label{subsec:weak_matrix_elements}

The matching factors of the weak matrix elements
were previously calculated
in Ref.~\cite{Kim:2004mk}, using the regularization independent (RI) 
scheme and a non-perturbative method~\cite{Martinelli:1994ty,Blum:2001sr}
at the scale $\mu = 1.44$ GeV.
The renormalization factors are summarized below,
\begin{equation}
Z_{ij} = \left(
\begin{array}{ccc}
0.8322(96) & 0 & 0\\
0 & 0.8940(75) & -0.0561(71)\\
0 & -0.0795(58) & 0.964(17) \\
\end{array}
\right), 
\end{equation}
with the operator indices $i,j = 27, 88, m88$.
In the following analysis we neglect the statistical errors of the renormalization constants, because
those of the matrix element are larger than the ones of the diagonal parts.
The renormalized weak matrix elements $|A_i^{\mathrm{RI}}|= Z_{ij} |A_j|$
for $i,j=27,88$ and $m88$ are summarized in table~\ref{tab:dcyamp_ri}.

%
%
\subsubsection{27 operator}
\label{subsucsec:A_27}

\begin{figure}[!h]
\centering
\scalebox{0.5}[0.5]{
\includegraphics*{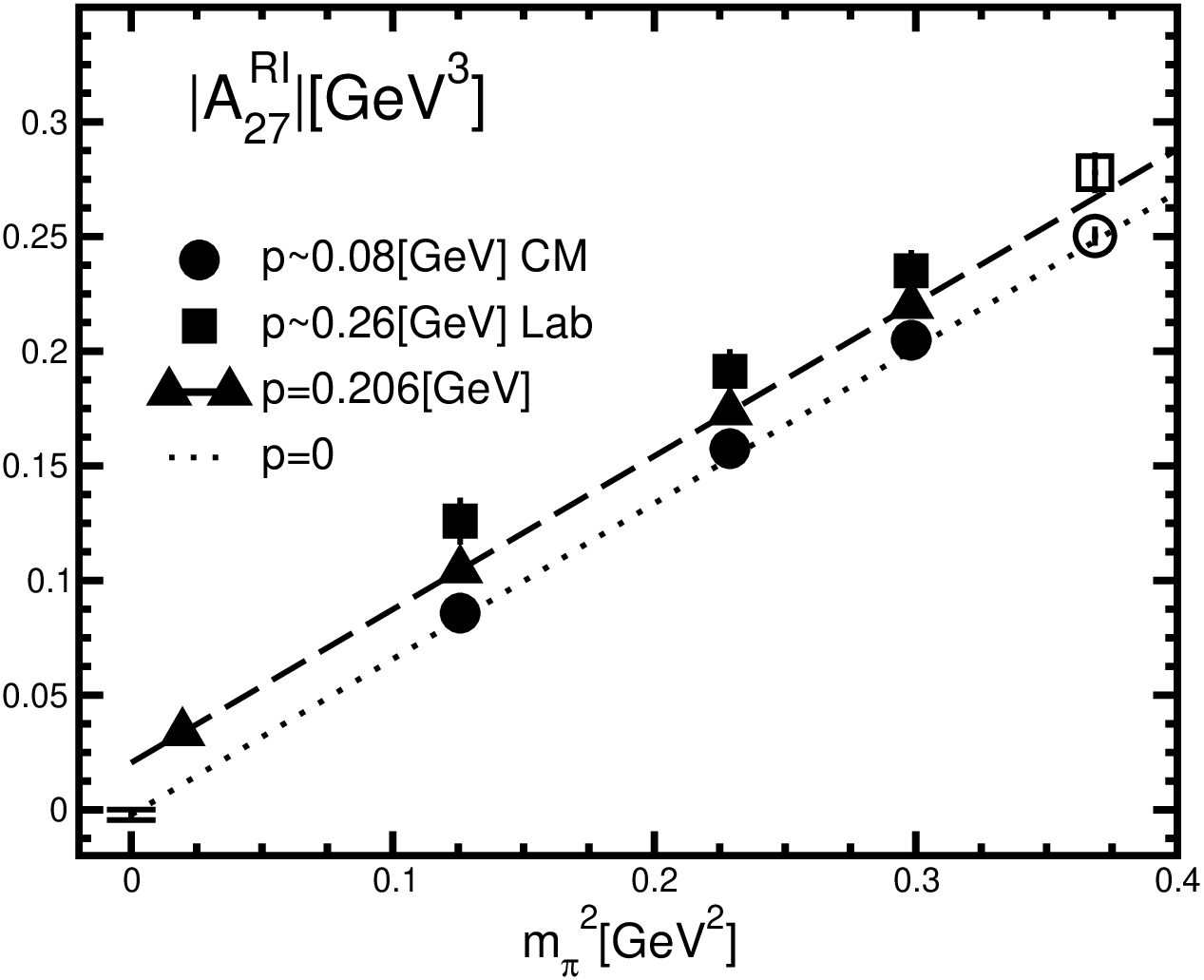}
}
\caption{
Weak matrix element of $27$ operator.
Dotted line is obtained from global fit with $p=0$.
Dashed line and triangle symbol denote global fit result with 
physical momentum $p=206$ MeV.
Open circle and square symbols are omitted in global fits.
}
\label{fig:lwme_ri_27}
\end{figure}

The weak matrix element of the 27 operator is shown in Fig.~\ref{fig:lwme_ri_27}.
The horizontal axis is the pion mass squared.
The matrix elements obtained from the CM and Lab calculations 
are clearly distinguished from each other.
This means that the relative momentum dependence is important in the matrix element.
The figure also shows that the matrix element decreases with the pion mass.
In order to investigate the dependences of the pion mass and relative momentum,
we carry out a global fit of the result with a naive polynomial form,
\begin{equation}
B_{00} + B_{10} m_\pi^2 + B_{01} p^2.
\label{eq:fit_func}
\end{equation}
We omit a $p^4$ term, because our result is obtained at only two different 
momentum in each pion mass, as discussed in Sec.~\ref{subsec:pipi_scattering}.
The value of $\chi^2/$d.o.f. is 8.6 when we fit all the four pion mass data,
so that we exclude the heaviest data to make the 
$\chi^2/$d.o.f. acceptable.
We obtain $B_{00}=-0.0022(23)$ which is consistent with zero within the statistical error.
This implies that the matrix element at $p=0$ vanishes in the chiral limit
as shown in Fig.~\ref{fig:lwme_ri_27} as a dotted line.
Thus, a fit form without the constant term $B_{00}$ also gives a reasonable 
$\chi^2/$d.o.f. as summarized in table~\ref{tab:fit_dcyamp_ri}.

This tendency is consistent with the prediction of 
leading order (LO) ChPT~\cite{Bernard:1985wf,Laiho:2002jq},
\begin{eqnarray}
|A_{27}^{\mathrm{RI}}| &=& 
-\alpha_{27} \frac{12\sqrt{3}}{f^3}(m_K^2 - m^2_\pi)
\label{eq:chpt_0_27}
\\
&=& 
-\alpha_{27} \frac{12\sqrt{3}}{f^3}(3 m^2_\pi + 4 p^2),
\label{eq:chpt_27}
\end{eqnarray}
where $\alpha_{27}$ is a constant and $f$ is the pion decay constant in the chiral limit.
In the second line we use the on-shell condition, 
\begin{equation}
m_K^2 = 4 ( m_\pi^2 + p^2 ),
\end{equation}
because our matrix element is calculated for on-shell states.
The normalization of eq.(\ref{eq:chpt_0_27}) is based on Ref.~\cite{Blum:2001xb}.
It should be noted that while the behavior of the matrix element
near the chiral limit agrees with the LO ChPT,
we omit logarithms predicted at NLO
due to the large pion mass used in our calculation.

We attempt to estimate a systematic error stemming from neglecting the 
NLO log terms in (quenched) ChPT formula~\cite{Lin:2002nq}.
Both the formulae are explained in the appendix, and the results are
summarized in table~\ref{tab:fit_dcyamp_ri_chpt}.
In the fits we assume that the formulae are valid in the lighter pion mass
region, $m_\pi < 0.5$ GeV, and then we use only the data 
at the two lightest pion masses. In this restricted pion mass region,
both the formulae work well.
We estimate a dimensionless quantity 
$\alpha_{27}/f^4$ with $f=0.133$ GeV~\cite{Aoki:2006ib}
using the full (quenched) ChPT formula, and then find that the absolute value is 52(37)\% 
smaller than the one estimated by $B_{10}$ of the linear fit.
This suggests that the determination of $\alpha_{27}/f^4$
has the large systematic error.
In order to remove the systematic error, we need to calculate more data
at the region where ChPT is valid.
While $\alpha_{27}/f^4$ varies largely with the fit form,
the linear fit and the two ChPT fits give consistent results 
at the physical point (see tables~\ref{tab:fit_dcyamp_ri} 
and \ref{tab:fit_dcyamp_ri_chpt}),  
\begin{equation}
m_\pi = 140\ \mathrm{MeV}, \ p = 206\ \mathrm{MeV}, 
\end{equation}
which is determined by the on-shell condition with
the physical kaon mass $m_K = 498$ MeV.

%
%
\subsubsection{88 and $m88$ operators}
\label{subsucsec:A_88_m88}

\begin{figure}[!h]
\centering
\scalebox{0.5}[0.5]{
\includegraphics*{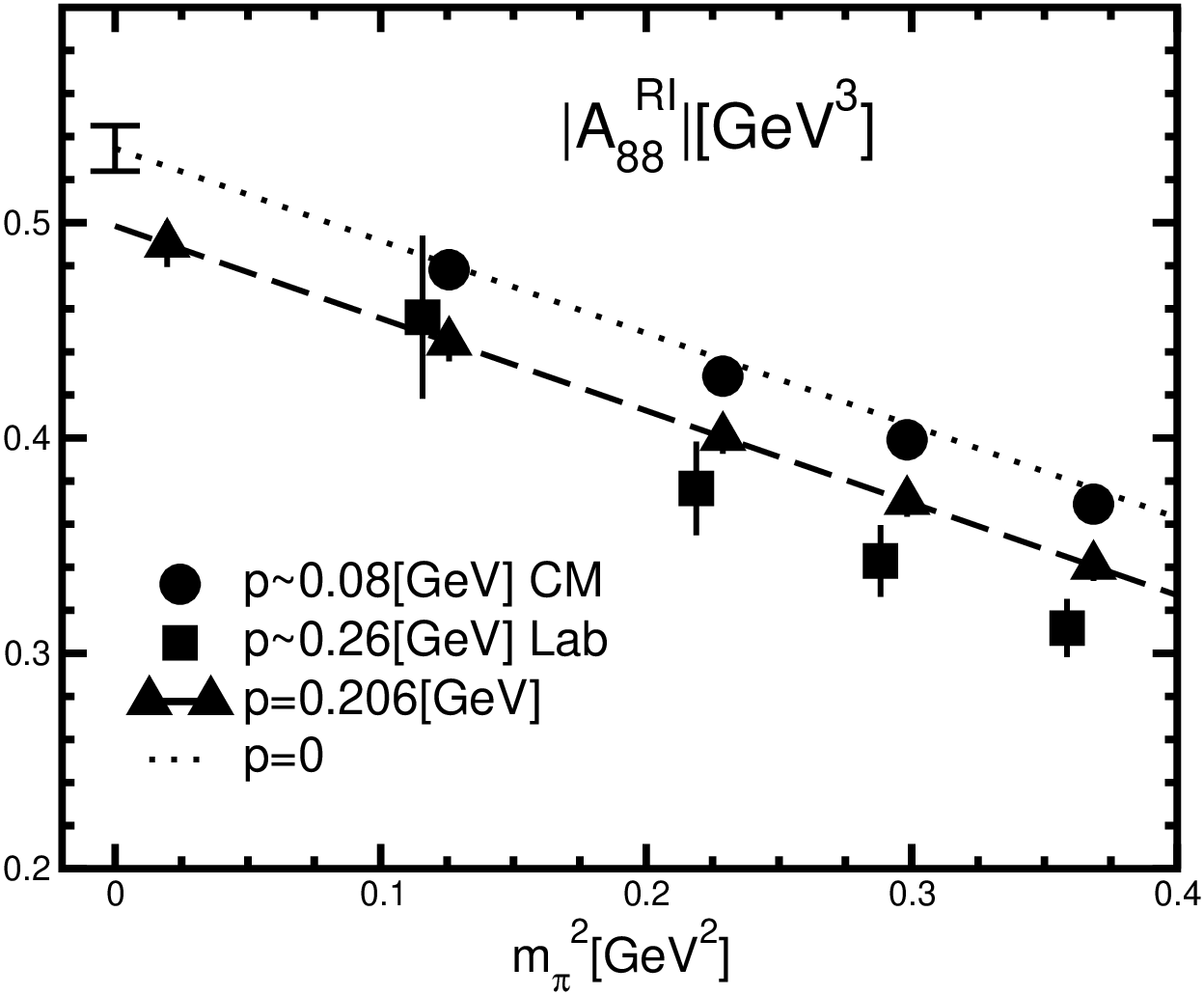}
}
\caption{
Weak matrix element of $88$ operator.
Dotted line is obtained from global fit with at $p=0$.
Dashed line and triangle symbol denote global fit result
with physical momentum $p=206$ MeV.
Square symbol is slightly shifted to minus direction in x-axis.
}
\label{fig:lwme_ri_88}
\end{figure}

\begin{figure}[!h]
\centering
\scalebox{0.5}[0.5]{
\includegraphics*{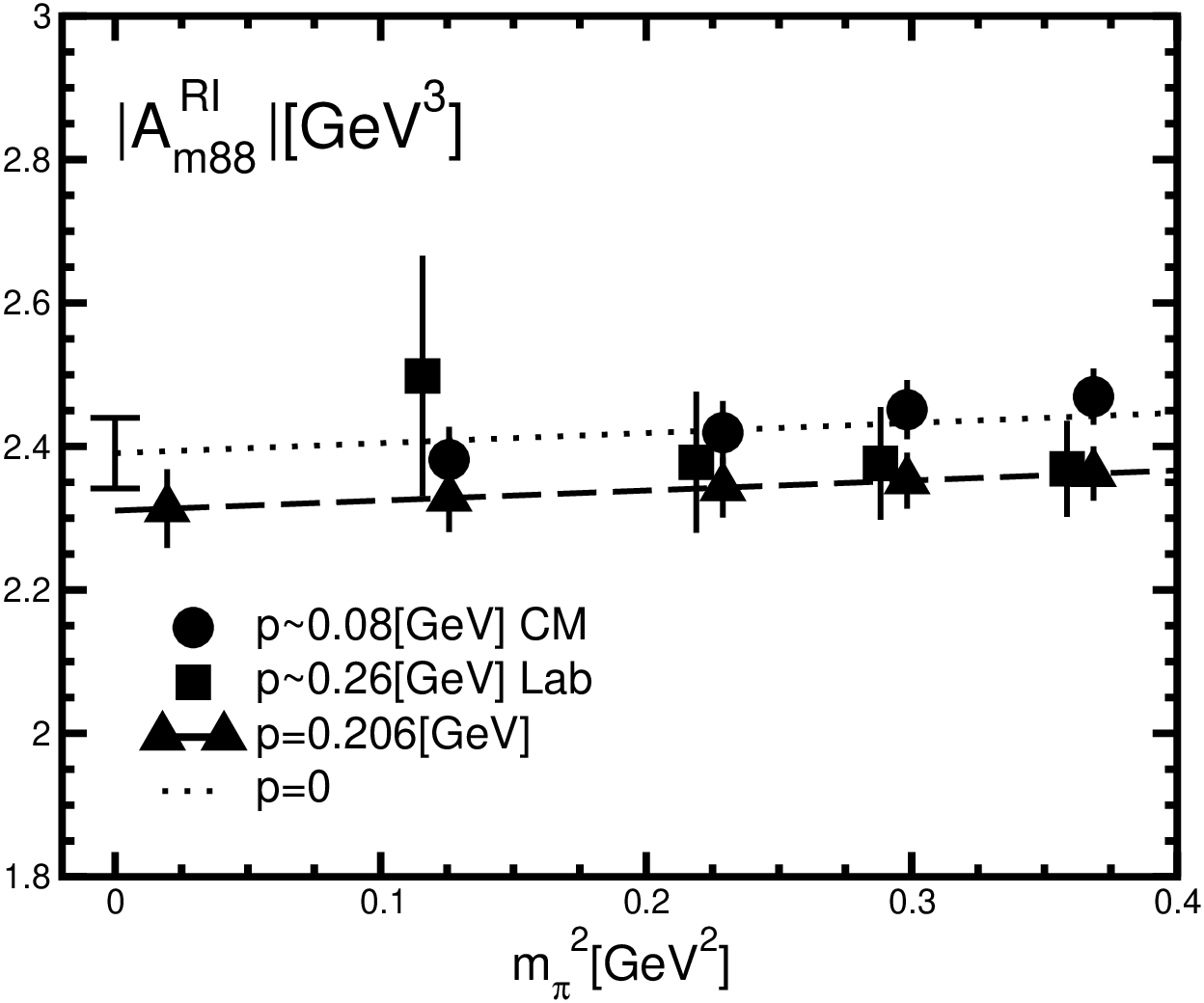}
}
\caption{
Weak matrix element of $m88$ operator.
Dotted line is obtained from global fit with $p=0$.
Dashed line and triangle symbol denote global fit result
with physical momentum $p=206$ MeV.
Square symbol is slightly shifted to minus direction in x-axis.
}
\label{fig:lwme_ri_m88}
\end{figure}

The weak matrix elements of the 88 and $m88$ operators are
presented in Figs.~\ref{fig:lwme_ri_88} and \ref{fig:lwme_ri_m88},
respectively.
The matrix elements of the 88 operator obtained from the different frames
are clearly separated and have appreciable slopes.
The pion mass and relative momentum dependences
are important in the matrix element as in the 27 operator case.
In contrast to the 88 operator, the results of the $m88$ operator
have neither strong pion mass nor relative momentum dependences.
Both the matrix elements do not have clear curvatures with respect
to the pion mass squared. 
Thus, we simply adopt the polynomial function in eq.(\ref{eq:fit_func})
for the global fit of the matrix elements.
In this fit we use the data at all four pion masses. 
The fit results are tabulated in table~\ref{tab:fit_dcyamp_ri}.
In the $m88$ case the $\chi^2$/d.o.f. is relatively larger than the other cases.
This might be due to the poor determination of the covariance matrix of the 
matrix element.
The matrix element at the physical point and the chiral limit is almost independent of fit form, 
because it does not have strong dependences for $m_\pi^2$ and $p^2$.
Therefore we choose to quote this fit as our result.

We plot the fit results at the chiral limit with $p=0$ 
for the $88$ and $m88$ operators
in Figs.~\ref{fig:lwme_ri_88} and \ref{fig:lwme_ri_m88}, respectively.
A constant remains in both the matrix elements at the limits $m_\pi=p=0$.
While the trend is quite different from the $27$ case,
it is consistent with the prediction of LO ChPT~\cite{Bijnens:1983ye,Cirigliano:1999pv,Laiho:2002jq,Laiho:2003uy},
\begin{equation}
|A_i^{\mathrm{RI}}| =  -\alpha_i \frac{24\sqrt{3}}{f^3},\ i = 88, m88,
\label{eq:chpt_88_m88}
\end{equation}
where $\alpha_i$ is a constant.
Again, although these trends agree with the behavior in the chiral limit of LO ChPT,
we omit logarithms which enter NLO in ChPT due to large pion mass.

A systematic error stemming from the fit without log terms 
is estimated using full ChPT formula~\cite{Lin:2002nq}, 
(see Appendix for details).
The quenched formula does not work in both the matrix elements as presented
in table~\ref{tab:fit_dcyamp_ri}. Therefore, 
we will not discuss the fit result obtained with the quenched formula.
We use the same assumption for the pion mass as in the 27 operator case, 
and then fit with only the two lightest pion mass data. 
It is found that a dimensionless quantity 
$\alpha_i/f^6$ ($i=88,m88$) in the ChPT fit is reasonably consistent with
that of the linear fit, determined by $B_{00}$.
The differences are less than 11\%. This means that the
low energy constants are less sensitive to the NLO logs 
than that in the 27 operator.
As presented in tables~\ref{tab:fit_dcyamp_ri} and
\ref{tab:fit_dcyamp_ri_chpt},
the polynomial and ChPT fit results at the physical point,
$m_\pi = 140$ MeV and $p = 206$ MeV, 
also reasonably agree with each other.

%
%
\subsubsection{Comparison between direct and indirect methods}
\label{subsucsec:comp}

We attempt to compare the results for the direct and indirect methods
through the dimensionless parameters in LO ChPT,
$\alpha_{27}/f^4$ and $\alpha_i/f^6$ ($i=88, m88$),
which have been already discussed in the previous sections.
The results are tabulated in table~\ref{tab:lochpt_constants} as well as
those obtained from the previous indirect calculation~\cite{Blum:2001xb}.
The constants of the previous work are estimated with 
$f = 0.137$ GeV~\cite{Blum:2000kn} and the renormalization
scale $\mu = 2.13$ GeV.
We find that the results for $\alpha_{88}$ and $\alpha_{m88}$
are almost consistent in both the methods,
while $\alpha_{27}$ differs by a factor of 2.6.
Our result of $\alpha_{27}$, however, contains a large systematic error
of the chiral extrapolation, and varies by a factor of 1/2 
as can be found in table~\ref{tab:lochpt_constants}.
Other possible systematic errors in this comparison are
different choice of parameters, such as lattice spacing and 
the renormalization scale, and different choice of gauge action.

These parameters in LO ChPT obtained from the direct method should be 
consistent with those from the indirect method if no systematic error is included.
This is because the final state interaction effect 
of the two pions in the matrix elements vanishes 
in both the limits $m_\pi^2 = p^2 = 0$.
However, more strict consistency check of these parameters 
is beyond this work,
because many systematic errors are included in the comparison 
discussed in the above.
For this check, we need to calculate the matrix elements
with both the direct and indirect methods on exactly same configurations
at the lighter pion masses.

%
%
%
\subsection{Re$A_2$ and Im$A_2$}

We calculate $\mathrm{Re}A_2$ and $\mathrm{Im}A_2$ from the weak matrix elements $|A_{27}^{\mathrm{RI}}|$,
$|A_{88}^{\mathrm{RI}}|$, and $|A_{m88}^{\mathrm{RI}}|$.
The definition of the decay amplitude $A_I$ is given by
\begin{eqnarray}
\langle (\pi\pi)_I | H_W | K^0 \rangle 
&\equiv&
A_I e^{i\delta_I} \\
&=&
\frac{G_F}{\sqrt{2}} V_{ud} V_{us}^{*} 
\left[
\sum_{i=1}^{10} ( z_i(\mu) + \tau y_i(\mu) ) \langle Q_i \rangle_I (\mu)
\right],
\label{eq:def_A_2}
\end{eqnarray}
where the index $I$ denotes the isospin,
$z_i(\mu)$ and $y_i(\mu)$ are the Wilson coefficients at the scale $\mu$,
and other parameters are presented in table~\ref{tab:input_param}.
The relations between the weak matrix elements $\langle Q_i \rangle_2$
and $|A_j^{\mathrm{RI}}|$ for $j=27,88,m88$ are listed below,
\begin{eqnarray}
|A_{27}^{\mathrm{RI}}| &=& 3 | \langle Q_1 \rangle_2 | = 3 | \langle Q_2 \rangle_2 |
= 2 | \langle Q_9 \rangle_2 | = 2 | \langle Q_{10} \rangle_2 | \\
|A_{88}^{\mathrm{RI}}| &=& 2 | \langle Q_7 \rangle_2 | \\
|A_{m88}^{\mathrm{RI}}| &=& 2 | \langle Q_8 \rangle_2 |.
\end{eqnarray}
The other weak matrix elements vanish in the $I=2$ case, {\it i.e.},
$\langle Q_3 \rangle_2 = \langle Q_4 \rangle_2 = \langle Q_5 \rangle_2 = \langle Q_6 \rangle_2 = 0$.
The detail of the calculation of the Wilson coefficients 
is in Ref.~\cite{Kim:2004mk}.
The coefficients are evaluated in the NDR scheme~\cite{Buchalla:1995vs},
which are converted to the RI scheme at the scale $\mu = 1.44$ GeV~\cite{Kim:2004mk}.
The values of the Wilson coefficients are summarized in table~\ref{tab:wilson_coef}.

%
%
%
\subsubsection{Re$A_2$}
\label{subsec:re_a2}

The real part of $A_2$ is evaluated through the following equation,
\begin{equation}
\mathrm{Re} A_2 = 
\frac{G_F}{\sqrt{2}} |V_{ud}| |V_{us}| 
\left[
\sum_{i=1}^{10} 
( z_i(\mu) + \mathrm{Re}(\tau) y_i(\mu) )
| \langle Q_i \rangle_2 | (\mu)
\right].
\end{equation}
$\mathrm{Re}A_2$ obtained in the CM and Lab frame calculations
is tabulated in table~\ref{tab:rea2}.

\begin{figure}[!h]
\centering
\scalebox{0.5}[0.5]{
\includegraphics*{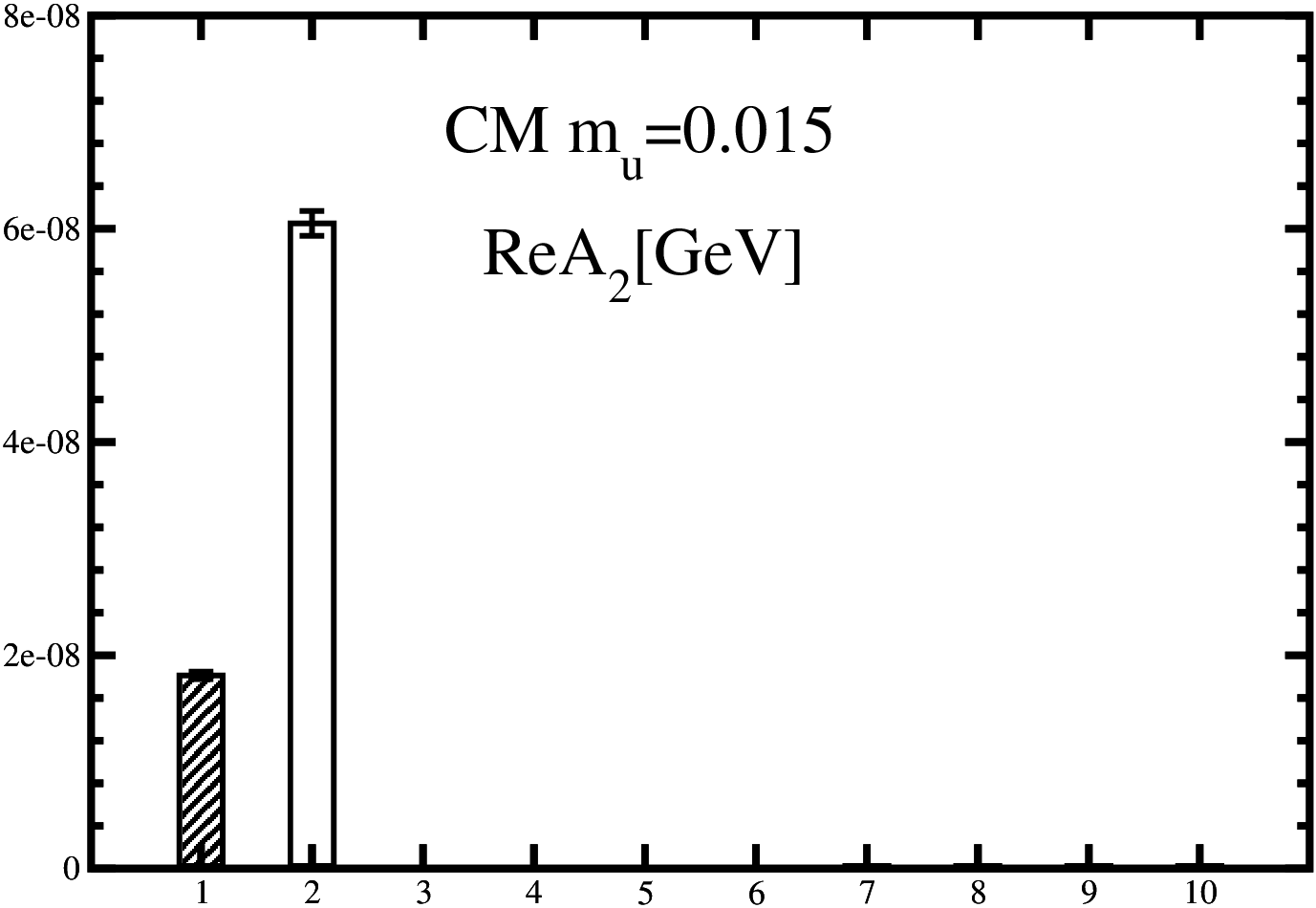}
}
\\
\scalebox{0.5}[0.5]{
\includegraphics*{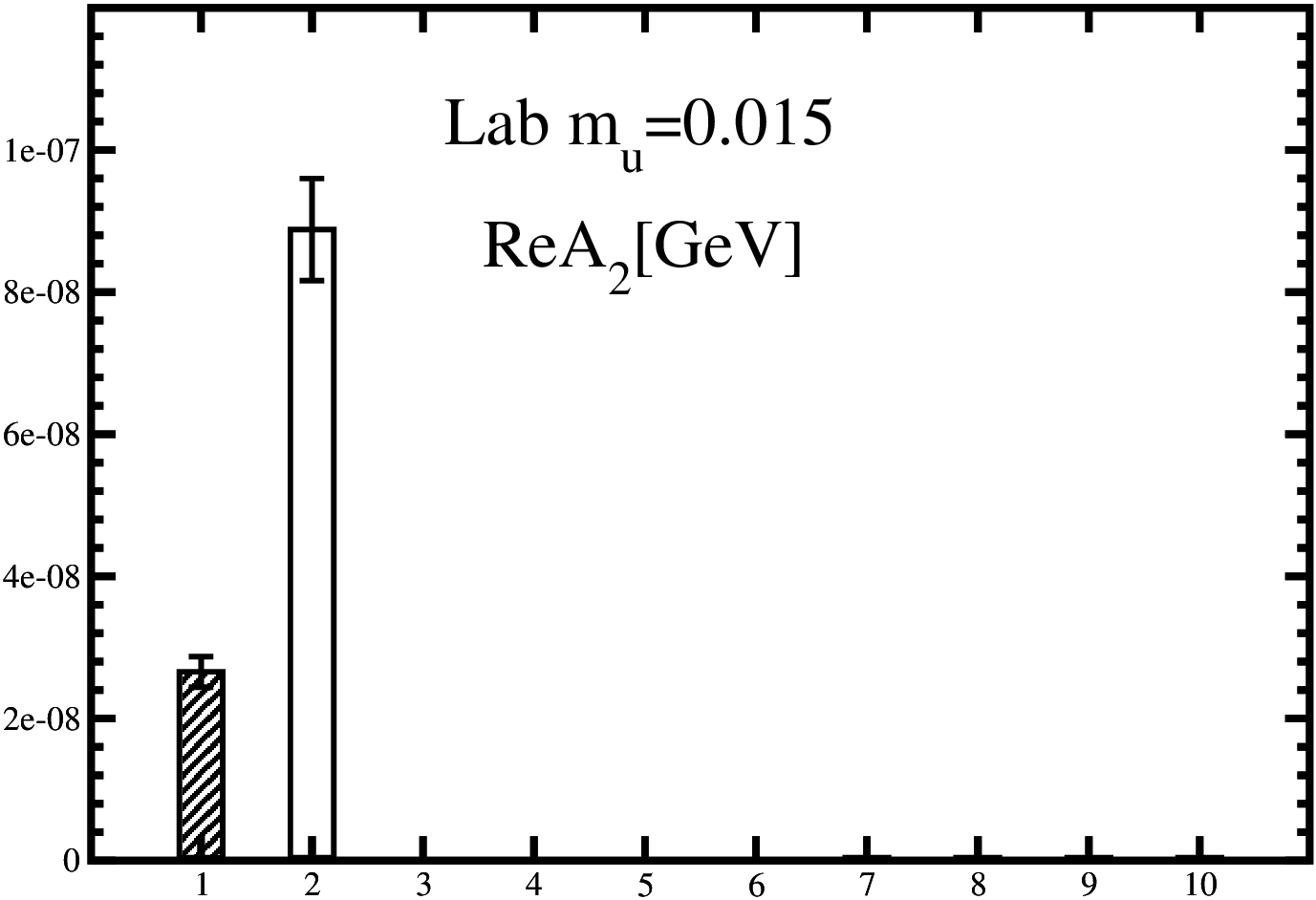}
}
\caption{
Absolute values of each weak operator contribution for Re$A_2$ at
lightest quark mass in CM(upper panel) and Lab(lower panel) calculations.
Open and stripe bars denote positive and negative contributions,
respectively.
Signs of operator 1 and 8 contributions are changed.
}
\label{fig:oprt_rea2_CM_Lab}
\end{figure}

The upper panel of Fig.~\ref{fig:oprt_rea2_CM_Lab} shows that the main contribution of $\mathrm{Re}A_2$ 
comes from $|\langle Q_1 \rangle_2|$ and $|\langle Q_2 \rangle_2|$ in the CM calculation with the
lightest pion mass (contributions come only from $|A_{27}^{\mathrm{RI}}|$).
This is consistent with previous results~\cite{Noaki:2001un,Blum:2001xb}.
The trend is not changed in the Lab calculation as shown in 
the lower panel of Fig.~\ref{fig:oprt_rea2_CM_Lab}.

We will not discuss the chiral extrapolations of Re$A_2$ in detail,
because it is essentially same as those of $|A_{27}^{\mathrm{RI}}|$
in Sec.~\ref{subsucsec:A_27}. In spite of this, we just summarize
the result of the chiral extrapolations 
in table~\ref{tab:fit_rea2}.
At the physical point, we obtain 
$\mathrm{Re}A_2 = 1.66(23)(^{+48}_{-03})\times 10^{-8}$ GeV,
where the first and second errors are statistical and systematic, respectively.
The central value and the first error are obtained from the simple polynomial
fit eq.(\ref{eq:fit_func}) with $B_{00}=0$.
The systematic error is estimated by comparing the central value with
the result from a fit with an added $m_\pi^2 p^2$ term,
\begin{equation}
B_{10} m_\pi^2 + B_{01} p^2 + B_{11} m_\pi^2 p^2,
\label{eq:fit_rea2_4}
\end{equation}
and also (quenched) ChPT fit using the two lightest pion mass data
as in Sec.~\ref{subsucsec:A_27}.
The larger systematic error stems from the polynomial fit result
with eq.~(\ref{eq:fit_rea2_4})
as shown in table~\ref{tab:fit_rea2}, while the ChPT fits give
consistent results with the one obtained from the simple polynomial fit,
eq.(\ref{eq:fit_func}) with $B_{00}=0$, as in $|A_{27}^{\mathrm{RI}}|$.

\begin{figure}[!h]
\centering
\scalebox{0.5}[0.5]{
\includegraphics*{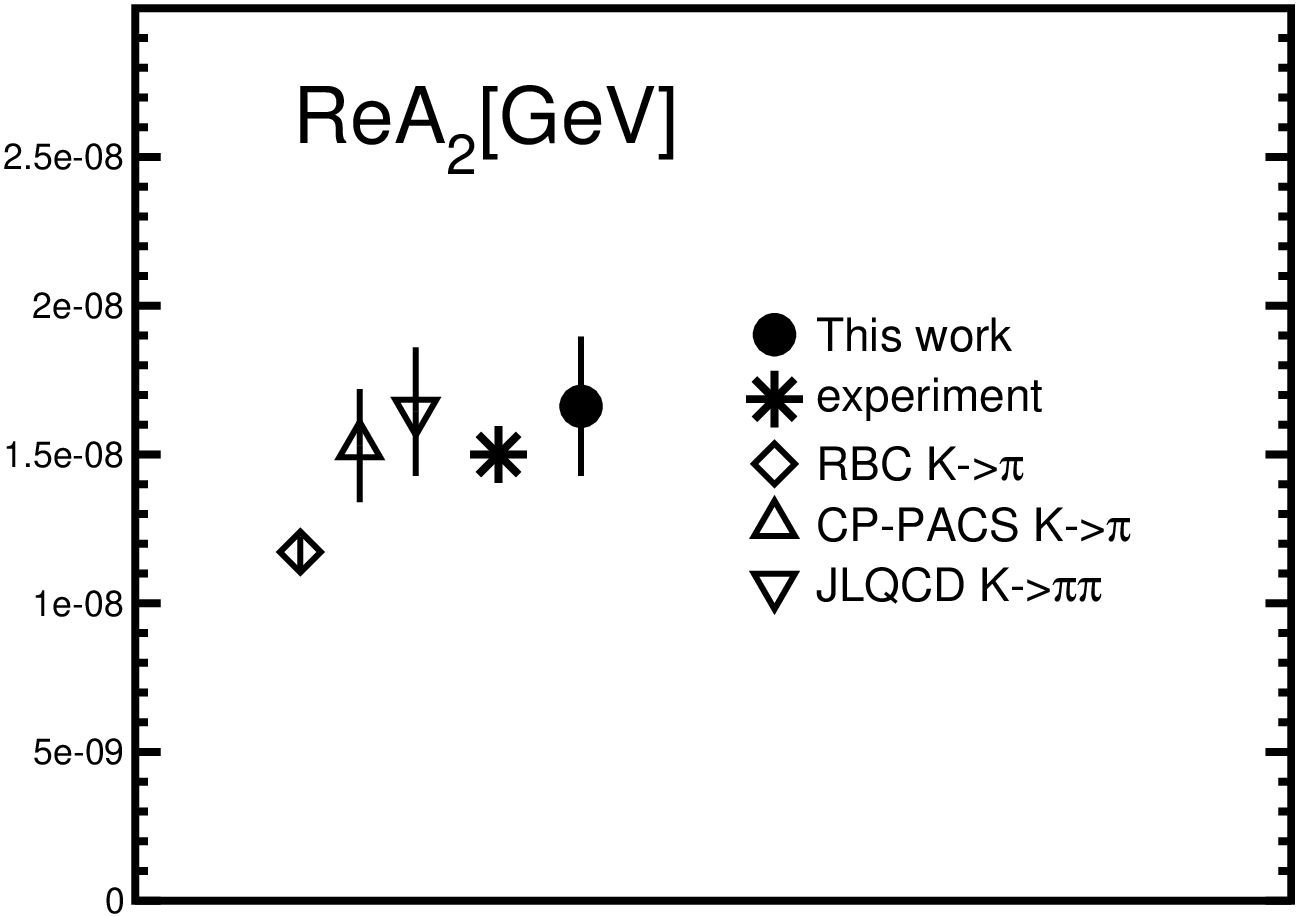}
}
\caption{
Re$A_2$ obtained from global fit results denoted by circle symbol with
only statistical error.
Results of previous works~\cite{Noaki:2001un,Blum:2001xb,Aoki:1997ev} 
are also plotted by open symbols.
Star symbol is experimental result.
}
\label{fig:rea2_comp}
\end{figure}

In Fig.~\ref{fig:rea2_comp} we plot the result at the physical point
as well as the previous results obtained with
the indirect method~\cite{Noaki:2001un,Blum:2001xb},
direct calculation with ChPT~\cite{Aoki:1997ev}, and the experiment.
Our result reasonably agrees with the experimental value and 
also the previous results
except the result in Ref.~\cite{Blum:2001xb} (choice 2 with $\mu = 2.13$ GeV).
The difference of Re$A_2$ between our result and the RBC result
is a factor of 1.4.
This difference is smaller than what we observed in $\alpha_{27}/f^4$ 
as described in Sec.~\ref{subsucsec:A_27}.
We may consider that Re$A_2 \propto \alpha_{27} ( 3m_\pi^2 + 4 p^2 )$
according to the LO ChPT relation eq.~(\ref{eq:chpt_27}),
so that the reduction of the discrepancy between our result and the
RBC result, found in $\alpha_{27}/f^4$ to that of Re$A_2$, seems inconsistent 
with the expectation derived from the LO relation.
Indeed, our polynomial fit with $B_{00} = 0$
is not consistent with LO ChPT, as 
$B_{10}/3 > B_{01}/4$ can be read off in table~\ref{tab:fit_rea2},
although both the quantities give the same value of $\alpha_{27}/f^4$ if LO ChPT is valid.
This inconsistency between $B_{10}/3$ and $B_{01}/4$ causes the smaller 
difference between ours and the RBC result of Re$A_2$ than that 
of $\alpha_{27}/f^4$.
Again, this is due to the large systematic uncertainty 
during the determination of $\alpha_{27}/f^4$.

%
%
%
\subsubsection{Im$A_2$}
\label{subsec:im_a2}

The imaginary part of $A_2$ is determined through the definition of the decay amplitude,
eq.~(\ref{eq:def_A_2}), as
\begin{equation}
\mathrm{Im} A_2 = 
\frac{G_F}{\sqrt{2}} |V_{ud}| |V_{us}| 
\left[
\sum_{i=7}^{10} 
\mathrm{Im}(\tau) y_i(\mu)
| \langle Q_i \rangle_2 | (\mu)
\right].
\end{equation}
The results at each pion mass and frame are 
tabulated in table~\ref{tab:ima2}.

\begin{figure}[!h]
\centering
\scalebox{0.5}[0.5]{
\includegraphics*{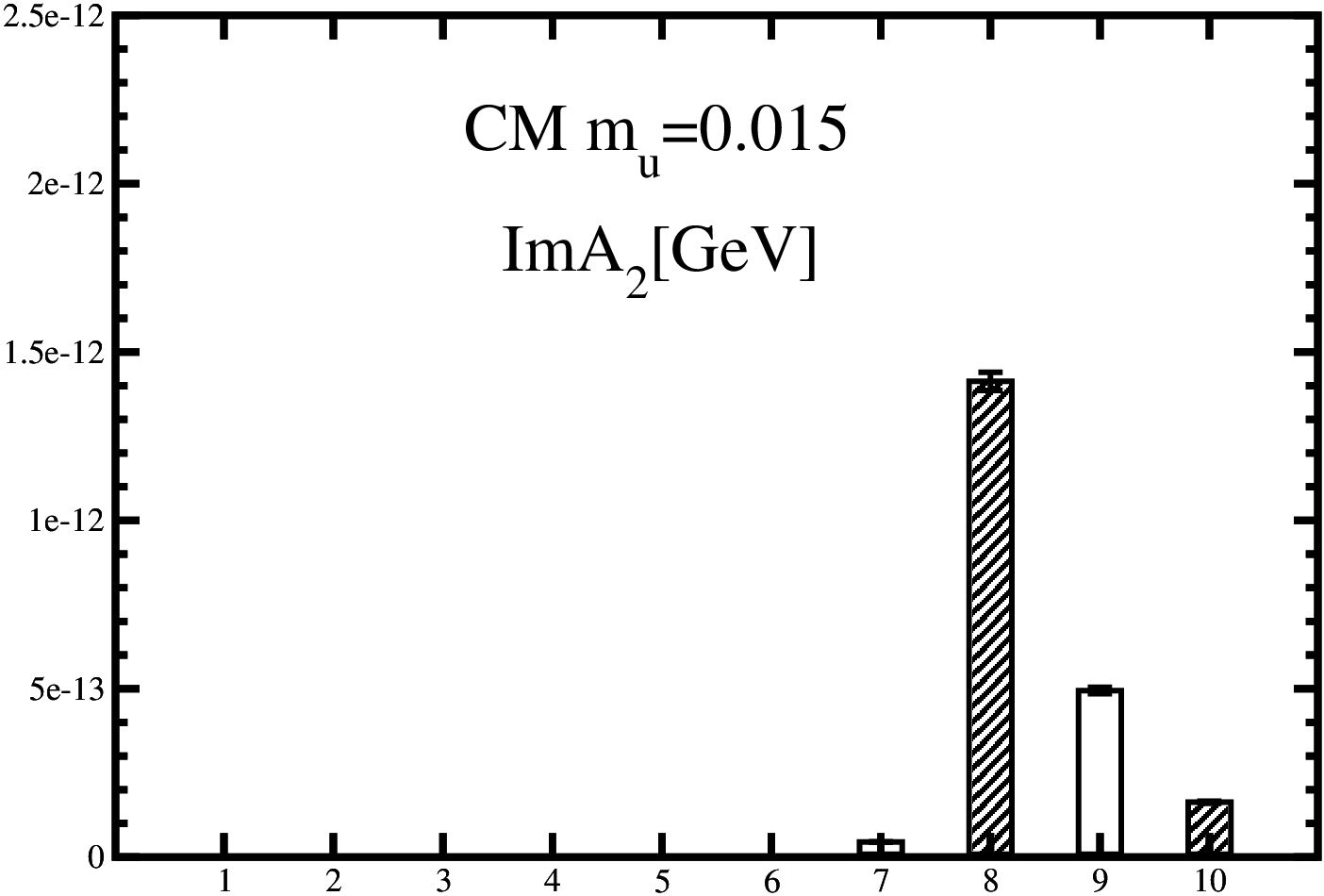}
}
\\
\scalebox{0.5}[0.5]{
\includegraphics*{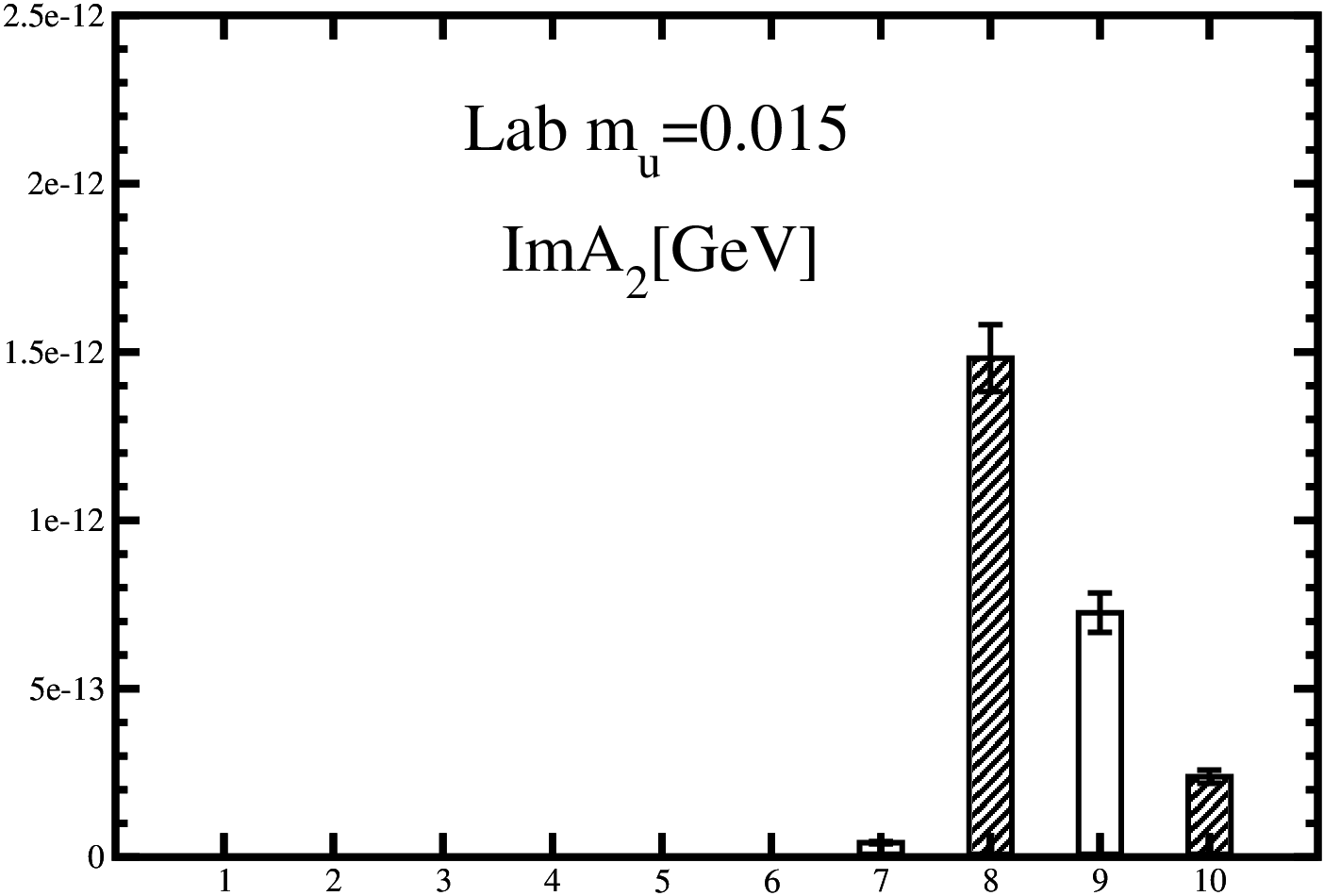}
}
\caption{
Absolute values of each weak operator $Q_i$ contribution for Im$A_2$ at
lightest quark mass in CM(upper panel) and Lab(lower panel) calculations.
Open and stripe bars denote positive and negative contributions,
respectively.
Signs of operator 8 and 10 contributions are changed.
}
\label{fig:oprt_ima2_CM_Lab}
\end{figure}

\begin{figure}[!h]
\centering
\scalebox{0.5}[0.5]{
\includegraphics*{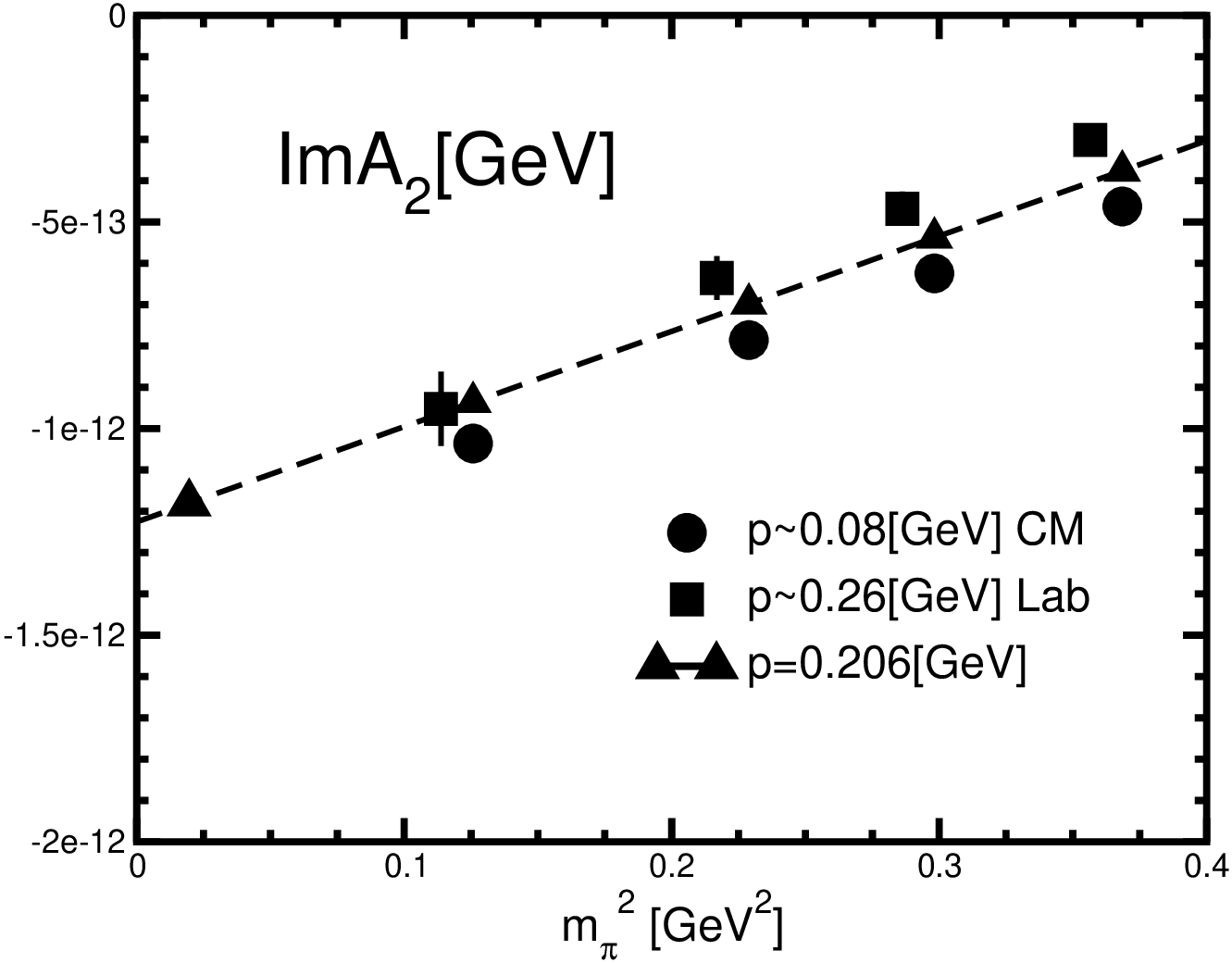}
}
\caption{
Im$A_2$ obtained from CM and Lab calculations
and its global fit.
Triangle symbol denotes fit result with physical momentum $p=206$ MeV.
Dashed line is chiral extrapolation to physical point.
Square symbol is slightly shifted to minus direction in x-axis.
}
\label{fig:ima2}
\end{figure}

\begin{figure}[!h]
\centering
\scalebox{0.5}[0.5]{
\includegraphics*{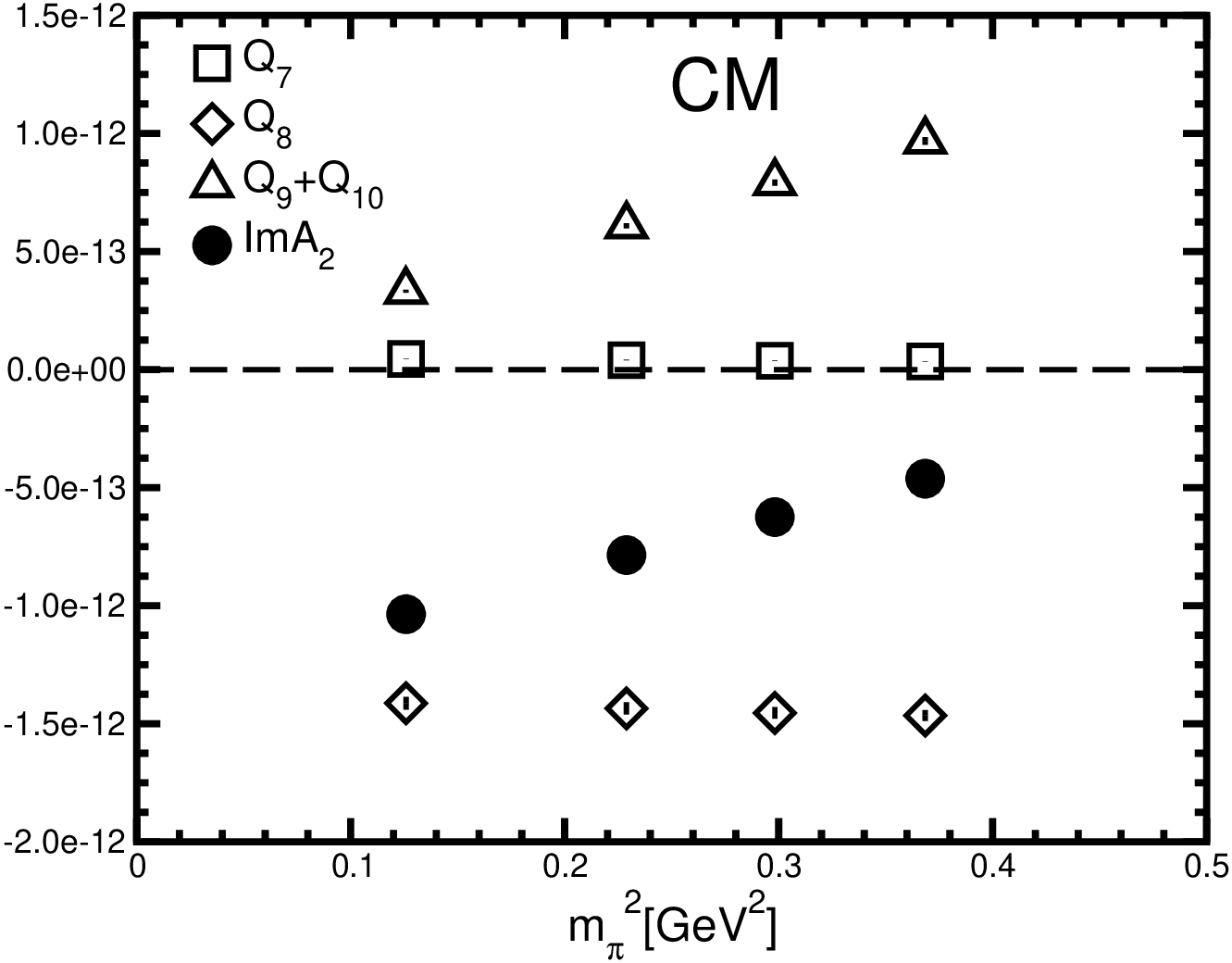}
}
\caption{
Pion mass dependence of each contribution for Im$A_2$ in CM calculation.
Open and closed symbol denote each and total contribution.
}
\label{fig:each_ima2}
\end{figure}

Figure~\ref{fig:oprt_ima2_CM_Lab} shows the contribution
of each operator to Im$A_2$ in the CM and Lab calculations at the lightest pion mass.
In both frames
the largest contribution to Im$A_2$ is $|\langle Q_8 \rangle_2|$,
which is constructed only from $|A^{\mathrm{RI}}_{m88}|$,
and the second largest one is $|\langle Q_9 \rangle_2|$ with the opposite sign
and comes from $|A^{\mathrm{RI}}_{27}|$.
Figure~\ref{fig:ima2} shows that Im$A_2$ has a significant slope for $m_\pi^2$,
while the leading contribution of Im$A_2$, $|A^{\mathrm{RI}}_{m88}|$ does
not have such a slope as can be seen in Fig.~\ref{fig:lwme_ri_m88}.
This slope is caused by the large pion mass dependence of 
$|\langle Q_9 \rangle_2|$ and $|\langle Q_{10} \rangle_2|$,
or $|A^{\mathrm{RI}}_{27}|$,
as shown in Fig.~\ref{fig:each_ima2} for the CM case.
This tendency does not change in the Lab calculation.

We estimate Im$A_2$ at the physical point using
the same polynomial fit form, eq.~(\ref{eq:fit_func}), 
because the largest contribution of 
Im$A_2$ is given by $|A^{\mathrm{RI}}_{m88}|$.
The result with the physical momentum and its chiral extrapolation are 
plotted in Fig.~\ref{fig:ima2}.
We obtain
Im$A_2 = -1.181(26)(^{+141}_{-014})\times 10^{-12}$ GeV at the
physical point.
Again, the central value and statistical error are obtained from a fit to
eq.~(\ref{eq:fit_func}), while the systematic error is determined by
comparing the central value with that from a fit form 
with an added $m_\pi^2 p^2$ term,
\begin{equation}
B_{00} + B_{10} m_\pi^2 + B_{01} p^2 + B_{11} m_\pi^2 p^2,
\label{eq:fit_ima2_3}
\end{equation}
and ChPT formula for 88 and $m88$ operators, explained in appendix, 
with the data at the two lightest pion masses as in Sec.~\ref{subsucsec:A_88_m88}.
In this case, the quenched ChPT formula does not give a reasonable $\chi^2/$d.o.f. as in
$|A_{88,m88}^{\mathrm{RI}}|$.
The larger systematic error comes from the fit with the ChPT formula.
These fit results are presented in table~\ref{tab:fit_ima2}.

\begin{figure}[!h]
\centering
\scalebox{0.5}[0.5]{
\includegraphics*{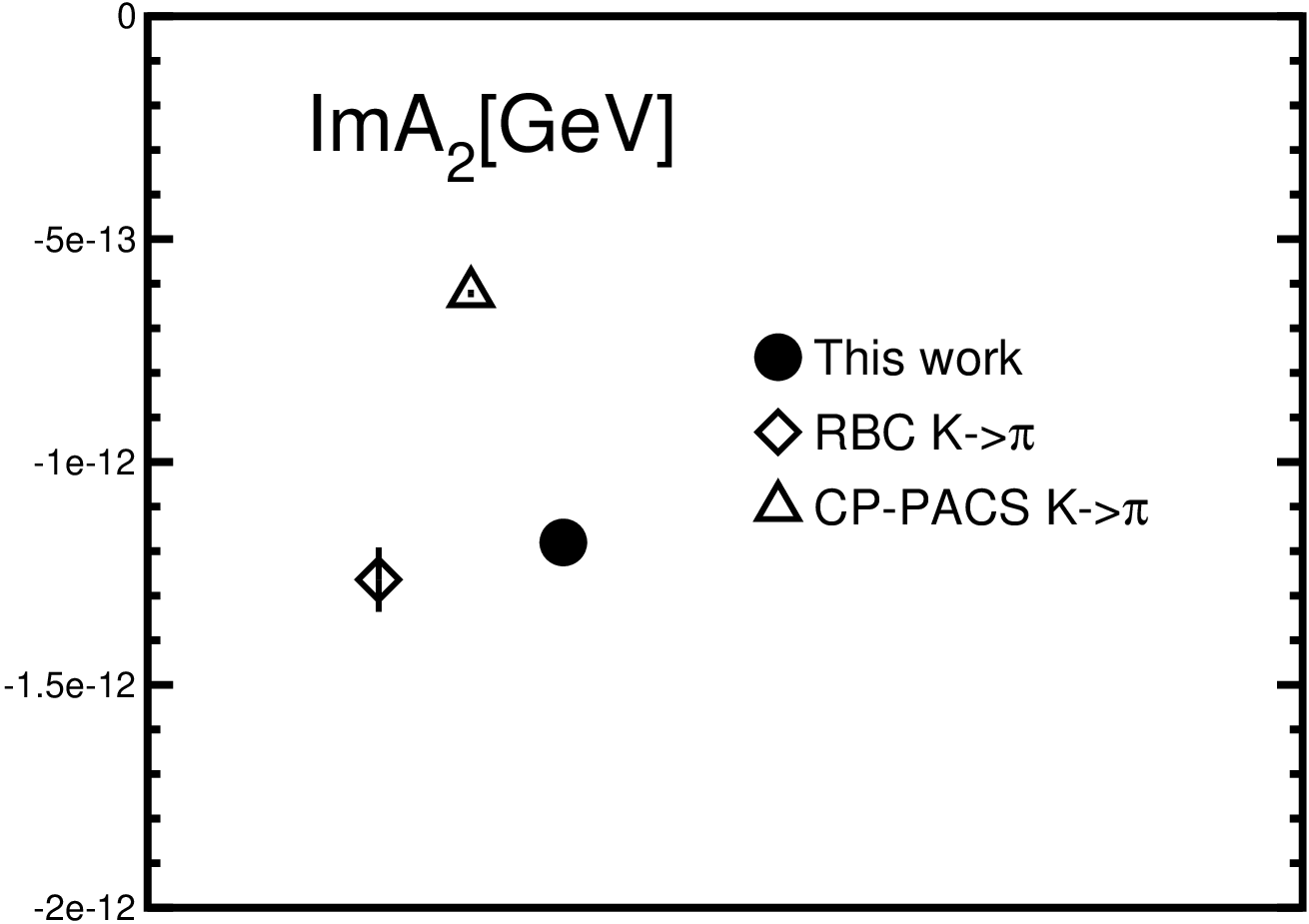}
}
\caption{
Circle symbol is Im$A_2$ obtained form global fit.
Error is only statistical.
Results of previous works~\cite{Noaki:2001un,Blum:2001xb} 
are also plotted by open symbols.
}
\label{fig:ima2_comp}
\end{figure}

The result at the physical point is compared with the previous 
indirect calculation results~\cite{Noaki:2001un,Blum:2001xb} in Fig.~\ref{fig:ima2_comp}.
The differences of our result from CP-PACS~\footnote{
CP-PACS result is estimated from $P^{(3/2)}$ 
on $24^3\times 32$ lattice of $\Lambda^{(4)}_{\overline{\mathrm{MS}}} = 325$
MeV with a linear chiral extrapolation
using the parameters in the reference.} 
and RBC (choice 2 with $\mu = 2.13$ GeV) results are a factor of 1.9
and 0.93, respectively.
However, from the comparison we cannot conclude that the final state interaction effect 
is appreciable or not in Im$A_2$,
because the previous works employed different simulation parameters from
our calculation, and also
even the two results with the indirect method are inconsistent.
Moreover our result has about 10\% systematic error of the chiral extrapolation
estimated in the above.
More a detailed comparison between the direct and indirect methods
is required for making a firm conclusion how much 
the final state interaction effect is in this quantity.

%
%
%
\subsubsection{Systematic error from definition of momentum}
\label{subsec:systematic_error_def_mom}

A systematic error for Re$A_2$ and Im$A_2$ arises from 
the violation of the naive dispersion relation
on lattice. This error would decrease toward the continuum limit.
To estimate the systematic error, 
we compare the results in the above Secs.~\ref{subsec:re_a2} and \ref{subsec:im_a2}
with the ones given in 
the analysis with the lattice dispersion relation
in Refs.~\cite{Rummukainen:1995vs,Aoki:2007rd},
\begin{equation}
\cosh(E^P_\pi) = \cosh(m_\pi) + 2 \sin^2 ( P / 2 ).
\end{equation}
In the analysis eqs.(\ref{eq:relative_mom}) and 
(\ref{eq:boost}) are replaced by
\begin{eqnarray}
2 \sin^2(p/2) &=& \cosh(E_{\pi\pi}/2) - \cosh(m_\pi),\\
\cosh(E_{\pi\pi}) &=& \cosh(E_{\pi\pi}^P) - 2 \sin^2 ( P / 2 ),
\end{eqnarray}
respectively. Basically, these equations provide a different
relative momentum $p$ from the naive analysis.

In the CM results, the change of the relative momentum is small, 0.6--1.8\%
in all the four masses, which increases as $m_\pi$ increases.
The changes for Re$A_2$ and Im$A_2$ are less than 1.5\% in this case.
In contrast to the CM case, $p$ differs by 2.7--5.2\% in the Lab calculation.
The difference also increases as $m_\pi$ increases.
At the heaviest pion mass the analysis gives 12(11)\% smaller(larger) 
value than the naive dispersion analysis in Re$A_2$(Im$A_2$).
The main source of this difference is the 
conversion factor whose difference is 8.6\% at the heaviest point.
At the lightest pion mass the difference in Re$A_2$(Im$A_2$)
is reduced to 5.8(5.2)\%, where the difference of the conversion
factor is 3.0\%.
This suggests that the systematic
error decreases as the pion mass gets lighter at fixed relative momentum,
or as the momentum decreases at fixed pion mass.

In this analysis we determine Re$A_2$ and Im$A_2$ at the physical point 
with the same fit forms as used in the above sections. For Re$A_2$ the fit
form, eq.(\ref{eq:fit_func}) with $B_{00}=0$,
no longer gives a reasonable $\chi^2$/d.o.f., because the Lab results at the heavier point are largely changed.
On the other hand, eq.(\ref{eq:fit_rea2_4})
gives a consistent result, 
$\mathrm{Re}A_2 = 2.19(36)\times 10^{-8}$ GeV, 
at the physical point compared to the one given by using the same form as in Sec.~\ref{subsec:re_a2}.
For Im$A_2$ the consistent result is obtained with the fit form eq.(\ref{eq:fit_func}),
while the result with eq.(\ref{eq:fit_ima2_3}) is larger,
$\mathrm{Im}A_2 = -1.137(37)\times 10^{-12}$ GeV,
than the corresponding one in Sec.~\ref{subsec:im_a2}.

We estimate the systematic errors for Re$A_2$ and Im$A_2$
at the physical point arising from this analysis
by comparing the central values with the ones obtained in the above sections.
We combine them with the previous systematic errors, 
and finally quote 
\begin{eqnarray}
\mathrm{Re}A_2 &=& 1.66(23)(^{+48}_{-03})(^{+53}_{-0})\times 10^{-8}\ \mathrm{GeV},\\
\mathrm{Im}A_2 &=& -1.181(26)(^{+141}_{-014})(^{+44}_{-0})\times 10^{-12}\ \mathrm{GeV},
\end{eqnarray}
where the first error is statistic, and the second and third are systematic ones.

%
%
%
\section{Conclusions}
\label{sec:conclusions}

In this article we have presented our results of the $\Delta I = 3 / 2$ kaon weak matrix elements 
calculated with non-zero total momentum using the quenched approximation on a coarse lattice; $a^{-1} = 1.31(4)$ GeV.
The calculation is carried out with an extension of the Lellouch and L\"uscher formula,
recently proposed by two groups, to obtain the infinite volume, on-shell, decay amplitude.
It is very encouraging that we have obtained the on-shell weak matrix elements, taking into account final state interactions properly,
with a reasonable statistical error.
We have found that our result of Re$A_2$ at the physical point is reasonably consistent with the experimental value, and Im$A_2$ is comparable with previous results using 
the indirect method.
While we have attempted to compare our results with those from the 
indirect method, many systematic errors are involved in the 
comparison, {\it e.g.,} chiral extrapolations and different simulation parameters 
from previous calculations. A more comprehensive investigation is required.

Another systematic uncertainty may come from the fact that we use the heavy strange quark masses, which we have used in order to evaluate
the decay amplitude closely at the on-shell kinematical point in this study.
The difference of the definition of the relative momentum on lattice may cause
an additional systematic error.
To get rid of all the systematic errors related to a rigorous satisfaction 
of the on-shell condition,
we need either the lighter pion mass or larger volume simulations to make the two-pion energy closer to the physical kaon mass, and also the simulations at finer lattice spacing
are preferable. 

We also note that although
chiral log behavior in the weak matrix elements are expected from ChPT, we have not seen such effects in our data.
The problem may stem from the large pion mass and the coarse lattice spacing
in our simulation because the prediction of ChPT is valid only
in small pion mass region and may also be modified by non-zero lattice spacing effects.
To confirm expected ChPT behavior, the simulations  at the smaller pion mass
and finer lattice spacing are again required. 

Apparently the systematic error due to quenching is uncontrolled
in this study, so that we should extend the present calculation in the dynamical lattice simulation as well. Besides the investigation of the unrevealed systematic errors,
further investigation of the final state interaction remains an important future work.
It is also important to calculate the $\Delta I = 1 / 2$ kaon weak matrix elements
with this method to evaluate the CP violation parameter 
$\varepsilon^\prime / \varepsilon$ and 
the $\Delta I = 1/2$ selection rule.
However, it is more difficult than the present $\Delta I= 3/2$ case
because the $I = 0$ $\pi\pi$ scattering contribution 
enters in the $\Delta I = 1 / 2$ case.

\section*{Acknowledgments}

I thank our colleagues in the RBC and UKQCD collaborations and especially 
N.~H.~Christ, C.~Kim, and A.~Soni for useful discussion
and comments on the first draft of the manuscript,
and T.~Blum and S.~Sasaki for their careful reading of the manuscript.
I thank Columbia University, University of Edinburgh, 
PPARC, RIKEN, BNL and the U.S. DOE 
for providing the QCDOC supercomputers used in this work.
I was supported by the U.S. DOE under contract
DE-FG02-92ER40716.

%
%
%
\appendix
\section{NLO ChPT formulae at physical kinematics}
\label{app:chpt_27_88}

The NLO ChPT formula for the 27 and 88 operators 
at the physical kinematics in full and quenched case 
is obtained in Ref.~\cite{Lin:2002nq}. Following the analysis in 
Ref.~\cite{Boucaud:2004aa}, we rewrite $f_\pi$ and $f_K$ in the
formula in terms of $f$.
Thus, we use the following fit forms in ChPT analysis,
\begin{eqnarray}
|A_{27}^{\mathrm{RI}}| &=& - \alpha_{27}\frac{12\sqrt{3}}{f^3}
\left[
(m_K^2-m_\pi^2)\left(
1 + \frac{m_K^2}{(4 \pi f)^2}( I_{zf} - 2 I_\pi - I_K )
\right)\right. \nonumber \\ 
&&+ 
\left.
\frac{m_K^4}{(4 \pi f)^2}( I_a + \mathrm{Re}[I_b] + I_{c+d}) +
\beta_{20} m_\pi^4 + \beta_{11} m_\pi^2 p^2
\right]
\label{eq:chpt_NLO_27}
\end{eqnarray}
for the 27 operator, and
\begin{eqnarray}
|A_{88}^{\mathrm{RI}}| &=& - \alpha_{88}\frac{24\sqrt{3}}{f^3}
\left[
1 + \frac{m_K^2}{(4 \pi f)^2}( I_{zf} - 2 I_\pi - I_K )
\right. \nonumber \\ 
&&+ 
\left.
\frac{m_K^2}{(4 \pi f)^2}( J_a + \mathrm{Re}[J_b] + J_{c+d}) +
\gamma_{10} m_\pi^2 + \gamma_{01} p^2
\right]
\label{eq:chpt_NLO_88}
\end{eqnarray}
for the 88 ($m88$) operator, where $I_{zf,a,b,c+d}$ and 
$J_{a,b,c+d}$ are presented 
in Ref.~\cite{Lin:2002nq}, and $I_{\pi,K}$ in Ref.~\cite{Golterman:1997wb}.
$I_i$ and $J_j$ contain log terms and the scale $\mu$. In our analysis
we set $\mu = 1$ GeV for simplicity. 
We omit a $p^4$ term in eq.(\ref{eq:chpt_NLO_27}) as explained in Sec.~\ref{subsucsec:A_27}.
The normalization of these equations are based on Ref.~\cite{Blum:2001xb}.

The quenched formula is obtained by replacing $I_{zf,a,b,c+d}$ and
$J_{a,b,c+d}$ by $I_{zf,a,b,c+d}^q$ and $J_{a,b,c+d}^q$ 
in Ref.~\cite{Lin:2002nq}, respectively, and $I_K$ by $\widetilde{I_K}$
in Ref.~\cite{Golterman:1997wb}, and setting $I_\pi = 0$
in the above equations.
In the quenched analysis, we need two parameters $\alpha$ and $m_0$.
We find that the result does not largely depend on the parameters
in the ranges, $0 \le \alpha \le 0.5$ and $0.1 \le m_0 \le 0.866$ GeV,
so that we use $\alpha = 0.1$ and $m_0 = 0.866$ GeV in our analysis.

%
%
%
\bibliography{paper}

\begin{thebibliography}{71}
\expandafter\ifx\csname natexlab\endcsname\relax\def\natexlab#1{#1}\fi
\expandafter\ifx\csname bibnamefont\endcsname\relax
  \def\bibnamefont#1{#1}\fi
\expandafter\ifx\csname bibfnamefont\endcsname\relax
  \def\bibfnamefont#1{#1}\fi
\expandafter\ifx\csname citenamefont\endcsname\relax
  \def\citenamefont#1{#1}\fi
\expandafter\ifx\csname url\endcsname\relax
  \def\url#1{\texttt{#1}}\fi
\expandafter\ifx\csname urlprefix\endcsname\relax\def\urlprefix{URL }\fi
\providecommand{\bibinfo}[2]{#2}
\providecommand{\eprint}[2][]{\url{#2}}

\bibitem[{\citenamefont{Batley et~al.}(2002)}]{Batley:2002gn}
\bibinfo{author}{\bibfnamefont{J.~R.} \bibnamefont{Batley}}
  \bibnamefont{et~al.} (\bibinfo{collaboration}{NA48}), \bibinfo{journal}{Phys.
  Lett.} \textbf{\bibinfo{volume}{B544}}, \bibinfo{pages}{97}
  (\bibinfo{year}{2002}), \eprint{hep-ex/0208009}.

\bibitem[{\citenamefont{Alavi-Harati et~al.}(2003)}]{AlaviHarati:2002ye}
\bibinfo{author}{\bibfnamefont{A.}~\bibnamefont{Alavi-Harati}}
  \bibnamefont{et~al.} (\bibinfo{collaboration}{KTeV}), \bibinfo{journal}{Phys.
  Rev.} \textbf{\bibinfo{volume}{D67}}, \bibinfo{pages}{012005}
  (\bibinfo{year}{2003}), \eprint{hep-ex/0208007}.

\bibitem[{\citenamefont{Maiani and Testa}(1990)}]{Maiani:1990ca}
\bibinfo{author}{\bibfnamefont{L.}~\bibnamefont{Maiani}} \bibnamefont{and}
  \bibinfo{author}{\bibfnamefont{M.}~\bibnamefont{Testa}},
  \bibinfo{journal}{Phys. Lett.} \textbf{\bibinfo{volume}{B245}},
  \bibinfo{pages}{585} (\bibinfo{year}{1990}).

\bibitem[{\citenamefont{Bernard
  et~al.}(1985{\natexlab{a}})\citenamefont{Bernard, Draper, Soni, Politzer, and
  Wise}}]{Bernard:1985wf}
\bibinfo{author}{\bibfnamefont{C.~W.} \bibnamefont{Bernard}},
  \bibinfo{author}{\bibfnamefont{T.}~\bibnamefont{Draper}},
  \bibinfo{author}{\bibfnamefont{A.}~\bibnamefont{Soni}},
  \bibinfo{author}{\bibfnamefont{H.}~\bibnamefont{Politzer}}, \bibnamefont{and}
  \bibinfo{author}{\bibfnamefont{M.~B.} \bibnamefont{Wise}},
  \bibinfo{journal}{Phys. Rev.} \textbf{\bibinfo{volume}{D32}},
  \bibinfo{pages}{2343} (\bibinfo{year}{1985}{\natexlab{a}}).

\bibitem[{\citenamefont{Bernard
  et~al.}(1985{\natexlab{b}})\citenamefont{Bernard, Draper, Hockney, Rushton,
  and Soni}}]{Bernard:1985tm}
\bibinfo{author}{\bibfnamefont{C.~W.} \bibnamefont{Bernard}},
  \bibinfo{author}{\bibfnamefont{T.}~\bibnamefont{Draper}},
  \bibinfo{author}{\bibfnamefont{G.}~\bibnamefont{Hockney}},
  \bibinfo{author}{\bibfnamefont{A.~M.} \bibnamefont{Rushton}},
  \bibnamefont{and} \bibinfo{author}{\bibfnamefont{A.}~\bibnamefont{Soni}},
  \bibinfo{journal}{Phys. Rev. Lett.} \textbf{\bibinfo{volume}{55}},
  \bibinfo{pages}{2770} (\bibinfo{year}{1985}{\natexlab{b}}).

\bibitem[{\citenamefont{Pekurovsky and Kilcup}(2001)}]{Pekurovsky:1998jd}
\bibinfo{author}{\bibfnamefont{D.}~\bibnamefont{Pekurovsky}} \bibnamefont{and}
  \bibinfo{author}{\bibfnamefont{G.}~\bibnamefont{Kilcup}},
  \bibinfo{journal}{Phys. Rev.} \textbf{\bibinfo{volume}{D64}},
  \bibinfo{pages}{074502} (\bibinfo{year}{2001}), \eprint{hep-lat/9812019}.

\bibitem[{\citenamefont{Lellouch and Lin}(1999)}]{Lellouch:1998sg}
\bibinfo{author}{\bibfnamefont{L.}~\bibnamefont{Lellouch}} \bibnamefont{and}
  \bibinfo{author}{\bibfnamefont{C.~J.~D.} \bibnamefont{Lin}}
  (\bibinfo{collaboration}{UKQCD}), \bibinfo{journal}{Nucl. Phys. Proc. Suppl.}
  \textbf{\bibinfo{volume}{73}}, \bibinfo{pages}{312} (\bibinfo{year}{1999}),
  \eprint{hep-lat/9809142}.

\bibitem[{\citenamefont{Donini et~al.}(1999)\citenamefont{Donini, Gimenez,
  Giusti, and Martinelli}}]{Donini:1999nn}
\bibinfo{author}{\bibfnamefont{A.}~\bibnamefont{Donini}},
  \bibinfo{author}{\bibfnamefont{V.}~\bibnamefont{Gimenez}},
  \bibinfo{author}{\bibfnamefont{L.}~\bibnamefont{Giusti}}, \bibnamefont{and}
  \bibinfo{author}{\bibfnamefont{G.}~\bibnamefont{Martinelli}},
  \bibinfo{journal}{Phys. Lett.} \textbf{\bibinfo{volume}{B470}},
  \bibinfo{pages}{233} (\bibinfo{year}{1999}), \eprint{hep-lat/9910017}.

\bibitem[{\citenamefont{Lee and Fleming}(2005)}]{Lee:2005sr}
\bibinfo{author}{\bibfnamefont{W.}~\bibnamefont{Lee}} \bibnamefont{and}
  \bibinfo{author}{\bibfnamefont{G.~T.} \bibnamefont{Fleming}},
  \bibinfo{journal}{PoS} \textbf{\bibinfo{volume}{LAT2005}},
  \bibinfo{pages}{339} (\bibinfo{year}{2005}), \eprint{hep-lat/0510005}.

\bibitem[{\citenamefont{Noaki et~al.}(2003)}]{Noaki:2001un}
\bibinfo{author}{\bibfnamefont{J.~I.} \bibnamefont{Noaki}} \bibnamefont{et~al.}
  (\bibinfo{collaboration}{CP-PACS}), \bibinfo{journal}{Phys. Rev.}
  \textbf{\bibinfo{volume}{D68}}, \bibinfo{pages}{014501}
  (\bibinfo{year}{2003}), \eprint{hep-lat/0108013}.

\bibitem[{\citenamefont{Blum et~al.}(2003)}]{Blum:2001xb}
\bibinfo{author}{\bibfnamefont{T.}~\bibnamefont{Blum}} \bibnamefont{et~al.}
  (\bibinfo{collaboration}{RBC}), \bibinfo{journal}{Phys. Rev.}
  \textbf{\bibinfo{volume}{D68}}, \bibinfo{pages}{114506}
  (\bibinfo{year}{2003}), \eprint{hep-lat/0110075}.

\bibitem[{\citenamefont{Kaplan}(1992)}]{Kaplan:1992bt}
\bibinfo{author}{\bibfnamefont{D.~B.} \bibnamefont{Kaplan}},
  \bibinfo{journal}{Phys. Lett.} \textbf{\bibinfo{volume}{B288}},
  \bibinfo{pages}{342} (\bibinfo{year}{1992}), \eprint{hep-lat/9206013}.

\bibitem[{\citenamefont{Shamir}(1993)}]{Shamir:1993zy}
\bibinfo{author}{\bibfnamefont{Y.}~\bibnamefont{Shamir}},
  \bibinfo{journal}{Nucl. Phys.} \textbf{\bibinfo{volume}{B406}},
  \bibinfo{pages}{90} (\bibinfo{year}{1993}), \eprint{hep-lat/9303005}.

\bibitem[{\citenamefont{Furman and Shamir}(1995)}]{Furman:1994ky}
\bibinfo{author}{\bibfnamefont{V.}~\bibnamefont{Furman}} \bibnamefont{and}
  \bibinfo{author}{\bibfnamefont{Y.}~\bibnamefont{Shamir}},
  \bibinfo{journal}{Nucl. Phys.} \textbf{\bibinfo{volume}{B439}},
  \bibinfo{pages}{54} (\bibinfo{year}{1995}), \eprint{hep-lat/9405004}.

\bibitem[{\citenamefont{Bernard and Soni}(1989)}]{Bernard:1988zj}
\bibinfo{author}{\bibfnamefont{C.~W.} \bibnamefont{Bernard}} \bibnamefont{and}
  \bibinfo{author}{\bibfnamefont{A.}~\bibnamefont{Soni}},
  \bibinfo{journal}{Nucl. Phys. Proc. Suppl.} \textbf{\bibinfo{volume}{9}},
  \bibinfo{pages}{155} (\bibinfo{year}{1989}).

\bibitem[{\citenamefont{Aoki et~al.}(1998)}]{Aoki:1997ev}
\bibinfo{author}{\bibfnamefont{S.}~\bibnamefont{Aoki}} \bibnamefont{et~al.}
  (\bibinfo{collaboration}{JLQCD}), \bibinfo{journal}{Phys. Rev.}
  \textbf{\bibinfo{volume}{D58}}, \bibinfo{pages}{054503}
  (\bibinfo{year}{1998}), \eprint{hep-lat/9711046}.

\bibitem[{\citenamefont{Ishizuka}(2003)}]{Ishizuka:2002nm}
\bibinfo{author}{\bibfnamefont{N.}~\bibnamefont{Ishizuka}},
  \bibinfo{journal}{Nucl. Phys. Proc. Suppl.} \textbf{\bibinfo{volume}{119}},
  \bibinfo{pages}{84} (\bibinfo{year}{2003}), \eprint{hep-lat/0209108}.

\bibitem[{\citenamefont{Luscher and Wolff}(1990)}]{Luscher:1990ck}
\bibinfo{author}{\bibfnamefont{M.}~\bibnamefont{Luscher}} \bibnamefont{and}
  \bibinfo{author}{\bibfnamefont{U.}~\bibnamefont{Wolff}},
  \bibinfo{journal}{Nucl. Phys.} \textbf{\bibinfo{volume}{B339}},
  \bibinfo{pages}{222} (\bibinfo{year}{1990}).

\bibitem[{\citenamefont{Kim}(2005)}]{Kim:2004mb}
\bibinfo{author}{\bibfnamefont{C.}~\bibnamefont{Kim}}, \bibinfo{journal}{Nucl.
  Phys. Proc. Suppl.} \textbf{\bibinfo{volume}{140}}, \bibinfo{pages}{381}
  (\bibinfo{year}{2005}).

\bibitem[{\citenamefont{Kim}(2004{\natexlab{a}})}]{Kim:2004mk}
\bibinfo{author}{\bibfnamefont{C.}~\bibnamefont{Kim}}
  (\bibinfo{year}{2004}{\natexlab{a}}), \bibinfo{note}{his Ph.D. Thesis.,
  UMI-31-47246}.

\bibitem[{\citenamefont{Boucaud et~al.}(2005)}]{Boucaud:2004aa}
\bibinfo{author}{\bibfnamefont{P.}~\bibnamefont{Boucaud}} \bibnamefont{et~al.},
  \bibinfo{journal}{Nucl. Phys.} \textbf{\bibinfo{volume}{B721}},
  \bibinfo{pages}{175} (\bibinfo{year}{2005}), \eprint{hep-lat/0412029}.

\bibitem[{\citenamefont{Rummukainen and Gottlieb}(1995)}]{Rummukainen:1995vs}
\bibinfo{author}{\bibfnamefont{K.}~\bibnamefont{Rummukainen}} \bibnamefont{and}
  \bibinfo{author}{\bibfnamefont{S.~A.} \bibnamefont{Gottlieb}},
  \bibinfo{journal}{Nucl. Phys.} \textbf{\bibinfo{volume}{B450}},
  \bibinfo{pages}{397} (\bibinfo{year}{1995}), \eprint{hep-lat/9503028}.

\bibitem[{\citenamefont{Yamazaki et~al.}(2004)}]{Yamazaki:2004qb}
\bibinfo{author}{\bibfnamefont{T.}~\bibnamefont{Yamazaki}} \bibnamefont{et~al.}
  (\bibinfo{collaboration}{CP-PACS}), \bibinfo{journal}{Phys. Rev.}
  \textbf{\bibinfo{volume}{D70}}, \bibinfo{pages}{074513}
  (\bibinfo{year}{2004}), \eprint{hep-lat/0402025}.

\bibitem[{\citenamefont{Lellouch and Luscher}(2001)}]{Lellouch:2000pv}
\bibinfo{author}{\bibfnamefont{L.}~\bibnamefont{Lellouch}} \bibnamefont{and}
  \bibinfo{author}{\bibfnamefont{M.}~\bibnamefont{Luscher}},
  \bibinfo{journal}{Commun. Math. Phys.} \textbf{\bibinfo{volume}{219}},
  \bibinfo{pages}{31} (\bibinfo{year}{2001}), \eprint{hep-lat/0003023}.

\bibitem[{\citenamefont{Kim et~al.}(2005)\citenamefont{Kim, Sachrajda, and
  Sharpe}}]{Kim:2005gf}
\bibinfo{author}{\bibfnamefont{C.~h.} \bibnamefont{Kim}},
  \bibinfo{author}{\bibfnamefont{C.~T.} \bibnamefont{Sachrajda}},
  \bibnamefont{and} \bibinfo{author}{\bibfnamefont{S.~R.}
  \bibnamefont{Sharpe}}, \bibinfo{journal}{Nucl. Phys.}
  \textbf{\bibinfo{volume}{B727}}, \bibinfo{pages}{218} (\bibinfo{year}{2005}),
  \eprint{hep-lat/0507006}.

\bibitem[{\citenamefont{Christ et~al.}(2005)\citenamefont{Christ, Kim, and
  Yamazaki}}]{Christ:2005gi}
\bibinfo{author}{\bibfnamefont{N.~H.} \bibnamefont{Christ}},
  \bibinfo{author}{\bibfnamefont{C.}~\bibnamefont{Kim}}, \bibnamefont{and}
  \bibinfo{author}{\bibfnamefont{T.}~\bibnamefont{Yamazaki}},
  \bibinfo{journal}{Phys. Rev.} \textbf{\bibinfo{volume}{D72}},
  \bibinfo{pages}{114506} (\bibinfo{year}{2005}), \eprint{hep-lat/0507009}.

\bibitem[{\citenamefont{Takaishi}(1996)}]{Takaishi:1996xj}
\bibinfo{author}{\bibfnamefont{T.}~\bibnamefont{Takaishi}},
  \bibinfo{journal}{Phys. Rev.} \textbf{\bibinfo{volume}{D54}},
  \bibinfo{pages}{1050} (\bibinfo{year}{1996}).

\bibitem[{\citenamefont{de~Forcrand et~al.}(2000)}]{deForcrand:1999bi}
\bibinfo{author}{\bibfnamefont{P.}~\bibnamefont{de~Forcrand}}
  \bibnamefont{et~al.} (\bibinfo{collaboration}{QCD-TARO}),
  \bibinfo{journal}{Nucl. Phys.} \textbf{\bibinfo{volume}{B577}},
  \bibinfo{pages}{263} (\bibinfo{year}{2000}), \eprint{hep-lat/9911033}.

\bibitem[{\citenamefont{Sachrajda and Villadoro}(2005)}]{Sachrajda:2004mi}
\bibinfo{author}{\bibfnamefont{C.~T.} \bibnamefont{Sachrajda}}
  \bibnamefont{and}
  \bibinfo{author}{\bibfnamefont{G.}~\bibnamefont{Villadoro}},
  \bibinfo{journal}{Phys. Lett.} \textbf{\bibinfo{volume}{B609}},
  \bibinfo{pages}{73} (\bibinfo{year}{2005}), \eprint{hep-lat/0411033}.

\bibitem[{\citenamefont{Kim}(2004{\natexlab{b}})}]{Kim:2003xt}
\bibinfo{author}{\bibfnamefont{C.}~\bibnamefont{Kim}}, \bibinfo{journal}{Nucl.
  Phys. Proc. Suppl.} \textbf{\bibinfo{volume}{129}}, \bibinfo{pages}{197}
  (\bibinfo{year}{2004}{\natexlab{b}}), \eprint{hep-lat/0311003}.

\bibitem[{\citenamefont{Yamazaki}(2005)}]{Yamazaki:2005eg}
\bibinfo{author}{\bibfnamefont{T.}~\bibnamefont{Yamazaki}}
  (\bibinfo{collaboration}{the RBC}), \bibinfo{journal}{PoS}
  \textbf{\bibinfo{volume}{LAT2005}}, \bibinfo{pages}{351}
  (\bibinfo{year}{2005}), \eprint{hep-lat/0509135}.

\bibitem[{\citenamefont{Yamazaki}(2006)}]{Yamazaki:2006ce}
\bibinfo{author}{\bibfnamefont{T.}~\bibnamefont{Yamazaki}}
  (\bibinfo{collaboration}{the RBC}), \bibinfo{journal}{PoS}
  \textbf{\bibinfo{volume}{LAT2006}}, \bibinfo{pages}{100}
  (\bibinfo{year}{2006}), \eprint{hep-lat/0610051}.

\bibitem[{\citenamefont{Luscher}(1991)}]{Luscher:1990ux}
\bibinfo{author}{\bibfnamefont{M.}~\bibnamefont{Luscher}},
  \bibinfo{journal}{Nucl. Phys.} \textbf{\bibinfo{volume}{B354}},
  \bibinfo{pages}{531} (\bibinfo{year}{1991}).

\bibitem[{\citenamefont{Aoki et~al.}(2004)}]{Aoki:2002vt}
\bibinfo{author}{\bibfnamefont{Y.}~\bibnamefont{Aoki}} \bibnamefont{et~al.},
  \bibinfo{journal}{Phys. Rev.} \textbf{\bibinfo{volume}{D69}},
  \bibinfo{pages}{074504} (\bibinfo{year}{2004}), \eprint{hep-lat/0211023}.

\bibitem[{\citenamefont{Aoki et~al.}(2007{\natexlab{a}})\citenamefont{Aoki,
  Dawson, Noaki, and Soni}}]{Aoki:2006ib}
\bibinfo{author}{\bibfnamefont{Y.}~\bibnamefont{Aoki}},
  \bibinfo{author}{\bibfnamefont{C.}~\bibnamefont{Dawson}},
  \bibinfo{author}{\bibfnamefont{J.}~\bibnamefont{Noaki}}, \bibnamefont{and}
  \bibinfo{author}{\bibfnamefont{A.}~\bibnamefont{Soni}},
  \bibinfo{journal}{Phys. Rev.} \textbf{\bibinfo{volume}{D75}},
  \bibinfo{pages}{014507} (\bibinfo{year}{2007}{\natexlab{a}}),
  \eprint{hep-lat/0607002}.

\bibitem[{\citenamefont{Blum et~al.}(2004)}]{Blum:2000kn}
\bibinfo{author}{\bibfnamefont{T.}~\bibnamefont{Blum}} \bibnamefont{et~al.},
  \bibinfo{journal}{Phys. Rev.} \textbf{\bibinfo{volume}{D69}},
  \bibinfo{pages}{074502} (\bibinfo{year}{2004}), \eprint{hep-lat/0007038}.

\bibitem[{\citenamefont{Lin et~al.}(2006)\citenamefont{Lin, Ohta, Soni, and
  Yamada}}]{Lin:2006vc}
\bibinfo{author}{\bibfnamefont{H.-W.} \bibnamefont{Lin}},
  \bibinfo{author}{\bibfnamefont{S.}~\bibnamefont{Ohta}},
  \bibinfo{author}{\bibfnamefont{A.}~\bibnamefont{Soni}}, \bibnamefont{and}
  \bibinfo{author}{\bibfnamefont{N.}~\bibnamefont{Yamada}},
  \bibinfo{journal}{Phys. Rev.} \textbf{\bibinfo{volume}{D74}},
  \bibinfo{pages}{114506} (\bibinfo{year}{2006}), \eprint{hep-lat/0607035}.

\bibitem[{\citenamefont{Aoki et~al.}(2005)}]{Aoki:2005uf}
\bibinfo{author}{\bibfnamefont{S.}~\bibnamefont{Aoki}} \bibnamefont{et~al.}
  (\bibinfo{collaboration}{CP-PACS}), \bibinfo{journal}{Phys. Rev.}
  \textbf{\bibinfo{volume}{D71}}, \bibinfo{pages}{094504}
  (\bibinfo{year}{2005}), \eprint{hep-lat/0503025}.

\bibitem[{\citenamefont{Sasaki and Ishizuka}(2008)}]{Sasaki:2008sv}
\bibinfo{author}{\bibfnamefont{K.}~\bibnamefont{Sasaki}} \bibnamefont{and}
  \bibinfo{author}{\bibfnamefont{N.}~\bibnamefont{Ishizuka}}
  (\bibinfo{year}{2008}), \eprint{arXiv:0804.2941 [hep-lat]}.

\bibitem[{\citenamefont{Ali~Khan et~al.}(2002)}]{AliKhan:2001tx}
\bibinfo{author}{\bibfnamefont{A.}~\bibnamefont{Ali~Khan}} \bibnamefont{et~al.}
  (\bibinfo{collaboration}{CP-PACS}), \bibinfo{journal}{Phys. Rev.}
  \textbf{\bibinfo{volume}{D65}}, \bibinfo{pages}{054505}
  (\bibinfo{year}{2002}), \eprint{hep-lat/0105015}.

\bibitem[{\citenamefont{Sharpe et~al.}(1992)\citenamefont{Sharpe, Gupta, and
  Kilcup}}]{Sharpe:1992pp}
\bibinfo{author}{\bibfnamefont{S.~R.} \bibnamefont{Sharpe}},
  \bibinfo{author}{\bibfnamefont{R.}~\bibnamefont{Gupta}}, \bibnamefont{and}
  \bibinfo{author}{\bibfnamefont{G.~W.} \bibnamefont{Kilcup}},
  \bibinfo{journal}{Nucl. Phys.} \textbf{\bibinfo{volume}{B383}},
  \bibinfo{pages}{309} (\bibinfo{year}{1992}).

\bibitem[{\citenamefont{Gupta et~al.}(1993)\citenamefont{Gupta, Patel, and
  Sharpe}}]{Gupta:1993rn}
\bibinfo{author}{\bibfnamefont{R.}~\bibnamefont{Gupta}},
  \bibinfo{author}{\bibfnamefont{A.}~\bibnamefont{Patel}}, \bibnamefont{and}
  \bibinfo{author}{\bibfnamefont{S.~R.} \bibnamefont{Sharpe}},
  \bibinfo{journal}{Phys. Rev.} \textbf{\bibinfo{volume}{D48}},
  \bibinfo{pages}{388} (\bibinfo{year}{1993}), \eprint{hep-lat/9301016}.

\bibitem[{\citenamefont{Kuramashi et~al.}(1993)\citenamefont{Kuramashi,
  Fukugita, Mino, Okawa, and Ukawa}}]{Kuramashi:1993ka}
\bibinfo{author}{\bibfnamefont{Y.}~\bibnamefont{Kuramashi}},
  \bibinfo{author}{\bibfnamefont{M.}~\bibnamefont{Fukugita}},
  \bibinfo{author}{\bibfnamefont{H.}~\bibnamefont{Mino}},
  \bibinfo{author}{\bibfnamefont{M.}~\bibnamefont{Okawa}}, \bibnamefont{and}
  \bibinfo{author}{\bibfnamefont{A.}~\bibnamefont{Ukawa}},
  \bibinfo{journal}{Phys. Rev. Lett.} \textbf{\bibinfo{volume}{71}},
  \bibinfo{pages}{2387} (\bibinfo{year}{1993}).

\bibitem[{\citenamefont{Fukugita et~al.}(1995)\citenamefont{Fukugita,
  Kuramashi, Okawa, Mino, and Ukawa}}]{Fukugita:1994ve}
\bibinfo{author}{\bibfnamefont{M.}~\bibnamefont{Fukugita}},
  \bibinfo{author}{\bibfnamefont{Y.}~\bibnamefont{Kuramashi}},
  \bibinfo{author}{\bibfnamefont{M.}~\bibnamefont{Okawa}},
  \bibinfo{author}{\bibfnamefont{H.}~\bibnamefont{Mino}}, \bibnamefont{and}
  \bibinfo{author}{\bibfnamefont{A.}~\bibnamefont{Ukawa}},
  \bibinfo{journal}{Phys. Rev.} \textbf{\bibinfo{volume}{D52}},
  \bibinfo{pages}{3003} (\bibinfo{year}{1995}), \eprint{hep-lat/9501024}.

\bibitem[{\citenamefont{Alford and Jaffe}(2000)}]{Alford:2000mm}
\bibinfo{author}{\bibfnamefont{M.~G.} \bibnamefont{Alford}} \bibnamefont{and}
  \bibinfo{author}{\bibfnamefont{R.~L.} \bibnamefont{Jaffe}},
  \bibinfo{journal}{Nucl. Phys.} \textbf{\bibinfo{volume}{B578}},
  \bibinfo{pages}{367} (\bibinfo{year}{2000}), \eprint{hep-lat/0001023}.

\bibitem[{\citenamefont{Juge}(2004)}]{Juge:2003mr}
\bibinfo{author}{\bibfnamefont{K.~J.} \bibnamefont{Juge}}
  (\bibinfo{collaboration}{BGR}), \bibinfo{journal}{Nucl. Phys. Proc. Suppl.}
  \textbf{\bibinfo{volume}{129}}, \bibinfo{pages}{194} (\bibinfo{year}{2004}),
  \eprint{hep-lat/0309075}.

\bibitem[{\citenamefont{Gattringer et~al.}(2005)\citenamefont{Gattringer,
  Hierl, and Pullirsch}}]{Gattringer:2004wr}
\bibinfo{author}{\bibfnamefont{C.}~\bibnamefont{Gattringer}},
  \bibinfo{author}{\bibfnamefont{D.}~\bibnamefont{Hierl}}, \bibnamefont{and}
  \bibinfo{author}{\bibfnamefont{R.}~\bibnamefont{Pullirsch}}
  (\bibinfo{collaboration}{Bern-Graz-Regensburg}), \bibinfo{journal}{Nucl.
  Phys. Proc. Suppl.} \textbf{\bibinfo{volume}{140}}, \bibinfo{pages}{308}
  (\bibinfo{year}{2005}), \eprint{hep-lat/0409064}.

\bibitem[{\citenamefont{Aoki et~al.}(2003)}]{Aoki:2002ny}
\bibinfo{author}{\bibfnamefont{S.}~\bibnamefont{Aoki}} \bibnamefont{et~al.}
  (\bibinfo{collaboration}{CP-PACS}), \bibinfo{journal}{Phys. Rev.}
  \textbf{\bibinfo{volume}{D67}}, \bibinfo{pages}{014502}
  (\bibinfo{year}{2003}), \eprint{hep-lat/0209124}.

\bibitem[{\citenamefont{Luscher}(1986)}]{Luscher:1986pf}
\bibinfo{author}{\bibfnamefont{M.}~\bibnamefont{Luscher}},
  \bibinfo{journal}{Commun. Math. Phys.} \textbf{\bibinfo{volume}{105}},
  \bibinfo{pages}{153} (\bibinfo{year}{1986}).

\bibitem[{\citenamefont{Aoki et~al.}(2002)}]{Aoki:2002in}
\bibinfo{author}{\bibfnamefont{S.}~\bibnamefont{Aoki}} \bibnamefont{et~al.}
  (\bibinfo{collaboration}{JLQCD}), \bibinfo{journal}{Phys. Rev.}
  \textbf{\bibinfo{volume}{D66}}, \bibinfo{pages}{077501}
  (\bibinfo{year}{2002}), \eprint{hep-lat/0206011}.

\bibitem[{\citenamefont{Liu et~al.}(2002)\citenamefont{Liu, Zhang, Chen, and
  Ma}}]{Liu:2001ss}
\bibinfo{author}{\bibfnamefont{C.}~\bibnamefont{Liu}},
  \bibinfo{author}{\bibfnamefont{J.-h.} \bibnamefont{Zhang}},
  \bibinfo{author}{\bibfnamefont{Y.}~\bibnamefont{Chen}}, \bibnamefont{and}
  \bibinfo{author}{\bibfnamefont{J.~P.} \bibnamefont{Ma}},
  \bibinfo{journal}{Nucl. Phys.} \textbf{\bibinfo{volume}{B624}},
  \bibinfo{pages}{360} (\bibinfo{year}{2002}), \eprint{hep-lat/0109020}.

\bibitem[{\citenamefont{Du et~al.}(2004)\citenamefont{Du, Meng, Miao, and
  Liu}}]{Du:2004ib}
\bibinfo{author}{\bibfnamefont{X.}~\bibnamefont{Du}},
  \bibinfo{author}{\bibfnamefont{G.-w.} \bibnamefont{Meng}},
  \bibinfo{author}{\bibfnamefont{C.}~\bibnamefont{Miao}}, \bibnamefont{and}
  \bibinfo{author}{\bibfnamefont{C.}~\bibnamefont{Liu}}, \bibinfo{journal}{Int.
  J. Mod. Phys.} \textbf{\bibinfo{volume}{A19}}, \bibinfo{pages}{5609}
  (\bibinfo{year}{2004}), \eprint{hep-lat/0404017}.

\bibitem[{\citenamefont{Li et~al.}(2007)}]{Li:2007ey}
\bibinfo{author}{\bibfnamefont{X.}~\bibnamefont{Li}} \bibnamefont{et~al.}
  (\bibinfo{collaboration}{CLQCD}), \bibinfo{journal}{JHEP}
  \textbf{\bibinfo{volume}{06}}, \bibinfo{pages}{053} (\bibinfo{year}{2007}),
  \eprint{hep-lat/0703015}.

\bibitem[{\citenamefont{Beane et~al.}(2006)\citenamefont{Beane, Bedaque,
  Orginos, and Savage}}]{Beane:2005rj}
\bibinfo{author}{\bibfnamefont{S.~R.} \bibnamefont{Beane}},
  \bibinfo{author}{\bibfnamefont{P.~F.} \bibnamefont{Bedaque}},
  \bibinfo{author}{\bibfnamefont{K.}~\bibnamefont{Orginos}}, \bibnamefont{and}
  \bibinfo{author}{\bibfnamefont{M.~J.} \bibnamefont{Savage}}
  (\bibinfo{collaboration}{NPLQCD}), \bibinfo{journal}{Phys. Rev.}
  \textbf{\bibinfo{volume}{D73}}, \bibinfo{pages}{054503}
  (\bibinfo{year}{2006}), \eprint{hep-lat/0506013}.

\bibitem[{\citenamefont{Beane et~al.}(2008)}]{Beane:2007xs}
\bibinfo{author}{\bibfnamefont{S.~R.} \bibnamefont{Beane}}
  \bibnamefont{et~al.}, \bibinfo{journal}{Phys. Rev.}
  \textbf{\bibinfo{volume}{D77}}, \bibinfo{pages}{014505}
  (\bibinfo{year}{2008}), \eprint{arXiv:0706.3026 [hep-lat]}.

\bibitem[{\citenamefont{Aoki et~al.}(2007{\natexlab{b}})}]{Aoki:2007rd}
\bibinfo{author}{\bibfnamefont{S.}~\bibnamefont{Aoki}} \bibnamefont{et~al.}
  (\bibinfo{collaboration}{CP-PACS}), \bibinfo{journal}{Phys. Rev.}
  \textbf{\bibinfo{volume}{D76}}, \bibinfo{pages}{094506}
  (\bibinfo{year}{2007}{\natexlab{b}}), \eprint{arXiv:0708.3705 [hep-lat]}.

\bibitem[{\citenamefont{Pislak et~al.}(2003)}]{Pislak:2003sv}
\bibinfo{author}{\bibfnamefont{S.}~\bibnamefont{Pislak}} \bibnamefont{et~al.},
  \bibinfo{journal}{Phys. Rev.} \textbf{\bibinfo{volume}{D67}},
  \bibinfo{pages}{072004} (\bibinfo{year}{2003}), \eprint{hep-ex/0301040}.

\bibitem[{\citenamefont{Colangelo et~al.}(2001)\citenamefont{Colangelo, Gasser,
  and Leutwyler}}]{Colangelo:2001df}
\bibinfo{author}{\bibfnamefont{G.}~\bibnamefont{Colangelo}},
  \bibinfo{author}{\bibfnamefont{J.}~\bibnamefont{Gasser}}, \bibnamefont{and}
  \bibinfo{author}{\bibfnamefont{H.}~\bibnamefont{Leutwyler}},
  \bibinfo{journal}{Nucl. Phys.} \textbf{\bibinfo{volume}{B603}},
  \bibinfo{pages}{125} (\bibinfo{year}{2001}), \eprint{hep-ph/0103088}.

\bibitem[{\citenamefont{Gasser and Leutwyler}(1984)}]{Gasser:1983yg}
\bibinfo{author}{\bibfnamefont{J.}~\bibnamefont{Gasser}} \bibnamefont{and}
  \bibinfo{author}{\bibfnamefont{H.}~\bibnamefont{Leutwyler}},
  \bibinfo{journal}{Ann. Phys.} \textbf{\bibinfo{volume}{158}},
  \bibinfo{pages}{142} (\bibinfo{year}{1984}).

\bibitem[{\citenamefont{Bernard and Golterman}(1996)}]{Bernard:1995ez}
\bibinfo{author}{\bibfnamefont{C.~W.} \bibnamefont{Bernard}} \bibnamefont{and}
  \bibinfo{author}{\bibfnamefont{M.~F.~L.} \bibnamefont{Golterman}},
  \bibinfo{journal}{Phys. Rev.} \textbf{\bibinfo{volume}{D53}},
  \bibinfo{pages}{476} (\bibinfo{year}{1996}), \eprint{hep-lat/9507004}.

\bibitem[{\citenamefont{Colangelo and Pallante}(1998)}]{Colangelo:1997ch}
\bibinfo{author}{\bibfnamefont{G.}~\bibnamefont{Colangelo}} \bibnamefont{and}
  \bibinfo{author}{\bibfnamefont{E.}~\bibnamefont{Pallante}},
  \bibinfo{journal}{Nucl. Phys.} \textbf{\bibinfo{volume}{B520}},
  \bibinfo{pages}{433} (\bibinfo{year}{1998}), \eprint{hep-lat/9708005}.

\bibitem[{\citenamefont{Sasaki and Yamazaki}(2006)}]{Sasaki:2006jn}
\bibinfo{author}{\bibfnamefont{S.}~\bibnamefont{Sasaki}} \bibnamefont{and}
  \bibinfo{author}{\bibfnamefont{T.}~\bibnamefont{Yamazaki}},
  \bibinfo{journal}{Phys. Rev.} \textbf{\bibinfo{volume}{D74}},
  \bibinfo{pages}{114507} (\bibinfo{year}{2006}), \eprint{hep-lat/0610081}.

\bibitem[{\citenamefont{Martinelli et~al.}(1995)\citenamefont{Martinelli,
  Pittori, Sachrajda, Testa, and Vladikas}}]{Martinelli:1994ty}
\bibinfo{author}{\bibfnamefont{G.}~\bibnamefont{Martinelli}},
  \bibinfo{author}{\bibfnamefont{C.}~\bibnamefont{Pittori}},
  \bibinfo{author}{\bibfnamefont{C.~T.} \bibnamefont{Sachrajda}},
  \bibinfo{author}{\bibfnamefont{M.}~\bibnamefont{Testa}}, \bibnamefont{and}
  \bibinfo{author}{\bibfnamefont{A.}~\bibnamefont{Vladikas}},
  \bibinfo{journal}{Nucl. Phys.} \textbf{\bibinfo{volume}{B445}},
  \bibinfo{pages}{81} (\bibinfo{year}{1995}), \eprint{hep-lat/9411010}.

\bibitem[{\citenamefont{Blum et~al.}(2002)}]{Blum:2001sr}
\bibinfo{author}{\bibfnamefont{T.}~\bibnamefont{Blum}} \bibnamefont{et~al.},
  \bibinfo{journal}{Phys. Rev.} \textbf{\bibinfo{volume}{D66}},
  \bibinfo{pages}{014504} (\bibinfo{year}{2002}), \eprint{hep-lat/0102005}.

\bibitem[{\citenamefont{Laiho and Soni}(2002)}]{Laiho:2002jq}
\bibinfo{author}{\bibfnamefont{J.}~\bibnamefont{Laiho}} \bibnamefont{and}
  \bibinfo{author}{\bibfnamefont{A.}~\bibnamefont{Soni}},
  \bibinfo{journal}{Phys. Rev.} \textbf{\bibinfo{volume}{D65}},
  \bibinfo{pages}{114020} (\bibinfo{year}{2002}), \eprint{hep-ph/0203106}.

\bibitem[{\citenamefont{Lin et~al.}(2003)\citenamefont{Lin, Martinelli,
  Pallante, Sachrajda, and Villadoro}}]{Lin:2002nq}
\bibinfo{author}{\bibfnamefont{C.~J.~D.} \bibnamefont{Lin}},
  \bibinfo{author}{\bibfnamefont{G.}~\bibnamefont{Martinelli}},
  \bibinfo{author}{\bibfnamefont{E.}~\bibnamefont{Pallante}},
  \bibinfo{author}{\bibfnamefont{C.~T.} \bibnamefont{Sachrajda}},
  \bibnamefont{and}
  \bibinfo{author}{\bibfnamefont{G.}~\bibnamefont{Villadoro}},
  \bibinfo{journal}{Nucl. Phys.} \textbf{\bibinfo{volume}{B650}},
  \bibinfo{pages}{301} (\bibinfo{year}{2003}), \eprint{hep-lat/0208007}.

\bibitem[{\citenamefont{Bijnens and Wise}(1984)}]{Bijnens:1983ye}
\bibinfo{author}{\bibfnamefont{J.}~\bibnamefont{Bijnens}} \bibnamefont{and}
  \bibinfo{author}{\bibfnamefont{M.~B.} \bibnamefont{Wise}},
  \bibinfo{journal}{Phys. Lett.} \textbf{\bibinfo{volume}{B137}},
  \bibinfo{pages}{245} (\bibinfo{year}{1984}).

\bibitem[{\citenamefont{Cirigliano and Golowich}(2000)}]{Cirigliano:1999pv}
\bibinfo{author}{\bibfnamefont{V.}~\bibnamefont{Cirigliano}} \bibnamefont{and}
  \bibinfo{author}{\bibfnamefont{E.}~\bibnamefont{Golowich}},
  \bibinfo{journal}{Phys. Lett.} \textbf{\bibinfo{volume}{B475}},
  \bibinfo{pages}{351} (\bibinfo{year}{2000}), \eprint{hep-ph/9912513}.

\bibitem[{\citenamefont{Laiho and Soni}(2005)}]{Laiho:2003uy}
\bibinfo{author}{\bibfnamefont{J.}~\bibnamefont{Laiho}} \bibnamefont{and}
  \bibinfo{author}{\bibfnamefont{A.}~\bibnamefont{Soni}},
  \bibinfo{journal}{Phys. Rev.} \textbf{\bibinfo{volume}{D71}},
  \bibinfo{pages}{014021} (\bibinfo{year}{2005}), \eprint{hep-lat/0306035}.

\bibitem[{\citenamefont{Buchalla et~al.}(1996)\citenamefont{Buchalla, Buras,
  and Lautenbacher}}]{Buchalla:1995vs}
\bibinfo{author}{\bibfnamefont{G.}~\bibnamefont{Buchalla}},
  \bibinfo{author}{\bibfnamefont{A.~J.} \bibnamefont{Buras}}, \bibnamefont{and}
  \bibinfo{author}{\bibfnamefont{M.~E.} \bibnamefont{Lautenbacher}},
  \bibinfo{journal}{Rev. Mod. Phys.} \textbf{\bibinfo{volume}{68}},
  \bibinfo{pages}{1125} (\bibinfo{year}{1996}), \eprint{hep-ph/9512380}.

\bibitem[{\citenamefont{Golterman and Leung}(1997)}]{Golterman:1997wb}
\bibinfo{author}{\bibfnamefont{M.~F.~L.} \bibnamefont{Golterman}}
  \bibnamefont{and} \bibinfo{author}{\bibfnamefont{K.~C.} \bibnamefont{Leung}},
  \bibinfo{journal}{Phys. Rev.} \textbf{\bibinfo{volume}{D56}},
  \bibinfo{pages}{2950} (\bibinfo{year}{1997}), \eprint{hep-lat/9702015}.

\end{thebibliography}

\clearpage

%
%
%

%
\begin{table}[!h]
\begin{tabular}{cccccc}\hline\hline
$t_K$ & $m_u$ &&& $m_s$ & \# of conf. \\\hline
16, 20 & 0.015 &&& 0.12, 0.18, 0.24 & 371 \\
       & 0.03, 0.04, 0.05 &&& 0.12, 0.18, 0.24, 0.28. 0.35, 0.44 & 252 \\
25 & 0.015, 0.03, 0.04, 0.05 &&& 0.12, 0.18, 0.24, 0.28. 0.35, 0.44 & 100 \\
\hline\hline
\end{tabular}
\caption{
Time slice of kaon operator $t_K$, $u,d$ quark mass $m_u$,
strange quark mass $m_s$, and number of configuration.
}
\label{tab:sim_para}
\end{table}

\begin{table}[!h]
\begin{tabular}{ccccc}\hline\hline
$m_u$ & 0.015 & 0.03 & 0.04 & 0.05 \\\hline
$m_\pi$[GeV] & 
0.35462(97) & 0.4784(10) & 0.54609(94) & 0.60703(89) \\
$E_{\pi\pi}$[GeV] &
0.7233(20) & 0.9698(20) & 1.1043(19) & 1.2254(18) \\
$p$[GeV] &
0.0710(19) & 0.0792(16) & 0.0816(16) & 0.0831(16) \\
$\delta(p)$[deg.] &
$-$2.55(19) & $-$3.43(19) & $-$3.72(19) & $-$3.90(19) \\
$T(p)$ &
$-$0.227(11) & $-$0.367(13) & $-$0.439(14) & $-$0.502(16) \\
$a_0/m_\pi$[1/GeV$^2$] &
$-$1.770(81) & $-$1.580(54) & $-$1.455(47) & $-$1.348(42) \\
$q \cdot ( \partial \phi(q) / \partial q )$ &
0.1215(83) & 0.1601(81) & 0.1724(82) & 0.1801(83) \\
$p \cdot ( \partial \delta(p) / \partial p )$ &
$-$0.0473(32) & $-$0.0611(29) & $-$0.0657(30) & $-$0.0685(31) \\
$F$ &
44.41(48) & 67.56(51) & 81.50(54) & 94.86(59) \\
$F/\overline{F}$ &
0.8708(89) & 0.8456(59) & 0.8363(52) & 0.8306(49) \\
\hline\hline
\end{tabular}
\caption{
Results of CM calculation.
Pion mass $m_\pi$, two-pion energy $E_{\pi\pi}$,
relative momentum $p$, scattering phase shift $\delta(p)$,
scattering length over pion mass $a_0/m_\pi$,
and derivatives $q \cdot ( \partial \phi_{\vec P}(q) / \partial q )$
and $p \cdot ( \partial \delta(p) / \partial p )$ are summarized.
$T(p)$ and $F$ are scattering amplitude and conversion factor, 
which are defined in eqs.(\ref{eq:T}) and (\ref{eq:F}).
$F/\overline{F}$ is ratio of conversion factor to one in non-interacting case.
Definition of $\overline{F}$ is given in eq.(\ref{eq:F0}).
}
\label{tab:twopi_CM}
\end{table}
\begin{table}[!h]
\begin{tabular}{ccccc}\hline\hline
$A$[1/GeV$^2$] & $B$ & $C$[GeV$^2$] & $\chi^2$/d.o.f. & 
$m_\pi^{phys}$ \\\hline
$-$2.10(13) & 2.78(69) & $-$1.93(92) & 0.11 & $-$2.05(11) \\
\hline\hline
\end{tabular}
\caption{
Fit result of scattering length $a_0/m_\pi$ [1/GeV$^2$]
with a quadratic function.
Results at physical pion mass are also listed.
}
\label{tab:fit_a0}
\end{table}
\begin{table}[!h]
\begin{tabular}{ccccc}\hline\hline
$f$[GeV] & $l(\mu)$ & $c_l$ & $\chi^2$/d.o.f. & 
$m_\pi^{phys}$ \\\hline
0.1330(56) & 0.85(14) & 0.68(23) & 0.22 & $-$2.13(15) \\
\hline\hline
\end{tabular}
\caption{
Fit result of scattering length $a_0/m_\pi$ [1/GeV$^2$]
with NLO ChPT formula.
Results at physical pion mass are also listed.
}
\label{tab:fit_a0_chpt}
\end{table}

\begin{table}[!h]
\begin{tabular}{ccccc}\hline\hline
$m_u$ & 0.015 & 0.03 & 0.04 & 0.05 \\\hline
$m_\pi$[GeV] & 
0.35462(97) & 0.4784(10) & 0.54609(94) & 0.60703(89) \\
$\Delta E^P_{\pi\pi}$[GeV] &
0.0250(36) & 0.0213(20) & 0.0196(15) & 0.0183(12) \\
$E^P_{\pi\pi}$[GeV] &
1.0045(39) & 1.2022(26) & 1.3159(22) & 1.4210(20) \\
$p$[GeV] &
0.2456(37) & 0.2575(24) & 0.2618(19) & 0.2649(16) \\
$\gamma$ &
1.1643(16) & 1.10641(54) & 1.08646(33) & 1.07277(23) \\
$\delta(p)$[deg.] &
$-$11.6(1.6) & $-$12.4(1.1) & $-$12.59(92) & $-$12.78(79) \\
$T(p)$ &
$-$0.362(48) & $-$0.462(40) & $-$0.517(36) & $-$0.567(33) \\
$q \cdot ( \partial \phi_{\vec P}(q) / \partial q )$ &
1.878(32) & 2.137(20) & 2.232(16) & 2.300(13) \\
$p \cdot ( \partial \delta(p) / \partial p )$ &
$-$0.261(57) & $-$0.266(42) & $-$0.261(35) & $-$0.253(30) \\
$F^P$ &
48.9(1.2) & 65.8(1.1) & 76.1(1.0) & 86.0(1.0) \\
$F^P/\overline{F}^P$ &
0.869(20) & 0.865(14) & 0.865(11) & 0.866(10) \\
\hline\hline
\end{tabular}
\caption{
Results of Lab calculation.
Pion mass $m_\pi$, energy shift $\Delta E^P_{\pi\pi}$, two-pion energy $E^P_{\pi\pi}$,
relative momentum $p$, boost factor $\gamma$, scattering phase shift $\delta(p)$,
and derivatives $q \cdot ( \partial \phi_{\vec P}(q) / \partial q )$
and $p \cdot ( \partial \delta(p) / \partial p )$ are summarized.
$T(p)$ and $F^P$ are scattering amplitude and conversion factor, 
which are defined in eqs.(\ref{eq:T}) and (\ref{eq:FP}).
$F^P/\overline{F}^P$ is ratio of conversion factor to one in non-interacting case.
Definition of $\overline{F}^P$ is given in eq.(\ref{eq:FP0}).
}
\label{tab:twopi_Lab}
\end{table}
\begin{table}[!h]
\begin{tabular}{cccccc}\hline\hline
$A_{10}$[1/GeV$^2$] & $A_{20}$[1/GeV$^4$] & $A_{30}$[1/GeV$^6$] & $A_{01}$[1/GeV$^2$] & $A_{11}$[1/GeV$^4$] 
& $\chi^2$/d.o.f. \\\hline
$-$2.04(13) & 2.59(71) & $-$1.78(93) & $-$2.42(81) & 3.1(1.8) & 0.34 \\
\hline\hline
\end{tabular}
\caption{
Fit results of scattering amplitude $T(p)$ with polynomial function eq.(\ref{eq:fit_sctamp}).
}
\label{tab:fit_sctamp}
\end{table}

\begin{table}[!h]
\begin{tabular}{ccccccc}\hline\hline
$m_K$[GeV] &\multicolumn{6}{c}{$m_s$} \\
$m_u$ & 0.12 & 0.18 & 0.24 & 0.28 & 0.35 & 0.44 \\\hline
0.015 &
0.7073(12) & 0.8507(13) & 0.9753(15) & ---        & ---        & ---        \\
0.03 &
0.7428(12) & 0.8814(12) & 1.0029(13) & 1.0763(14) & 1.1914(16) & 1.3163(18) \\
0.04 &
0.7666(11) & 0.9023(11) & 1.0221(12) & 1.0946(13) & 1.2086(14) & 1.3324(15) \\
0.05 &
0.7901(10) & 0.9232(11) & 1.0413(11) & 1.1129(12) & 1.2258(12) & 1.3487(13) \\
\hline\hline
\end{tabular}
\caption{
Kaon mass for each light and strange quark masses in $t_K = 20$ case.
}
\label{tab:mk}
\end{table}
\begin{table}[!h]
\begin{tabular}{ccccccc}\hline\hline
$E_K^P$[GeV] &\multicolumn{6}{c}{$m_s$} \\
$m_u$ & 0.12 & 0.18 & 0.24 & 0.28 & 0.35 & 0.44 \\\hline
0.015 &
0.8673(25) & 0.9847(24) & 1.0914(25) & ---        & ---        & ---        \\
0.03 &
0.8948(21) & 1.0111(19) & 1.1168(19) & 1.1818(20) & 1.2856(21) & 1.4002(22) \\
0.04 &
0.9146(18) & 1.0293(17) & 1.1339(16) & 1.1984(17) & 1.3014(17) & 1.4152(18) \\
0.05 &
0.9343(16) & 1.0476(15) & 1.1511(14) & 1.2150(14) & 1.3172(15) & 1.4303(16) \\
\hline\hline
\end{tabular}
\caption{
Kaon energy with momentum $P = 2\pi/L$ for all light and strange quark masses
in $t_K = 20$ case.
}
\label{tab:ek}
\end{table}
\begin{table}[!h]
\begin{tabular}{cccccccc}\hline\hline
$|M_{27}|$[$10^{-3}$] & \multicolumn{6}{c}{$m_s$} & \\
$m_u$ & 0.12 & 0.18 & 0.24 & 0.28 & 0.35 & 0.44 & On-shell \\\hline
0.015 &
1.026(16) & 1.082(18) & 1.118(21) & ---       & ---       & ---       &
1.033(16) \\
0.03 &
1.200(17) & 1.232(18) & 1.246(20) & 1.245(21) & 1.227(22) & 1.174(24) &
1.246(19) \\
0.04 &
1.328(17) & 1.349(18) & 1.353(20) & 1.347(20) & 1.320(21) & 1.255(22) &
1.343(20) \\
0.05 &
1.453(18) & 1.465(19) & 1.462(20) & 1.451(21) & 1.416(21) & 1.340(22) &
1.410(21) \\
\hline\hline
\end{tabular}
\caption{
Decay amplitude of $27$ operator for all light and strange quark masses
in CM frame. Result is in lattice unit.
Results of on-shell amplitude are also included.
}
\label{tab:dcyamp_27_CM}
\end{table}
\begin{table}[!h]
\begin{tabular}{cccccccc}\hline\hline
$|M_{88}|$[$10^{-3}$] & \multicolumn{6}{c}{$m_s$} & \\
$m_u$ & 0.12 & 0.18 & 0.24 & 0.28 & 0.35 & 0.44 & On-shell \\\hline
0.015 &
7.00(12)  & 6.56(11)  & 6.22(12)  & ---       & ---       & ---       &
6.95(11)  \\
0.03 &
4.709(76) & 4.391(75) & 4.133(74) & 3.972(74) & 3.693(74) & 3.320(72) &
4.216(74) \\
0.04 &
3.996(60) & 3.721(58) & 3.494(57) & 3.353(56) & 3.108(56) & 2.786(54) &
3.326(56) \\
0.05 &
3.511(51) & 3.268(48) & 3.065(47) & 2.937(46) & 2.718(45) & 2.430(43) & 
2.705(45) \\
\hline\hline
\end{tabular}
\caption{
Decay amplitude of $88$ operator for all light and strange quark masses
in CM frame. Result is in lattice unit.
Results of on-shell amplitude are also included.
}
\label{tab:dcyamp_88_CM}
\end{table}
\begin{table}[!h]
\begin{tabular}{cccccccc}\hline\hline
$|M_{m88}|$[$10^{-2}$] & \multicolumn{6}{c}{$m_s$} & \\
$m_u$ & 0.12 & 0.18 & 0.24 & 0.28 & 0.35 & 0.44 & On-shell \\\hline
0.015 &
2.548(40) & 2.417(39) & 2.314(40) & ---       & ---       & ---       &
2.532(39) \\
0.03 &
1.837(28) & 1.740(27) & 1.660(27) & 1.608(27) & 1.513(27) & 1.379(26) &
1.687(27) \\
0.04 &
1.632(23) & 1.546(22) & 1.473(21) & 1.425(21) & 1.339(21) & 1.219(20) &
1.415(21) \\
0.05 &
1.499(20) & 1.421(19) & 1.354(18) & 1.309(18) & 1.229(18) & 1.118(17) &
1.224(17) \\
\hline\hline
\end{tabular}
\caption{
Decay amplitude of $m88$ operator for all light and strange quark masses
in CM frame. Result is in lattice unit.
Results of on-shell amplitude are also included.
}
\label{tab:dcyamp_m88_CM}
\end{table}
\begin{table}[!h]
\begin{tabular}{cccccccc}\hline\hline
$|M_{27}^P|$[$10^{-3}$] & \multicolumn{6}{c}{$m_s$} & \\
$m_u$ & 0.12 & 0.18 & 0.24 & 0.28 & 0.35 & 0.44 & On-shell \\\hline
0.015 &
1.31(11)  & 1.37(11)  & 1.41(12)  & ---       & ---       & ---       &
1.38(11)  \\
0.03 &
1.500(70) & 1.545(70) & 1.562(72) & 1.559(74) & 1.531(76) & 1.457(77) &
1.554(74) \\
0.04 &
1.630(56) & 1.679(56) & 1.696(57) & 1.693(58) & 1.663(59) & 1.583(59) &
1.651(59) \\
0.05 &
1.774(49) & 1.824(49) & 1.841(49) & 1.837(50) & 1.802(51) & 1.715(51) &
1.726(51) \\
\hline\hline
\end{tabular}
\caption{
Decay amplitude of $27$ operator for all light and strange quark masses
in Lab frame. Result is in lattice unit.
Results of on-shell amplitude are also included.
}
\label{tab:dcyamp_27_Lab}
\end{table}
\begin{table}[!h]
\begin{tabular}{cccccccc}\hline\hline
$|M_{88}^P|$[$10^{-3}$] & \multicolumn{6}{c}{$m_s$} & \\
$m_u$ & 0.12 & 0.18 & 0.24 & 0.28 & 0.35 & 0.44 & On-shell \\\hline
0.015 &
6.31(56) & 6.19(48) & 6.05(45) & ---      & ---       & ---       &
6.16(48)  \\
0.03 &
4.41(24) & 4.26(22) & 4.09(21) & 3.96(20) & 3.72(20)  & 3.38(19)  &
3.92(20)  \\
0.04 &
3.85(16) & 3.70(15) & 3.54(14) & 3.43(14) & 3.21(13)  & 2.90(13)  &
3.16(13)  \\
0.05 &
3.47(12) & 3.33(11) & 3.18(10) & 3.07(10) & 2.870(97) & 2.586(92) &
2.614(93) \\
\hline\hline
\end{tabular}
\caption{
Decay amplitude of $88$ operator for all light and strange quark masses
in Lab frame. Result is in lattice unit.
Results of on-shell amplitude are also included.
}
\label{tab:dcyamp_88_Lab}
\end{table}
\begin{table}[!h]
\begin{tabular}{cccccccc}\hline\hline
$|M_{m88}^P|$[$10^{-2}$] & \multicolumn{6}{c}{$m_s$} & \\
$m_u$ & 0.12 & 0.18 & 0.24 & 0.28 & 0.35 & 0.44 & On-shell \\\hline
0.015 &
2.44(19)  & 2.42(16)  & 2.38(15)  & ---       & ---       & ---       &
2.41(16)  \\
0.03 &
1.843(84) & 1.808(75) & 1.757(70) & 1.717(68) & 1.632(65) & 1.503(62) &
1.701(67) \\
0.04 &
1.688(58) & 1.652(51) & 1.603(48) & 1.564(47) & 1.484(45) & 1.363(43) &
1.467(44) \\
0.05 &
1.595(45) & 1.559(40) & 1.511(37) & 1.473(36) & 1.396(35) & 1.280(33) &
1.292(33) \\
\hline\hline
\end{tabular}
\caption{
Decay amplitude of $m88$ operator for all light and strange quark masses
in Lab frame. Result is in lattice unit.
Results of on-shell amplitude are also included.
}
\label{tab:dcyamp_m88_Lab}
\end{table}
\begin{table}[!h]
\begin{tabular}{cccccccccc}\hline\hline
& \multicolumn{3}{c}{CM} &&& \multicolumn{3}{c}{Lab} \\
$m_u$ & 
$|A_{27}|$[GeV$^3$] & $|A_{88}|$[GeV$^3$] & $|A_{m88}|$[GeV$^3$] &&&
$|A_{27}|$[GeV$^3$] & $|A_{88}|$[GeV$^3$] & $|A_{m88}|$[GeV$^3$] \\\hline
0.015 &
0.1031(20) & 0.694(14) & 2.528(48) &&& 0.151(12) & 0.676(53) & 2.65(18)  \\
0.03 &
0.1892(33) & 0.640(13) & 2.562(47) &&& 0.230(12) & 0.579(30) & 2.51(10)  \\
0.04 &
0.2461(41) & 0.609(11) & 2.593(44) &&& 0.283(11) & 0.541(23) & 2.510(83) \\
0.05 &
0.3007(50) & 0.577(10) & 2.609(41) &&& 0.334(11) & 0.506(19) & 2.499(71) \\
\hline\hline
\end{tabular}
\caption{
Decay amplitudes in infinite volume for $27, 88, m88$ operators in CM and Lab calculations.
}
\label{tab:dcyamp_infinit_v}
\end{table}
\begin{table}[!h]
\begin{tabular}{cccccccccc}\hline\hline
& \multicolumn{3}{c}{CM} &&& \multicolumn{3}{c}{Lab} \\
$m_u$ & 
$|A_{27}^{\rm RI}|$[GeV$^3$] & $|A_{88}^{\rm RI}|$[GeV$^3$] & 
$|A_{m88}^{\rm RI}|$[GeV$^3$] &&&
$|A_{27}^{\rm RI}|$[GeV$^3$] & $|A_{88}^{\rm RI}|$[GeV$^3$] & 
$|A_{m88}^{\rm RI}|$[GeV$^3$] \\\hline
0.015 &
0.0858(17) & 0.4782(96) & 2.382(46) &&& 0.126(10)  & 0.456(38) & 2.50(17)  \\
0.03 &
0.1575(28) & 0.4288(88) & 2.419(44) &&& 0.1913(97) & 0.377(22) & 2.378(98) \\
0.04 &
0.2048(34) & 0.3992(78) & 2.451(41) &&& 0.2351(90) & 0.343(17) & 2.376(78) \\
0.05 &
0.2502(41) & 0.3693(70) & 2.470(39) &&& 0.2778(89) & 0.312(14) & 2.369(67) \\
\hline\hline
\end{tabular}
\caption{
Weak matrix elements in regularization independent scheme
for $27, 88, m88$ operators in CM and Lab calculations.
}
\label{tab:dcyamp_ri}
\end{table}
\begin{table}[!h]
\begin{tabular}{cccccc}\hline\hline
opr. &$B_{00}$[GeV$^3$] & $B_{10}$[GeV] & $B_{01}$[GeV] & $\chi^2$/d.o.f. 
& phys.[GeV$^3$] \\\hline
27&
---       & 0.6687(95)    & 0.48(11)       & 1.26 & 0.0336(47) \\
88&
0.534(11) & $-$0.429(22) & $-$0.86(12)     & 0.22 & 0.490(11) \\
$m88$&
2.391(49) & 0.14(10)     & $-$1.89(63)     & 2.48 & 2.313(55) \\
\hline\hline
\end{tabular}
\caption{
Fit results for weak matrix elements of each operator 
in regularization independent scheme,
$|A_i^{\mathrm{RI}}|$ for $i=27,88,m88$.
Fit function is defined in eq.(\ref{eq:fit_func}).
Results at physical point, $m_\pi=140$ MeV and $p=206$ MeV, are also
tabulated.
}
\label{tab:fit_dcyamp_ri}
\end{table}
\begin{table}[!h]
\begin{tabular}{cccccc}\hline\hline
opr. &$\alpha_{27}/f^4$ & $\beta_{20}$[1/GeV$^2$] & $\beta_{11}$[1/GeV$^2$] & $\chi^2$/d.o.f. 
& phys.[GeV$^3$] \\\hline
27(Full)&
$-$0.0391(15) & $-20.4$(6.3) & 8.5(1.5) & 0.02 & 0.0331(13) \\
27(Quen.)&
$-$0.0509(20) & $-11.7$(4.8) & 5.2(1.2) & 0.08 & 0.0374(15) \\
\hline
opr. &$\alpha_{i}/f^6$ & $\gamma_{10}$[1/GeV$^2$] & $\gamma_{01}$[1/GeV$^2$] & $\chi^2$/d.o.f. 
& phys.[GeV$^3$] \\\hline
88(Full)&
$-$5.17(19)  & $-$0.51(69) &    0.04(16) & 1.44 & 0.432(20) \\
88(Quen.)&
$-$8.20(31)  &    0.28(44) & $-$0.30(10) & 5.39 & 0.554(24) \\
$m$88(Full)&
$-$22.37(90) & $-$0.78(70) &    0.38(16) & 2.33 & 2.039(94) \\
$m$88(Quen.)&
$-$35.5(1.5) &    0.15(44) & $-$0.09(10) & 7.24 & 2.56(11)  \\
\hline\hline
\end{tabular}
\caption{
Fit results with full and quenched NLO ChPT formulae
for weak matrix elements of each operator in regularization independent scheme,
$|A_i^{\mathrm{RI}}|$ for $i=27,88,m88$.
Fit functions are defined in appendix.
Results at physical point, $m_\pi=140$ MeV and $p=206$ MeV, are also
tabulated.
}
\label{tab:fit_dcyamp_ri_chpt}
\end{table}
\clearpage
\begin{table}[!h]
\begin{tabular}{cccccc}\hline\hline
&$a^{-1}$[GeV] & $f$[GeV]
&$\alpha_{27}/f^4$
&$\alpha_{88}/f^6$
&$\alpha_{m88}/f^6$\\\hline
direct(Lin.) &1.31&0.133&
$-$0.0809(11)&
$-$5.52(11)&
$-$24.68(51)\\
direct(ChPT) &1.31&0.133&
$-$0.0391(15)&
$-$5.17(19)&
$-$22.37(90)\\
indirect &1.92&0.137&
$-$0.0306(13)&
$-$5.89(30)&
$-$21.4(1.2)\\\hline\hline
\end{tabular}
\caption{
Comparison of dimensionless constant of LO ChPT.
This work and previous work~\cite{Blum:2001xb} are denoted by
``direct'' and ``indirect'', respectively.
}
\label{tab:lochpt_constants}
\end{table}
\begin{table}[!h]
\begin{tabular}{lc}\hline\hline
$G_F$ & 1.166$\times$10$^{-5}$ GeV$^{-2}$\\
$|V_{us}|$ & 0.2237 \\
$|V_{ud}|$ & 0.9747 \\
$|V_{ts}|$ & 0.0410 \\
$V_{td}$   & 0.00708$-$0.00297$i$ \\
$\tau$     & 0.00133$-$0.000559$i$ \\
\hline\hline
\end{tabular}
\caption{
Parameters for weak matrix element~\cite{Blum:2001xb}.
}
\label{tab:input_param}
\end{table}
\begin{table}[!h]
\begin{tabular}{cr@{.}lr@{.}l}\hline\hline
 $i$ & \multicolumn{2}{c}{$z_i(\mu)$} & \multicolumn{2}{c}{$y_i(\mu)$}\\ \hline
  1  & $-$0&3522       &    0&0 \\
  2  &    1&17721      &    0&0 \\
  3  &    0&00446831   &    0&0241094 \\
  4  & $-$0&0140925    & $-$0&0503954 \\
  5  &    0&00506909   &    0&00563178 \\
  6  & $-$0&015967     & $-$0&0928098 \\
  7  &    0&0000502692 & $-$0&000186283 \\
  8  & $-$0&0000134347 &    0&00118057 \\
  9  &    0&0000428969 & $-$0&0114749 \\
 10  &    0&0000117198 &    0&0037748 \\
\hline\hline
\end{tabular}
\caption{
Results of Wilson coefficients~\cite{Kim:2004mk}.
}
\label{tab:wilson_coef}
\end{table}
\begin{table}[!h]
\begin{tabular}{ccccc}\hline\hline
$m_u$ & 0.015     & 0.03     & 0.04      & 0.05      \\ \hline
CM    & 4.241(81) & 7.78(14) & 10.12(17) & 12.37(20) \\
Lab   & 6.22(50)  & 9.45(48) & 11.62(45) & 13.74(44) \\
\hline\hline
\end{tabular}
\caption{Results of Re$A_2$[$10^{-8}$GeV] in CM and Lab calculations.
}
\label{tab:rea2}
\end{table}
\begin{table}[!h]
\begin{tabular}{cccccc}\hline\hline
Poly. &
$B_{10}$[$10^{-7}$/GeV] &
$B_{01}$[$10^{-7}$/GeV] &
$B_{11}$[$10^{-7}$/GeV$^3$] &
$\chi^2$/d.o.f.                      &
phys.[$10^{-8}$GeV]           \\\hline
&3.306(47) & 2.39(56) & ---         & 1.26 & 1.66(23) \\
&3.304(47) & 3.60(99) & $-4.0(2.7)$ & 0.90 & 2.14(40) \\ \hline
ChPT &{$12\sqrt{3}\alpha_{_{\mathrm{Re}A_2}}/f^3$[$10^{-8}$/GeV]} & $\beta_{20}$[1/GeV$^2$] &
$\beta_{11}$[1/GeV$^2$] & $\chi^2$/d.o.f. & phys.[$10^{-8}$GeV]\\\hline
Full&
{$-$5.33(21)} & $-$20.4(6.3) & 8.5(1.5) & 0.02 & 1.636(65) \\
Quen.&
{$-$6.93(27)} & $-$11.7(4.8) & 5.2(1.2) & 0.08 & 1.847(72) \\
\hline\hline
\end{tabular}
\caption{
Fit results with polynomial and NLO full (quenched) ChPT formula of Re$A_2$,
and result at physical point, $m_\pi=140$ MeV and $p=206$ MeV.
}
\label{tab:fit_rea2}
\end{table}
\begin{table}[!h]
\begin{tabular}{ccccc}\hline\hline
$m_u$ & 0.015        & 0.03        & 0.04        & 0.05        \\ \hline
CM    & $-$1.036(22) & $-$0.786(19) & $-$0.624(17) & $-$0.462(16) \\
Lab   & $-$0.952(90) & $-$0.635(53) & $-$0.468(42) & $-$0.301(35) \\
\hline\hline
\end{tabular}
\caption{Results of Im$A_2$[$10^{-12}$GeV] in CM and Lab calculations.
}
\label{tab:ima2}
\end{table}
\begin{table}[!h]
\begin{tabular}{ccccccc}\hline\hline
Poly. &
$B_{00}$[$10^{-12}$GeV] &
$B_{10}$[$10^{-12}$/GeV] &
$B_{01}$[$10^{-12}$/GeV] &
$B_{11}$[$10^{-12}$/GeV$^3$] &
$\chi^2$/d.o.f.                      &
phys.[$10^{-12}$GeV]           \\\hline
&
$-$1.333(25) & 2.307(56) & 2.54(36) & ---      & 0.15 & $-$1.181(26) \\
&
$-$1.332(25) & 2.302(57) & 2.2(1.0) & 0.7(1.9) & 0.15 & $-$1.194(43) \\
\hline
ChPT &\multicolumn{2}{c}{$24\sqrt{3}\alpha_{_{\mathrm{Im}A_2}}/f^3$[$10^{-12}$GeV]} & $\gamma_{10}$[1/GeV$^2$] &
$\gamma_{01}$[1/GeV$^2$] & $\chi^2$/d.o.f. & phys.[$10^{-12}$GeV]\\\hline
&
\multicolumn{2}{c}{1.250(42)} & $-$1.24(66) & $-$0.02(16) & 1.24 & $-$1.040(46) \\
\hline\hline
\end{tabular}
\caption{
Fit results with polynomial and NLO ChPT formula of Im$A_2$,
and result at physical point, $m_\pi=140$ MeV and $p=206$ MeV.
}
\label{tab:fit_ima2}
\end{table}

\clearpage

%
%
%

\clearpage

\end{document}